\documentclass[11pt,onecolumn,showpacs,floatfix]{revtex4}
\usepackage{bm}% bold math
\usepackage{ifpdf}

\ifpdf
\usepackage[pdftex]{graphicx}
\usepackage{epstopdf}
\else
\usepackage{graphicx}
\fi

\usepackage[center]{subfigure}
\usepackage{amsmath}
\usepackage{amssymb}
\usepackage[ruled]{algorithm}
\usepackage{algorithmic}

\graphicspath{{./Figures/}{./algorithms/}}

\begin{document}
\bibliographystyle{apsrev}

\title{Adaptive mesh computation of polycrystalline pattern formation
using a renormalization-group reduction of the phase-field crystal model}

\author{Badrinarayan P. Athreya$^1$,
Nigel Goldenfeld$^2$, Jonathan A. Dantzig$^1$, Michael
Greenwood$^3$, and Nikolas Provatas$^3$}

\affiliation{$^1$Department of Mechanical Science and Engineering,
University of Illinois at Urbana-Champaign, 1206 W.~Green Street,
Urbana, IL 61801, USA\\
$^2$Department of Physics, University of Illinois at Urbana-Champaign,
1110 W.~Green Street, Urbana, IL 61801, USA\\
$^3$Department of Materials Science and Engineering, McMaster
University, 1280 Main Street West, Hamilton, Ontario, L8S 4L7,
Canada}

\begin{abstract}

We implement an adaptive mesh algorithm for calculating the space and
time dependence of the atomic density field during materials
processing.  Our numerical approach uses the systematic
renormalization-group formulation of the phase field crystal model to
provide the underlying equations for the complex amplitude of the
atomic density field---a quantity that is spatially uniform except near
topological defects, grain boundaries and other lattice imperfections.
Our algorithm is a hybrid formulation of the amplitude equations,
combining Cartesian and polar decompositions of the complex amplitude.
We show that this approach leads to an acceleration by three orders of
magnitude in model calculations of polycrystalline domain formation in
two dimensions.

\end{abstract}

%Relevant Pacs numbers (up to four allowed, in order of relevance)
%
%05.70.Ln Nonequilibrium and irreversible thermodynamics
%81.16.Rf Nanoscale pattern formation
%05.10.Cc Renormalization group methods
%81.15.Aa Theory and models of film growth
%61.72.Cc Kinetics of defect formation and annealing
%64.60.Cn Order-disorder transformations; statistical mechanics of model systems

%%%%%%%%%%%%%%%%%%%%%%%%%%%%%%%%%%%%%%%%%%%%%%%%%%%%%%%%
 \maketitle
\section{Introduction}

A fundamental theoretical and computational challenge in materials
modeling is that of concurrently treating phenomena over a wide range
of length and time scales. For example, in studying the mechanical
response of polycrystalline materials, one must take into account the
dynamics and interactions of vacancies, impurities, dislocations and
grain boundaries, on time scales ranging from atomic vibrations to
system-wide diffusion times.

Numerous approaches to handling the wide range of length scales have
been proposed \cite{Phillipsbook}, including quasi-continuum methods
\cite{Tadmor, Shenoy, Ortiz, Miller}, the heterogeneous multiscale
method \cite{Weinan1, Weinan2}, multi-scale molecular dynamics
\cite{Rudd, Kaxiras, Robbins, CURT02}, multigrid variants \cite{Fish}
and phase field models \cite{Langer86,Karma,Beckermann1,warren03}.  In
general one can classify different techniques as being either atomistic
or continuum, and differentiate them further by the characteristic time
scale: density functional theory (DFT), for a quantum mechanical
description of processes at the atomic time scale; molecular dynamics
(MD) or Monte Carlo (MC) methods, appropriate for collective dynamics;
and coarse-grained descriptions involving continuum fields at the
mesoscale on diffusive time scales. The difficulty of merging
descriptions at different length and time scales limits the effective
application of most of these methods. Lack of a continuous transition
between scales can induce artifacts, such as spurious reflections in a
transition region between two levels \cite{VVED04, Weinan2}. Further,
any method using molecular dynamics is typically restricted to
sub-nanosecond time scales, whereas many interesting phenomena during
materials processing, such as microstructural pattern formation,
recrystallization, heat and solute diffusion, dislocation glide, etc.,
occur over time scales which are typically greater than $10^{-6}$s.

One continuum approach that has been used successfully, especially in
the multiscale modeling of solidification problems
\cite{provatasreview2005}, is the \emph{phase-field method}
\cite{Langer86}. Through the effective use of asymptotics \cite{Karma}
and adaptive mesh refinement \cite{Provatas1998,Jeong2}, the
phase-field method has been used to span several orders of magnitude in
length, from microns to centimeters. Extensions of the method by
Kobayashi and co-workers \cite{Kobayashi98,Kobayashi00}, and Warren
\cite{warren03} also make it possible to model polycrystalline
systems. Special forms of the free energy that incorporate strain
energy have been used to model the qualitative features of
strain-induced phase transformations
\cite{onuki89_1,onuki89_2,mg99,kmmkk01,karmafracture2001,Haa05}.  The
phase-field method represents a coarse-graining in space to length
scales much greater than those of the interfaces and defects of
interest in this work. As a result, the kinetic coefficients that
emerge in the final continuum equations are phenomenological, and can
be related to experimentally-measurable parameters only after a
suitable asymptotic matching of the phase field equations with
corresponding sharp-interface models
\cite{Karma2001,Echebarria04,provatasreview2005}. As such, traditional
phase field models do not fundamentally embody the emergent kinetic and
elasto-plastic mechanisms that originate at the atomic scale.  Perhaps
the most important limitation of phase field models is that, in
general, they do not preserve any record of the underlying crystal
lattice, so that ad hoc approaches must be used to model the variety of
phenomena which result from lattice interactions.

The phase field crystal (PFC) model \cite{ekhg02,eg04} is a promising
extension of the phase field model approach, in which the equilibrium
configuration of an atomic density field is constructed to be periodic,
rather than uniform in space.  The conserved dynamics of the PFC model
then naturally reproduce many of the non-equilibrium processing
dynamics arising in real polycrystalline materials. The PFC model is
founded on the insight that a free energy functional that is minimized
by a periodic field natively includes elastic energy, anisotropy and
symmetry properties of that field. Thus the model naturally
incorporates all properties of a crystal that are determined by
symmetry, as well as vacancies, dislocations, and other defects.
Moreover, the PFC model represents the evolution of the system over a
time scale that is much longer than the vibrational period of atoms
(${\cal{O}}(10^{-15}\mbox{s})$), but much shorter than the time scale
of diffusive processes in the system, such as the viscous glide of
dislocations, which typically occur over a time scale of
${\cal{O}}(10^{-6}\mbox{s})$. The PFC model yields a relatively simple
and well-behaved partial differential equation (PDE) for the evolution
of the time-averaged  density, giving it access to phenomena occurring
on atomic length scales, but over diffusive time scales. The PFC method
is thus able to capture atomic-scale elasticity and the interaction of
topological defects on the same time scales that govern diffusive
processes during phase transformations in pure materials
\cite{eg04,BGE05,Stefanovic06} and alloys \cite{Eld06}.

As with any model that resolves at the atomic scale, the PFC model is
limited in its ability to model systems of realistic dimensions,
because the computational grid must resolve the periodicity of the
field. For grid converged results, a minimum of 9 grid points per
period are required. In a physical system, the periodicity represents
interatomic distance, ${\cal{O}}(10^{-10}{\rm m})$. Thus, to simulate a
system having a characteristic dimension of one micron would require
about $10^5$ degrees of freedom per spatial dimension on a uniform
computational mesh. This would be a heroic computation in 2-D, and well
beyond reach in 3D, even with the use of massive parallelization.
Furthermore, the periodic lattice precludes the effective use of
adaptive mesh refinement (AMR) algorithms.

The first three authors have recently described an approach to overcome
this difficulty \cite{GAD05_1,GAD05_2}, using the perturbative
renormalization group (RG) method \cite{CGO2,Nozaki01} to
systematically coarse-grain the PFC equation\cite{AGD06_1}.  The basic
idea is to obtain renormalization group equations of motion for the
complex amplitude of the periodic density field, a quantity whose
modulus and phase are spatially uniform except near regions of lattice
disruption, such as at grain boundaries and at topological defects.
From the complex amplitude, it is possible to reconstruct the
atomic-scale density field at least within the one-mode approximation,
and to compute non-trivial materials properties and dynamics to high
accuracy (within one percent)\cite{GAD05_1, GAD05_2}. This approach,
which we will sometimes denote as the PFC-RG method, is much faster
than solving the PFC equation directly, because the complex amplitude
varies on much larger spatial length scales than the density itself,
thus permitting the use of an adaptively-generated coarse mesh over
much of the computational domain\cite{GAD05_1}.  It is important to
appreciate that the equations of motion for the complex amplitude must
be rotationally-covariant, in order that a polycrystalline material or
heterogeneous microstructure can be represented without any preferred
orientations imposed; this is readily achieved using renormalization
group methods\cite{AGD06_1}.  However, in a practical computation, the
reciprocal lattice vectors of the equilibrium crystal structure are
represented within a particular basis, and there is the potential for
interference between the density Fourier components and the
basis\cite{GAD05_1}, giving rise to artifactual \lq\lq fringes" or
\lq\lq beats" in the corresponding Fourier components.  The overall
density does not, of course, exhibit these interference fringes, but
their presence in the individual components means that to be properly
resolved, an adaptive mesh algorithm must generate grid refinement in
their vicinity.  As a result, efficient computation becomes
compromised.

The purpose of this paper is to develop a computationally-efficient
formulation of the PFC-RG method that enables the implementation of an
AMR algorithm up to micro- and meso- length scales, without being
deflected by artifacts arising from the choice of basis set.  The
approach is to use a hybrid representation of the complex amplitude,
switching between Cartesian and polar coordinates as appropriate in a
seamless fashion to avoid beating and coordinate singularities.  The
resultant description is fast, accurate and provides mesh refinement
and coarsening in the physically correct locations, without artifacts
arising from choice of basis or other implementation-dependent details.
As such, our work represents a first step towards providing a
systematic description of materials processing using continuum fields
across all relevant length scales.

The remainder of this paper is organized as follows: We introduce the
complex amplitude equations (interchangeably called the RG equations)
in Section \ref{sec:cartesian_rg} and illustrate the interference or
beat problem in the Cartesian representation of these equations that
limits the effectiveness of AMR techniques. In Section
\ref{sec:polar_rg} we introduce a polar formulation of the equations
that addresses the problem of beats, but also exhibits coordinate
singularities that make these equations unwieldy for numerical
solution. We then present a new hybrid formulation in Section
\ref{sec:hybrid_rg}, which is a procedure for solving the Cartesian
equations of Section \ref{sec:cartesian_rg} concurrently with a reduced
form of the polar equations of Section \ref{sec:polar_rg} in different
parts of the computational domain. In Section \ref{sec:AMR} the hybrid
formulation of the RG equations is demonstrated to be amenable to
solution using a new finite-difference-based AMR algorithm specifically
developed for our RG equations. Section \ref{sec:results} presents
numerical simulations and results, including efficiency benchmarks that
clearly demonstrate the computational advantage of our AMR-RG approach.
Section \ref{sec:conc} concludes and presents directions for future
work.

%%%%%%%%%%%%%%%%%%%%%%%%%%%%%%%%%%%%%%%%%%%%%%%%%%%%%%%%%%%%%%%%%%
\section{\label{sec:cartesian_rg} The complex amplitude equations}

\subsection{Governing equations}

In the PFC model, the evolution of the density $\rho$ is given by
\begin{equation}
\label{eq:eom} \frac{\partial \rho}{\partial t} =
\Gamma \nabla^2 \left( {\delta {\cal F}
\over \delta \rho} \right) + \eta
\end{equation}
where ${\cal F}$ is the free energy functional, which can be written as
${\cal F} = \int d\vec{r} \left(f\left(\rho,\nabla^2\rho,\ldots
\right)\right)$, where $f$ is the local free energy density, $\Gamma$ is a
constant and $\eta$ is a stochastic noise with zero mean and correlations
$\langle \eta(\vec{r},t) \eta(\vec{r}',t') \rangle =-\Gamma k_B T
\nabla^2\delta(\vec{r}-\vec{r}')\delta(t-t')$. The specific form of ${\cal
F}$ is chosen such that at high temperatures ${\cal F}$ is minimized by a
spatially uniform liquid state, and at low temperatures by a spatially
periodic ``crystalline'' phase. Furthermore, $f$ must be chosen such that
${\cal F}$ is independent of crystal orientation. These constraints
naturally incorporate both elastic and plastic deformations.

A free energy form that satisfies these criteria naturally produces
mobile regions of liquid/solid coexistence separated by free surfaces,
i.e., phase transformations. Elastic energy and defects in the
crystalline phase arise from the requirement that ${\cal F}$ be
minimized by a spatially periodic density field that is independent of
crystal orientation. Elder et al. \cite{ekhg02,eg04} demonstrated these
properties of the model for a variety of applications, including
studies of grain boundary energy, liquid phase epitaxial growth  and
the yield strength of nanocrystalline materials. The particular model
they used made the following choice for the function $f$:
\begin{equation}
\label{eq:simpH}
f = \rho\left(\alpha\Delta T +
\lambda\left(q_o^2+\nabla^2\right)^2\right)\rho/2+u\rho^4/4
\end{equation}
where $\alpha$, $\lambda$, $q_o$ and $u$ are model parameters that can
be specified to match some specific material properties, such as
Young's modulus and lattice spacing \cite{ekhg02,eg04}. In order to
discuss the dynamical behavior of the PFC model, it is useful to
rewrite the free energy in dimensionless units: $\vec{x} \equiv
\vec{r}q_o$, $\psi \equiv \rho \sqrt{u/\lambda q_o^4}$, $r\equiv
a\Delta T/\lambda q_o^4$, $\tau \equiv  \Gamma\lambda q_o^6 t$ and
$F\equiv {\cal F}u/\lambda^2q_o^{8-d}$ so that
\begin{equation}
\label{eq:simpF}
F = \int d\vec{x} \left[\psi\left(r+(1+\nabla^2)^2\right)\psi/2
+ \psi^4/4\right]
\end{equation}
In these units the conservation law of Eq.~(\ref{eq:simpH}) becomes
\begin{equation}
\label{eq:dyn}
\frac{\partial \psi}{\partial t} =
\nabla^2\left[\left(r+(1+\nabla^2)^2\right)\psi
+\psi^3\right]+\zeta
\end{equation}
where $\langle \zeta(\vec{r}_1,t_1)\zeta(\vec{r}_2,\tau_2) \rangle = {\cal
E} \nabla^2\delta(\vec{r}_1-\vec{r}_2)\delta(\tau_1-\tau_2)$ and ${\cal E}
\equiv  uk_BTq_o^{d-4}/\lambda^2$.   Eq.~(\ref{eq:dyn}), introduced by Elder
et al \cite{ekhg02,eg04}, will be referred to as the PFC equation in what
follows.  This equation can be used in any dimension by simply introducing
the appropriate form for the Laplacian operators.

The spatial density $\psi$ can be approximated in terms of the complex
amplitude $A_j$ as
\begin{equation}
\label{eq:1mode}
\psi\approx\sum_{j=1}^3A_{j}e^{i\mathbf{k}_j\cdot\mathbf{x}} +
\sum_{j=1}^3A_{j}^*e^{-i\mathbf{k}_j\cdot\mathbf{x}} + \bar{\psi},
\end{equation}
where
\begin{eqnarray}\label{eq:latticevectors}
\mathbf{k}_1 &=& k_0 ({-\vec{\mbox{i}}\sqrt{3}/2} -
{\vec{\mbox{j}}/2})\nonumber\\
\mathbf{k}_2 &=& k_0{{\vec{\mbox{j}}}}\nonumber\\
\mathbf{k}_3 &=& k_0 ({\vec{\mbox{i}}\sqrt{3}/2} -
{\vec{\mbox{j}}/2})
\end{eqnarray}
are the reciprocal lattice vectors of a crystal with hexagonal
symmetry, and $k_0$ is the dominant wavenumber of the pattern. For
all the calculations shown in this paper length has been scaled such
that $k_0=1$, which corresponds to an interatomic spacing of
$a_0=2\pi/(\sqrt{3}/2)$.
The complex amplitude equations, which constitute a coarse-grained
approximation to the PFC equation were shown in our earlier work
\cite{GAD05_1,GAD05_2} to be given by
\begin{equation}\label{eq:cmplx_rg}
\frac{\partial A_j}{\partial t} = \widetilde{\mathcal{L}}_{j}A_j -
3A_j|A_j|^2-6A_j\sum_{k:k\ne j}|A_k|^2-6\bar{\psi}\prod_{k:k\ne
j}A_k^*
\end{equation}
where $j,k\in[1,3]$ and
\begin{equation} \label{eq:cmplx_rg_operator} \widetilde{\mathcal{L}}_{j}
= \left[1 - \nabla^2 -
2i{\mathbf{k}_{j}}\cdot\nabla\right] \left[-r-3\bar{\psi}^2-
\left\{\nabla^2 + 2i{\mathbf{k}_{j}}\cdot\nabla\right\}^2\right]
\end{equation}
is a rotationally covariant operator. The superscript ``$*$'' denotes
complex conjugation. The parameters $r\;(\le0)$ and $\bar{\psi}\;(\ge0)$
control the bifurcation from a uniform liquid phase to a crystalline phase
with hexagonal symmetry. Specifically, $r$ is proportional to the
temperature quench from a critical temperature $T_c$, while $\bar{\psi}$
is the mean density in the system. We refer to this form as the
\emph{Cartesian} representation because the amplitudes are expressed along
each coordinate direction.

\begin{figure}[htb]
\begin{center}
\subfigure[\label{fig:ramp} $\Re(A_1)$]
{\includegraphics[height=2.5in,angle=0]{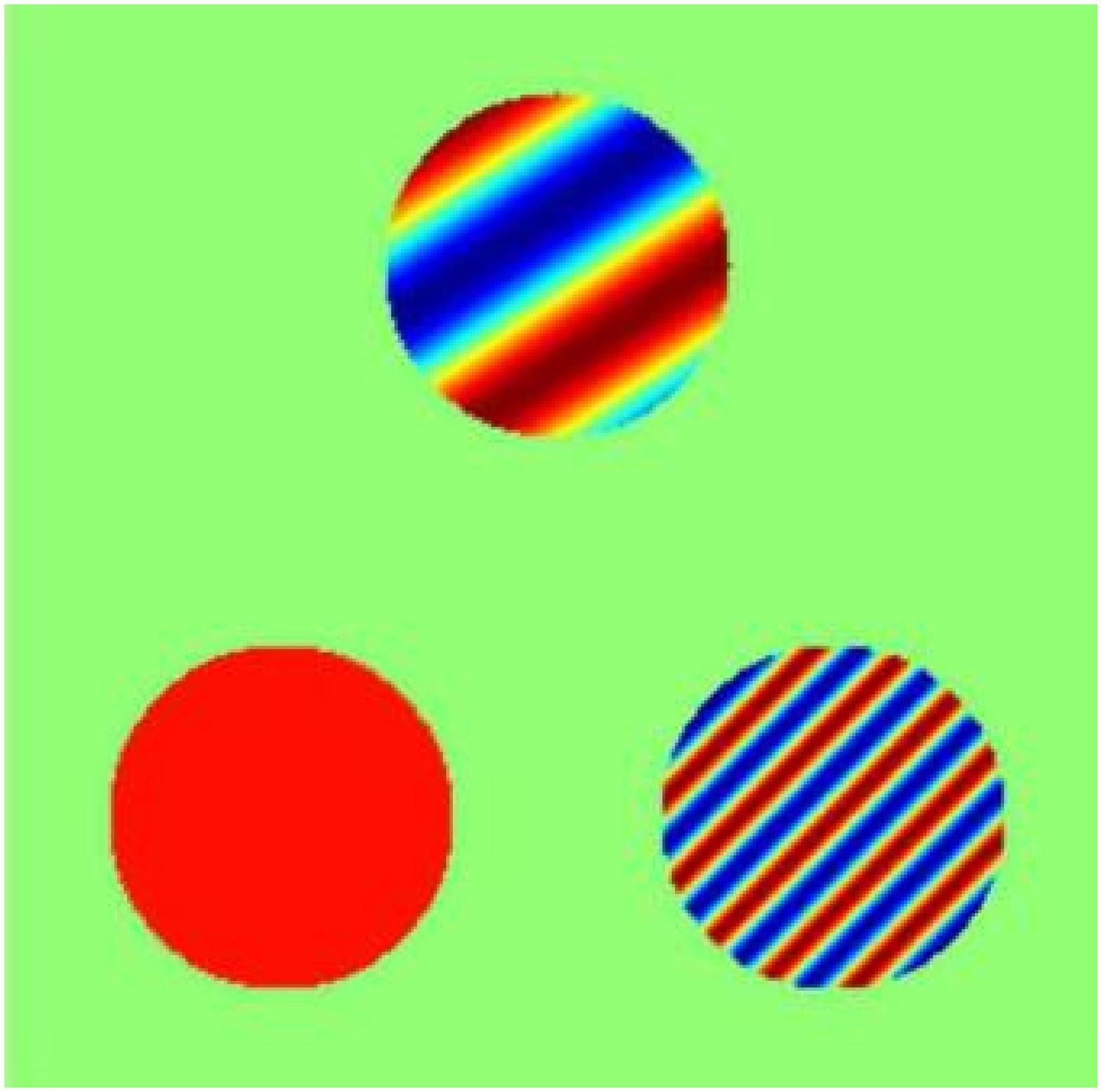}}
\hspace{5mm}
\subfigure[\label{fig:psi} $\psi$]
{\includegraphics[height=2.5in,angle=0]{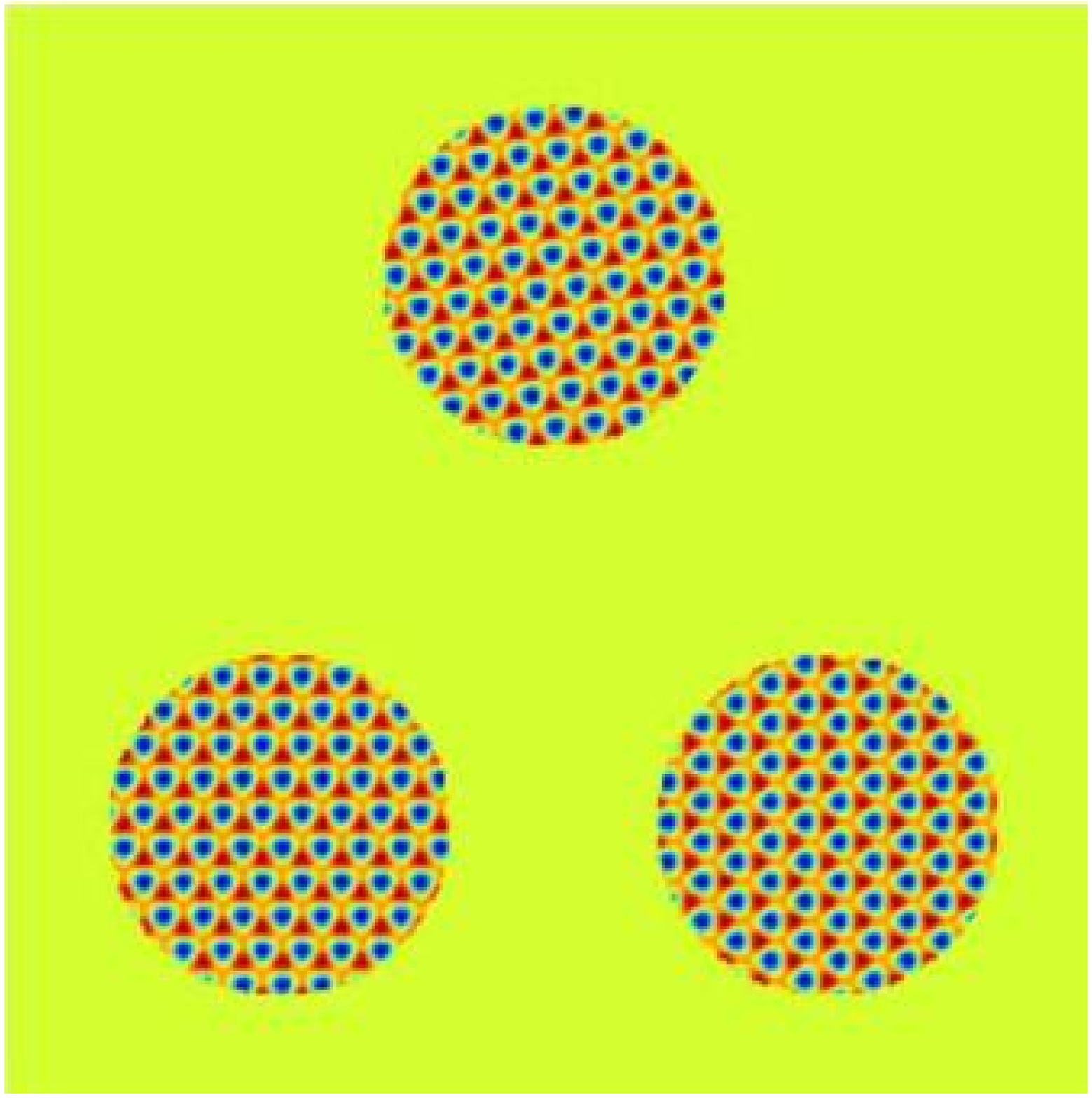}}
\caption{\label{fig:beats} (a) Real component of the complex
amplitude $A_1$. As the grain in the bottom-left corner is aligned
with the basis $\mathbf{k}_j$ in Eq.~(\ref{eq:latticevectors}) its
amplitude is constant, while amplitudes of the remaining misoriented
grains have ``beats''. (b) Density field $\psi$ reconstructed using
Eq.~(\ref{eq:1mode_theta3}). Clockwise from the lower left corner,
$\theta=0$, $\pi/24$ and $\pi/6$.}
\end{center}
\end{figure}

The rotational covariance of the operator $\widetilde{\cal{L}}$ permits
the incorporation of multiple crystal orientations using only the basis
vectors in Eq.~(\ref{eq:latticevectors}). To see this consider a density
field $\psi$ defined by Eq.~(\ref{eq:1mode}) with triangular lattice basis
vectors $\mathbf{k}_j(\theta)$ (where $|\mathbf{k}_j(\theta)|=1$) that are
rotated by an angle $\theta$ from the basis vectors $\mathbf{k}_j$ in
Eq.~(\ref{eq:latticevectors}), i.e.
\begin{equation}
\label{eq:1mode_theta1}
\psi(\theta)=\sum_{j=1}^3A_{j}e^{i\mathbf{k}_j(\theta)\cdot\mathbf{x}}
+ \sum_{j=1}^3A_{j}^*e^{-i\mathbf{k}_j(\theta)\cdot\mathbf{x}} +
\bar{\psi}.
\end{equation}
Equation (\ref{eq:1mode_theta1}) describes the density field of a grain
misoriented with respect to the basis vectors. Writing the basis vectors
as $\mathbf{k}_j(\theta)=\mathbf{k}_j+\delta\mathbf{k}_j(\theta)$, where
the vector $\delta\mathbf{k}_j(\theta)$ measures the rotation of each
lattice vector, we obtain
\begin{equation}
\label{eq:1mode_theta2}
\psi(\theta)=\sum_{j=1}^3A_{j}e^{i\delta\mathbf{k}_j(\theta)\cdot\mathbf{x}}
e^{i\mathbf{k}_j\cdot\mathbf{x}}
+
\sum_{j=1}^3A_{j}^*e^{-i\delta\mathbf{k}_j(\theta)\cdot\mathbf{x}}
e^{-i\mathbf{k}_j\cdot\mathbf{x}}
+ \bar{\psi},
\end{equation}
or
\begin{equation}
\label{eq:1mode_theta3}
\psi(\theta)=\sum_{j=1}^3A_{j}^{\theta}e^{i\mathbf{k}_j\cdot\mathbf{x}}
+ \sum_{j=1}^3{A_{j}^{\theta}}^*e^{-i\mathbf{k}_j\cdot\mathbf{x}} +
\bar{\psi},
\end{equation}
where
\begin{equation}\label{eq:amp_theta}
A_j^{\theta}=A_je^{i\delta\mathbf{k}_j(\theta)\cdot\mathbf{x}}.
\end{equation}
Thus grains arbitrarily misoriented from the global basis
$\mathbf{k}_j$ can still be described in terms of $\mathbf{k}_j$ by
suitably representing the complex amplitude $A_j$ in polar form
according to Eq.~(\ref{eq:amp_theta}). A straightforward way to
include differently oriented grains in the system is to specify an
initial condition via Eq.~(\ref{eq:1mode_theta3}). By making the
amplitude a non-uniform complex function with a periodic structure,
multiple grain orientations are automatically included.
Fig.~\ref{fig:beats} illustrates this idea. Fig.~\ref{fig:ramp}
shows the real component of one of the three complex amplitude
functions $A_j$, specified by Eq.~(\ref{eq:amp_theta}), and
Fig.~\ref{fig:psi} shows the corresponding density field constructed
using Eq.~(\ref{eq:1mode_theta3}). Since Eq.~(\ref{eq:cmplx_rg}) is
rotationally covariant, it allows these ``beat'' structures in
the amplitudes (and therefore the corresponding orientation of the
grain) to be preserved as the system evolves, thereby enabling the
representation of polycrystalline systems with a single set of basis
vectors.

\subsection{Limitations of the {Cartesian} representation of Equation
(\ref{eq:cmplx_rg}) \label{sec:polycryst_cart}}

A straightforward approach to solving Eq.~(\ref{eq:cmplx_rg}) is to
determine the real and imaginary parts of  the complex amplitudes $A_j$
directly, using the {Cartesian} definition. This leads to {six} equations
that can be evolved concurrently using a suitable time integration scheme.
The second order finite difference spatial discretizations of the
Laplacian and gradient operators are given in the Appendix. This approach
leads to limited success of AMR techniques because of the beats.

To illustrate this effect, we simulated heterogeneous nucleation and
growth of a two-dimensional film, randomly placing twelve randomly
oriented crystals of initial radius $8\pi$ in a square domain of side
$256\pi$ with periodic boundary conditions. The largest misorientation
angle between grains was $\theta=\pi/12$. The amplitude equations in
Cartesian form were solved using an adaptively evolving mesh (described
in detail below). The model parameters were $r=-0.25$ and
$\bar{\psi}=0.285$, the smallest mesh spacing was $\Delta
x_{min}=\pi/2$, while the largest mesh spacing at any given time was
$\Delta x_{max} = 2^4(\Delta x_{min})$ corresponding to 5 levels of
refinement. On a uniform grid, this simulation requires
$1025\times1025=1,050,625$ nodes with the PFC equation, and
$513\times513=263,169$ nodes with the amplitude equations. A time step
of $\Delta t = 0.04$ was used.

\begin{figure}[htb]
    \centering
    \begin{subfigure}[ t=0]
    {\includegraphics[width=0.3\textwidth]{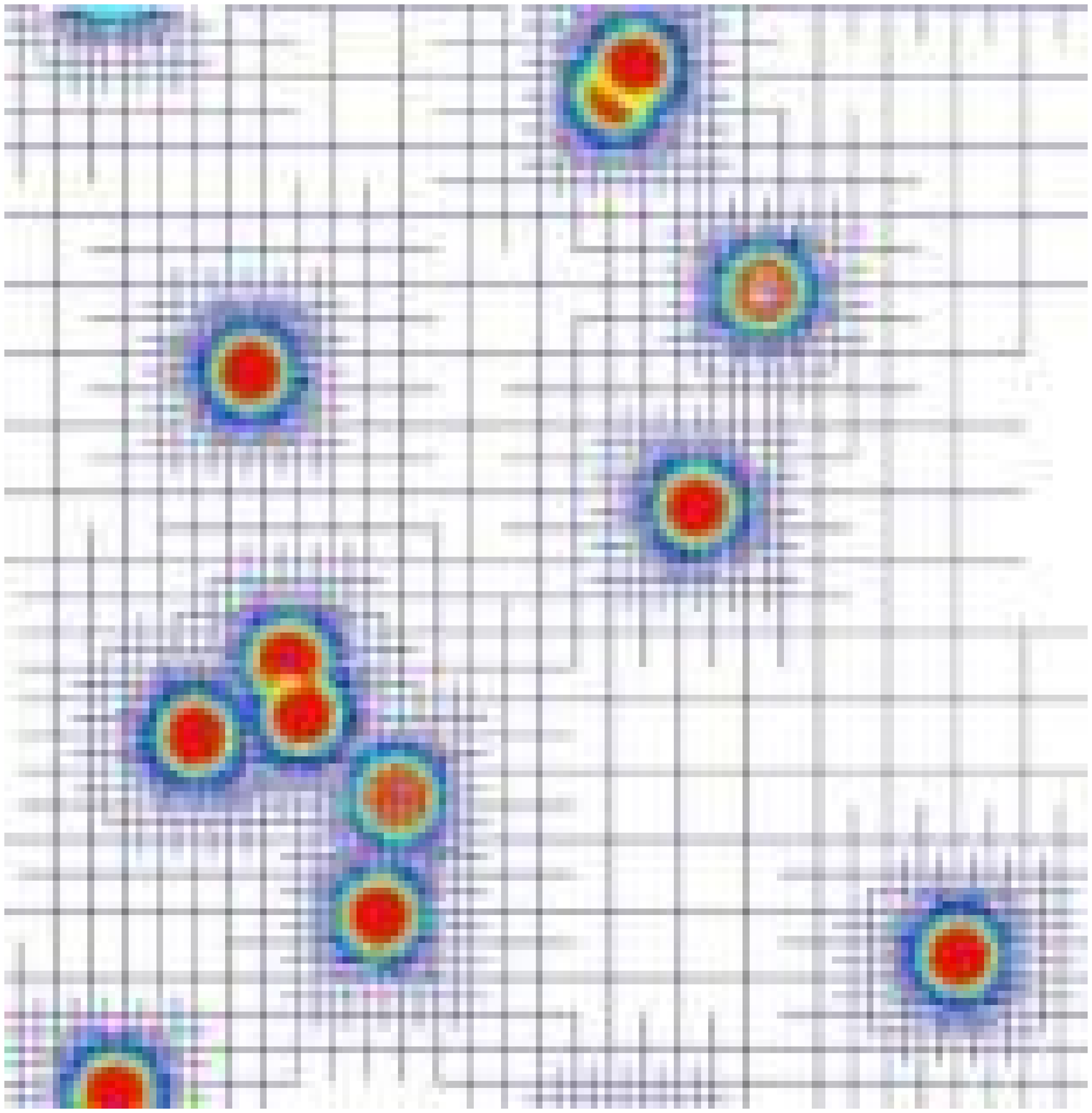}}
    \end{subfigure}
    \begin{subfigure}[ t=88]
    {\includegraphics[width=0.3\textwidth]{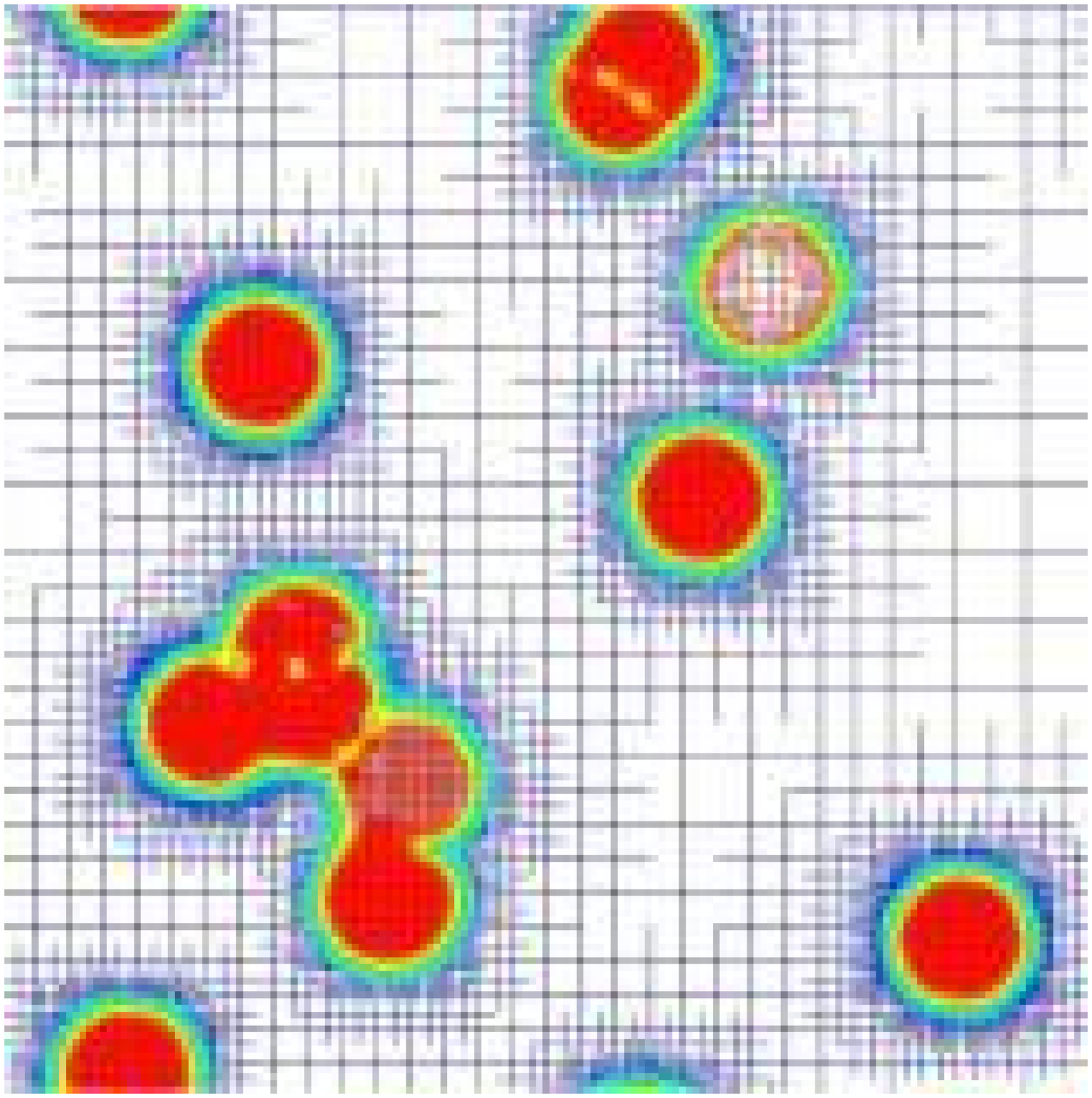}}
    \end{subfigure}
    \begin{subfigure}[ t=168]
    {\includegraphics[width=0.3\textwidth]{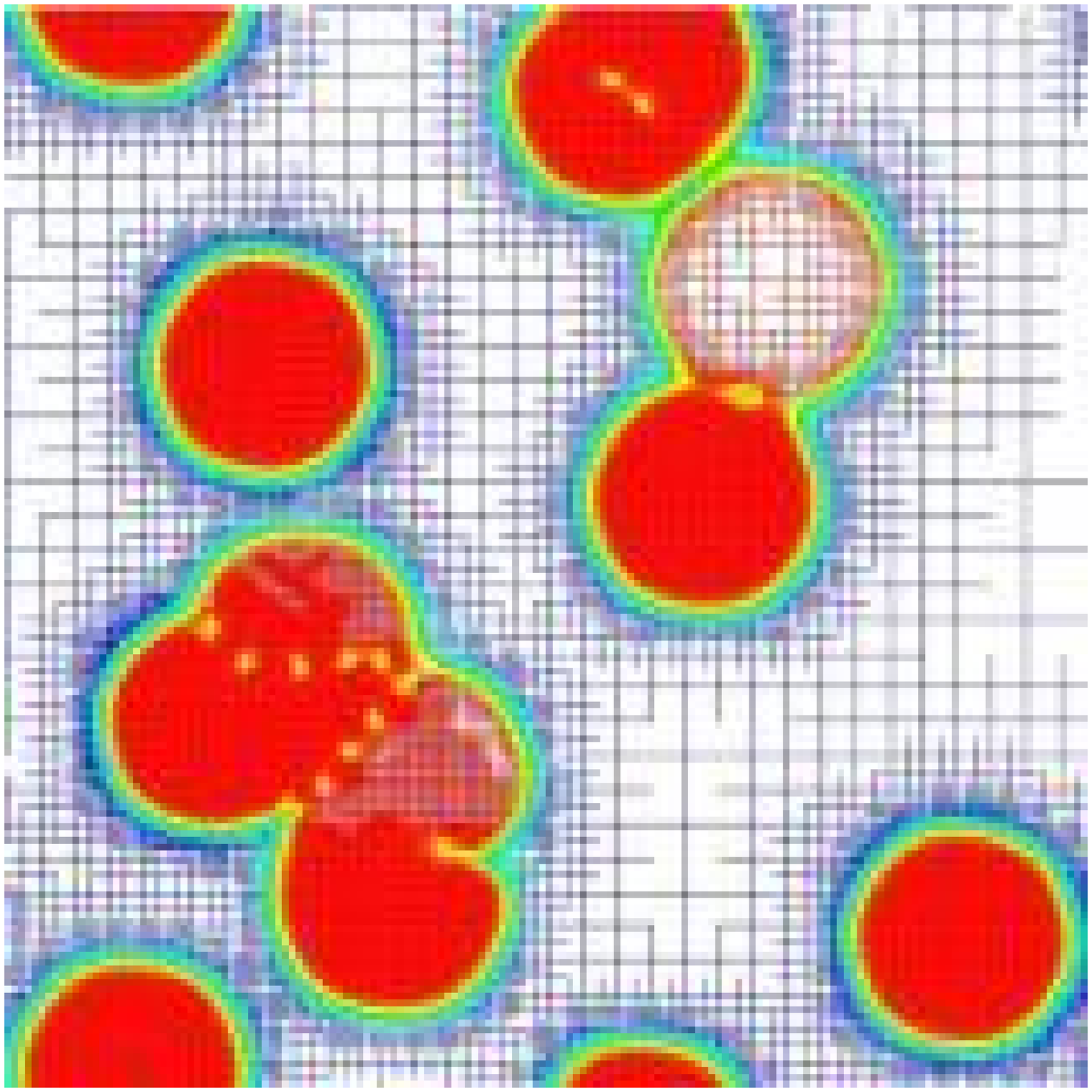}}
    \end{subfigure}
    \begin{subfigure}[ t=248]
    {\includegraphics[width=0.3\textwidth]{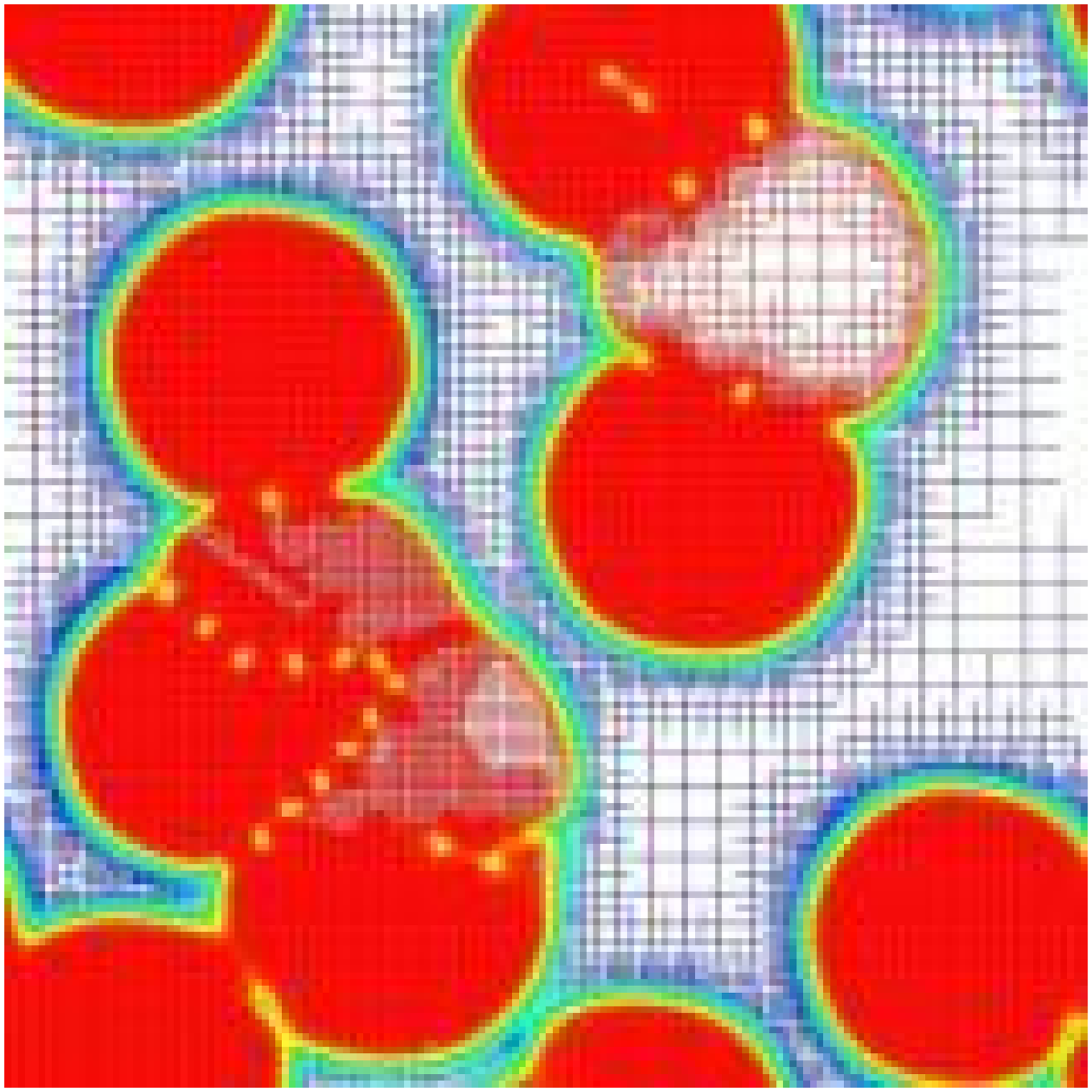}}
    \end{subfigure}
    \begin{subfigure}[ t=320]
    {\includegraphics[width=0.3\textwidth]{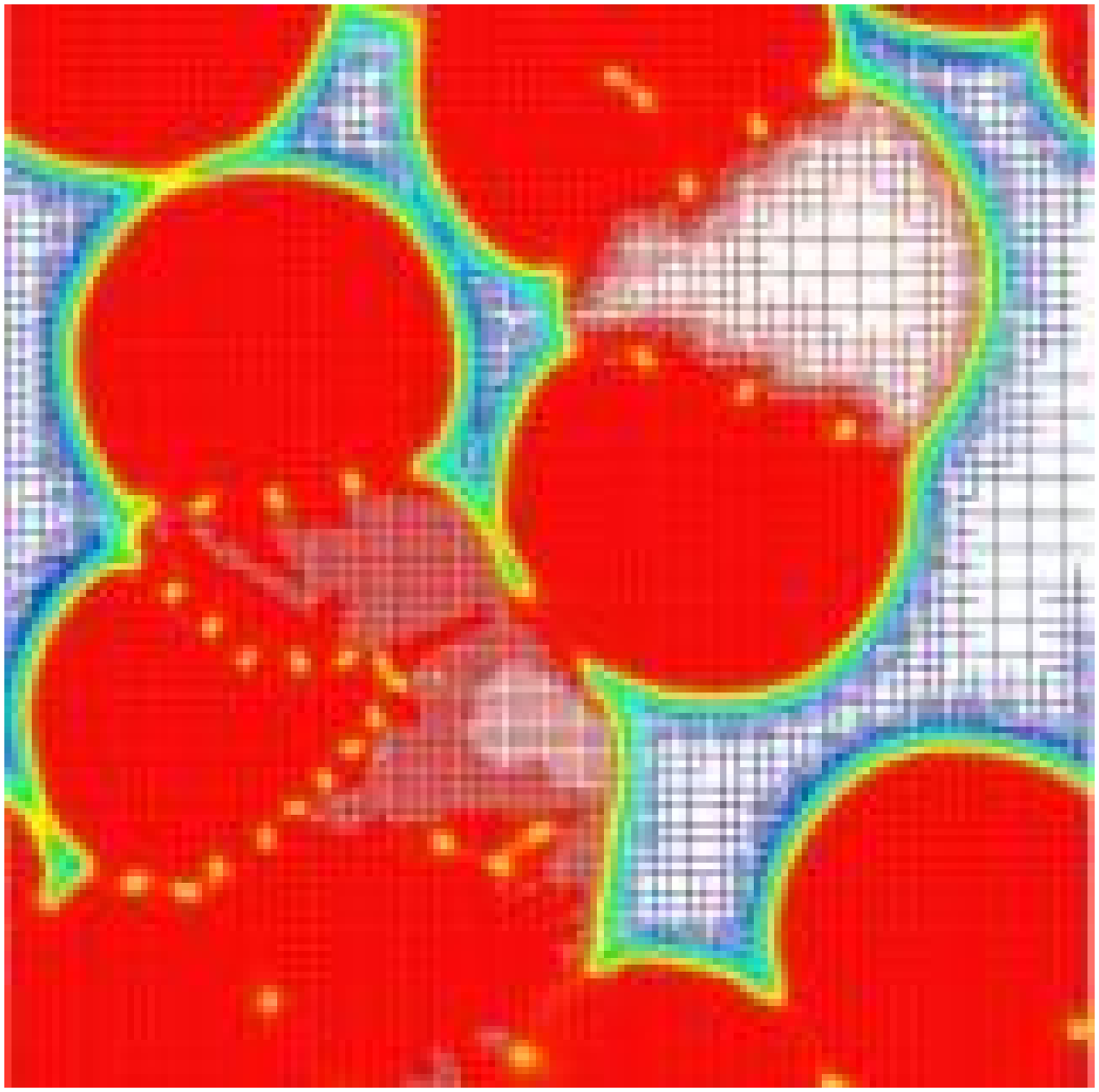}}
    \end{subfigure}
    \begin{subfigure}[ t=552]
    {\includegraphics[width=0.3\textwidth]{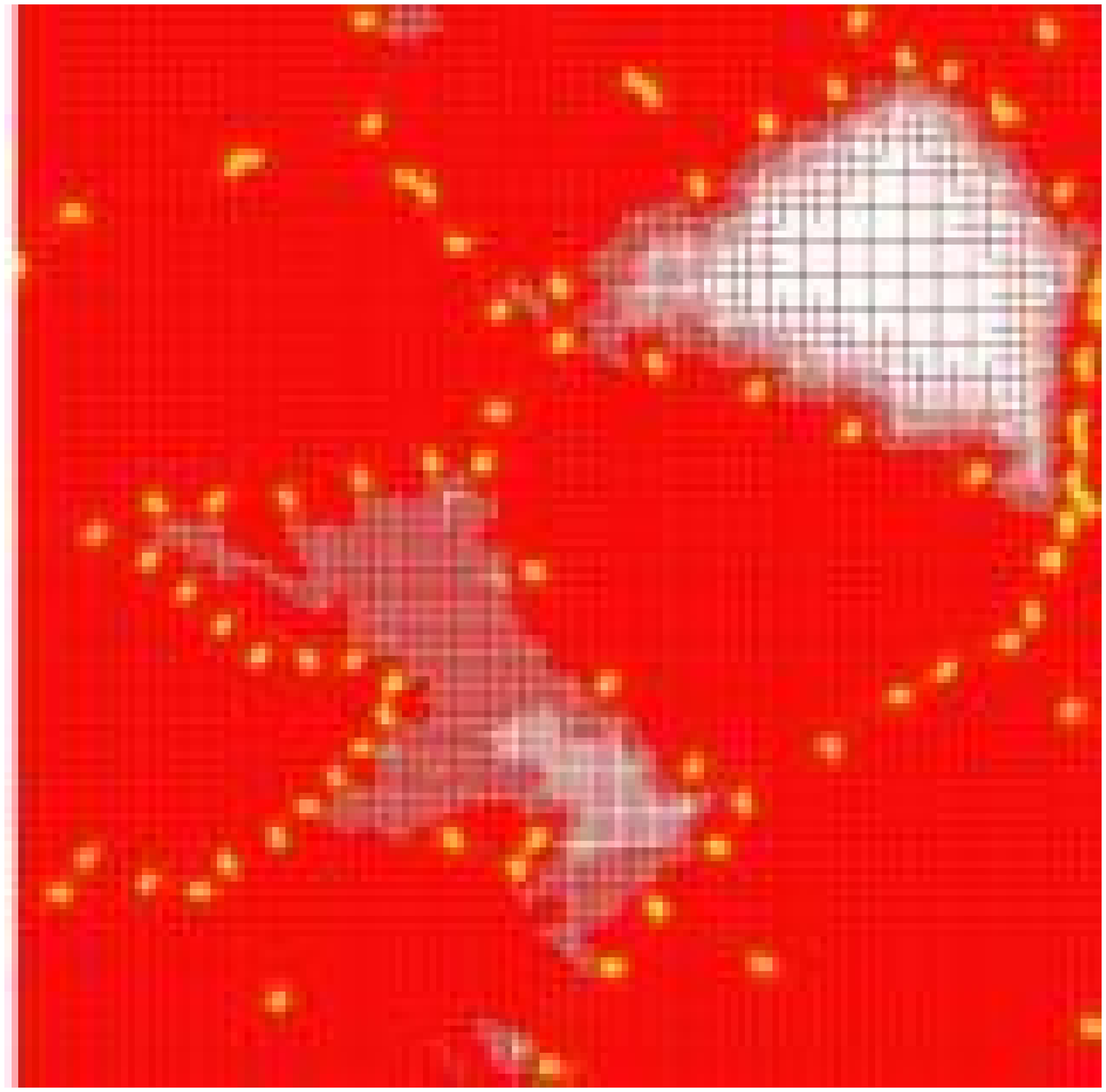}}
    \end{subfigure}
    \caption{\label{fig:cmplx_adaptive} (Color online) Evolution of a polycrystalline
    film simulated with complex amplitude equations,
    Eq.~(\ref{eq:cmplx_rg}), on an adaptive grid. Note that the grid does
    not coarsen inside many of the grains (misoriented with respect to
    $\mathbf{k}_j$) because of the fine scale structure of the ``beats''.
    The colored field plotted is the average amplitude modulus, which is
    ``red'' inside the crystal phase, ``blue'' in the liquid phase,
    ``green'' at the crystal/liquid interface, and ``yellow'' near
    defects. }
\end{figure}

Fig.~\ref{fig:cmplx_adaptive} shows the crystal boundaries and grid
structure at various times during the simulation. The field plotted is
the average amplitude modulus, $\sum_{j=1}^3A_j/3$. Although the grid
starts out quite coarse ($t=0$ and $t=88$) at several locations in the
computational domain because of the large liquid fraction, this
advantage falls off dramatically once the crystals evolve, collide, and
start to form grain boundaries. In particular, once all the liquid
freezes, only a few grains that are favorably oriented with respect to
$\mathbf{k}_j$ show any kind of grid coarsening at all. Those that are
greatly misorientated with respect to $\mathbf{k}_j$ lead to frequency
beating, causing the number of nodes in the adaptive grid to increase
rather than decrease. The polycrystal mesh shown in
Fig.~\ref{fig:cmplx_adaptive}(f) has $219,393$ nodes, which is very
near that on a uniform grid. Therefore the adaptive refinement
algorithm applied to a Cartesian formulation of Eq.~\ref{eq:cmplx_rg}
gives at best a marginal improvement over a fixed grid implementation.
The main purpose of this paper is to present a methodology for
overcoming this problem.

\section{\label{sec:polar_rg}Complex amplitude equations in a polar
representation}
\subsection{Governing equations}

We find that the computational benefits of AMR are potentially
greater if, instead of solving for the real and imaginary components
of $A_j$, we solve for the amplitude moduli $\Psi_j=|A_j|$, and the
phase angles $\Phi_j=\arctan(\Im(A_j)/\Re(A_j))$, which are
spatially uniform fields irrespective of crystal orientation.
Together these two fields constitute a \emph{polar representation}
of $A_j$.

In this section we derive evolution equations for $\Psi_j$ and $\Phi_j$
directly from Eq.~(\ref{eq:cmplx_rg}), by applying Euler's formula for
a complex number, i.e.~$A_j=\Psi_je^{i\Phi_j}$, and then by equating
corresponding real and imaginary parts on the left- and right-hand
sides of the resulting equations. In this manner we get the coupled
system of equations,
\begin{eqnarray}\label{eq:modamp}
\frac{\partial\Psi_j}{\partial t} &=& (r+3\bar{\psi}^2)\left[-\Psi_j+
\mathcal{C}^{\Re}(\Psi_j,\Phi_j)\right]-
\left[\mathcal{C}^{\Re\Re}(\Psi_j,\Phi_j)-
\mathcal{C}^{\Im\Im}(\Psi_j,\Phi_j)\right]\nonumber\\
  & & +\left[\mathcal{C}^{\Re\Re\Re}(\Psi_j,\Phi_j)-
  \mathcal{C}^{\Re\Im\Im}(\Psi_j,\Phi_j)-
  \mathcal{C}^{\Im\Im\Re}(\Psi_j,\Phi_j)-
  \mathcal{C}^{\Im\Re\Im}(\Psi_j,\Phi_j)\right]\nonumber\\
  & & -3\Psi_j\left(\Psi_j^2+2\sum_{k\ne j}
  \Psi_k^2\right)- 6\frac{\bar{\psi}}{\Psi_j}
  \left(\prod_{k}\Psi_k\right)\cos\left(\sum_k\Phi_k\right)
\end{eqnarray}
and
\begin{eqnarray}
\frac{\partial\Phi_j}{\partial t}\label{eq:phase}
&=&\left\{(r+3\bar{\psi}^2)\mathcal{C}^{\Im}(\Psi_j,\Phi_j)-
\left[\mathcal{C}^{\Re\Im}(\Psi_j,\Phi_j)+
\mathcal{C}^{\Im\Re}(\Psi_j,\Phi_j)\right]\right.\nonumber\\
& &\left.+\left[\mathcal{C}^{\Im\Re\Re}(\Psi_j,\Phi_j)-
\mathcal{C}^{\Im\Im\Im}(\Psi_j,\Phi_j)+
\mathcal{C}^{\Re\Im\Re}(\Psi_j,\Phi_j)+
\mathcal{C}^{\Re\Re\Im}(\Psi_j,\Phi_j)\right]\right\}/\Psi_j\nonumber\\
& &+
6\frac{\bar{\psi}}{\Psi_j^2}\left(\prod_{k}\Psi_k\right)\sin\left
(\sum_k\Phi_k\right)
\end{eqnarray}
where
\begin{eqnarray}
\mathcal{C}^{\Re}(\Psi_j,\Phi_j) &=&
\Re\left\{\frac{\left[\nabla^2+2i\mathbf{k}_j\cdot\nabla\right](\Psi_j
e^{i\Phi_j})}{e^{i\Phi_j}}\right\}\nonumber\\
\mathcal{C}^{\Im}(\Psi_j,\Phi_j) &=&
\Im\left\{\frac{\left[\nabla^2+2i\mathbf{k}_j\cdot\nabla\right](\Psi_j
e^{i\Phi_j})}{e^{i\Phi_j}}\right\}\nonumber\\
\mathcal{C}^{\Re\Re}(\Psi_j,\Phi_j)
&=&\Re\left\{\frac{\left[\nabla^2+2i\mathbf{k}_j\cdot\nabla\right]
\left(\mathcal{C}^{\Re}(\Psi_j,\Phi_j)
e^{i\Phi_j}\right)}{e^{i\Phi_j}}\right\}\nonumber\\
\mathcal{C}^{\Im\Re}(\Psi_j,\Phi_j)
&=&\Im\left\{\frac{\left[\nabla^2+2i\mathbf{k}_j\cdot\nabla\right]
\left(\mathcal{C}^{\Re}(\Psi_j,\Phi_j)
e^{i\Phi_j}\right)}{e^{i\Phi_j}}\right\}\nonumber\\
\end{eqnarray}
and so on for the remaining ${\mathcal{C}}$'s. From here on we refer to the
evolution equations for $\Psi_j$ and $\Phi_j$ as the phase/amplitude
equations, whereas Eq.~(\ref{eq:cmplx_rg}) will be referred to as the
complex amplitude equation. Unfortunately, the phase/amplitude equations
in Eqs.~(\ref{eq:modamp}) and (\ref{eq:phase}) turn out to be quite
difficult to solve globally. The principal difficulties are summarized
below.

\begin{figure}
    \centering
    \begin{subfigure}[ Contours of $\Psi_1$]
    {\includegraphics[ height=2.6in]{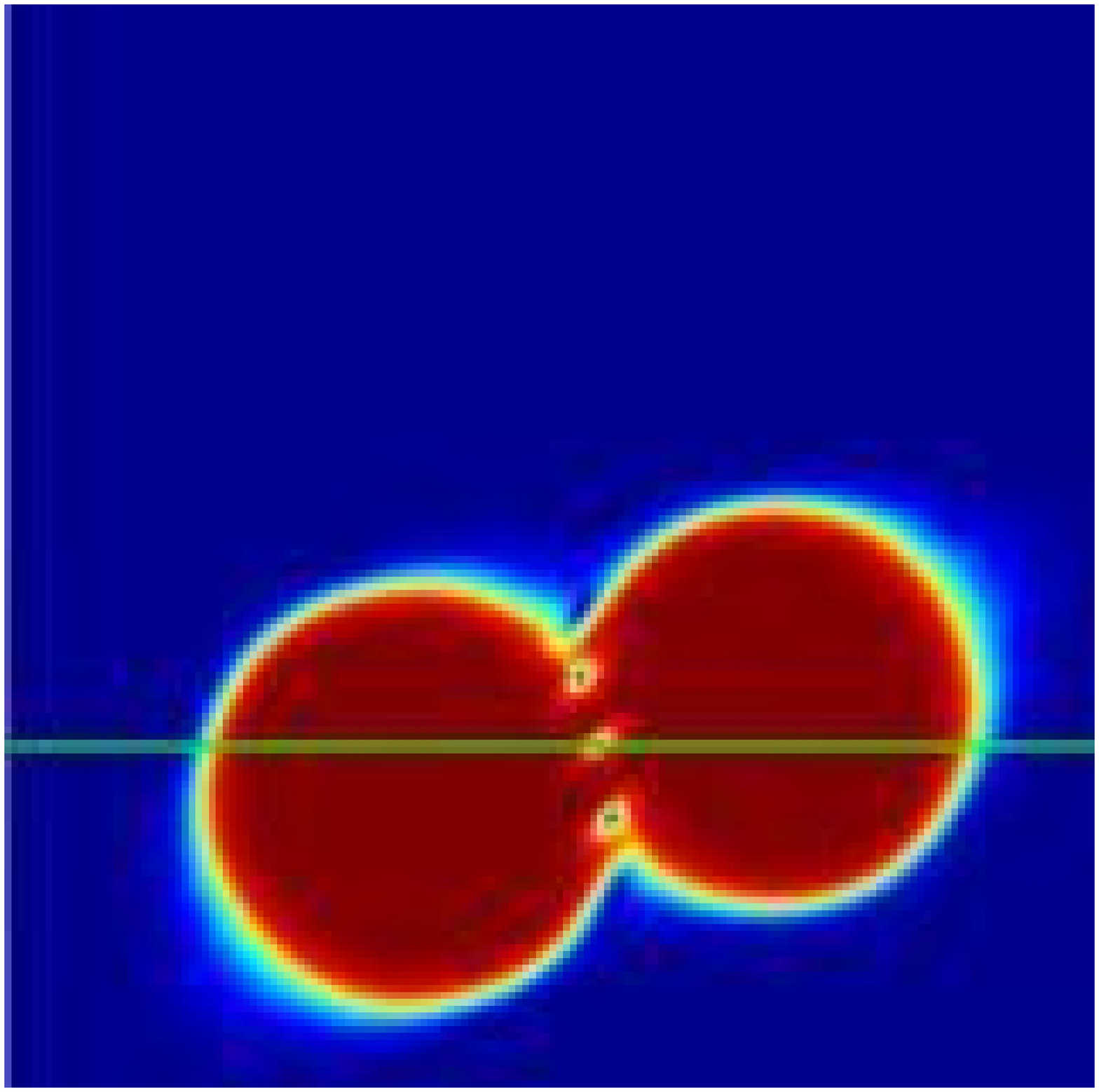}}
    \end{subfigure}
    \begin{subfigure}[ Contours of $\Phi_1$]
    {\includegraphics[ height=2.6in]{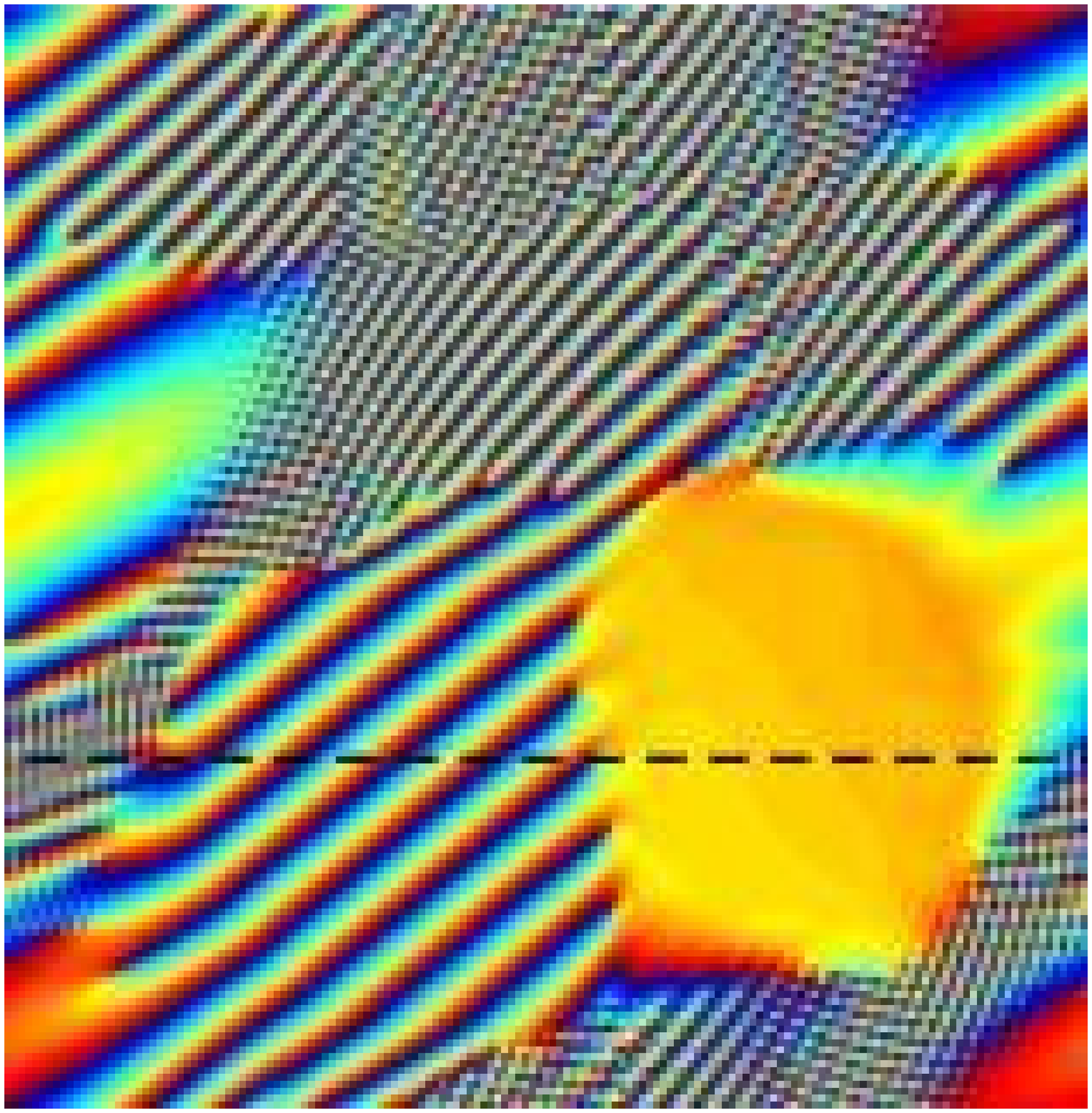}}
    \end{subfigure} \\
    \begin{subfigure}[ Lineplot of $\Psi_1$ along solid line in (a)]
    {\includegraphics[ height=2.6in]{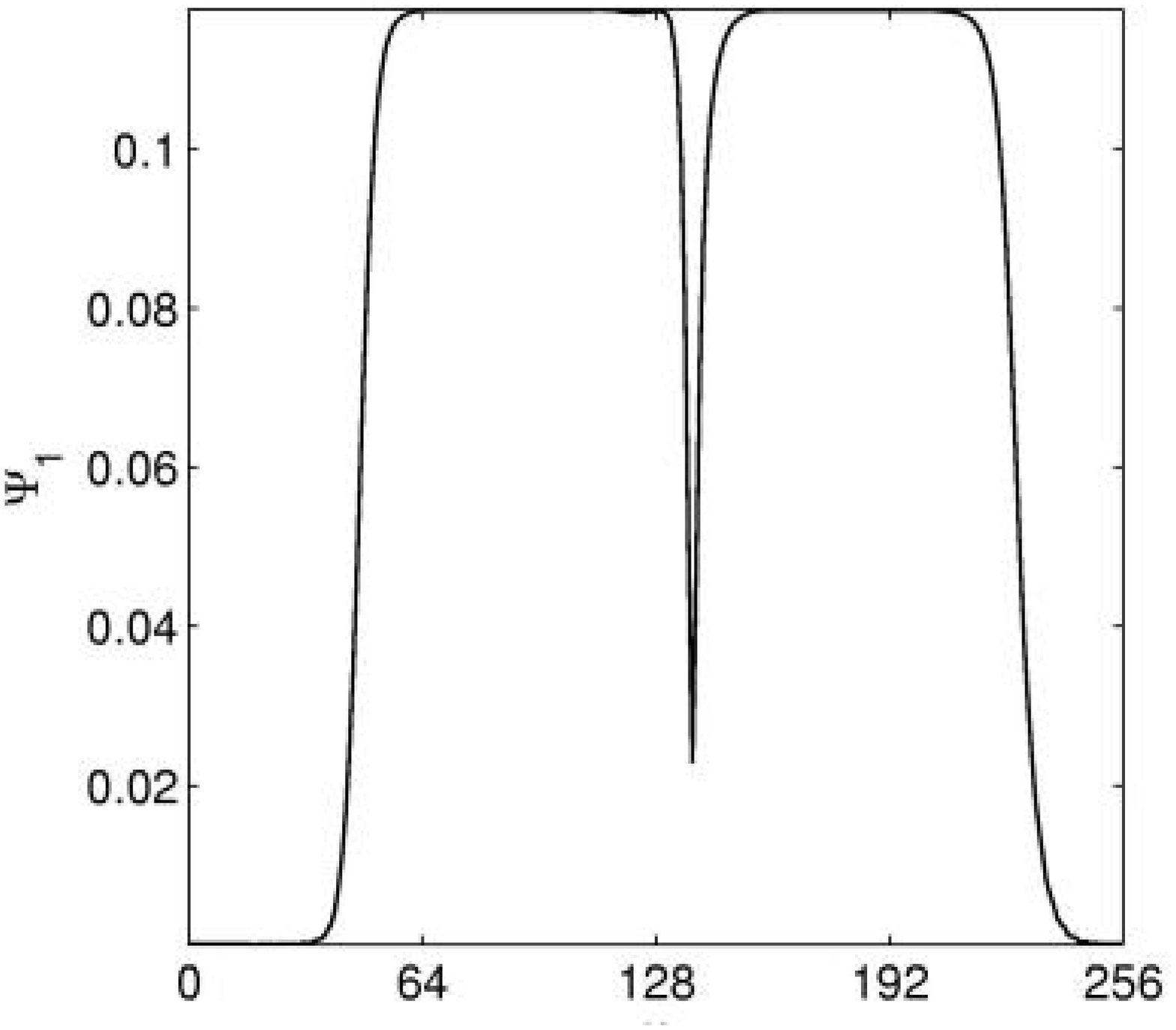}}
    \end{subfigure}
    \begin{subfigure}[ Lineplot of $\Phi_1$ along dashed line in (c)]
    {\includegraphics[ height=2.6in]{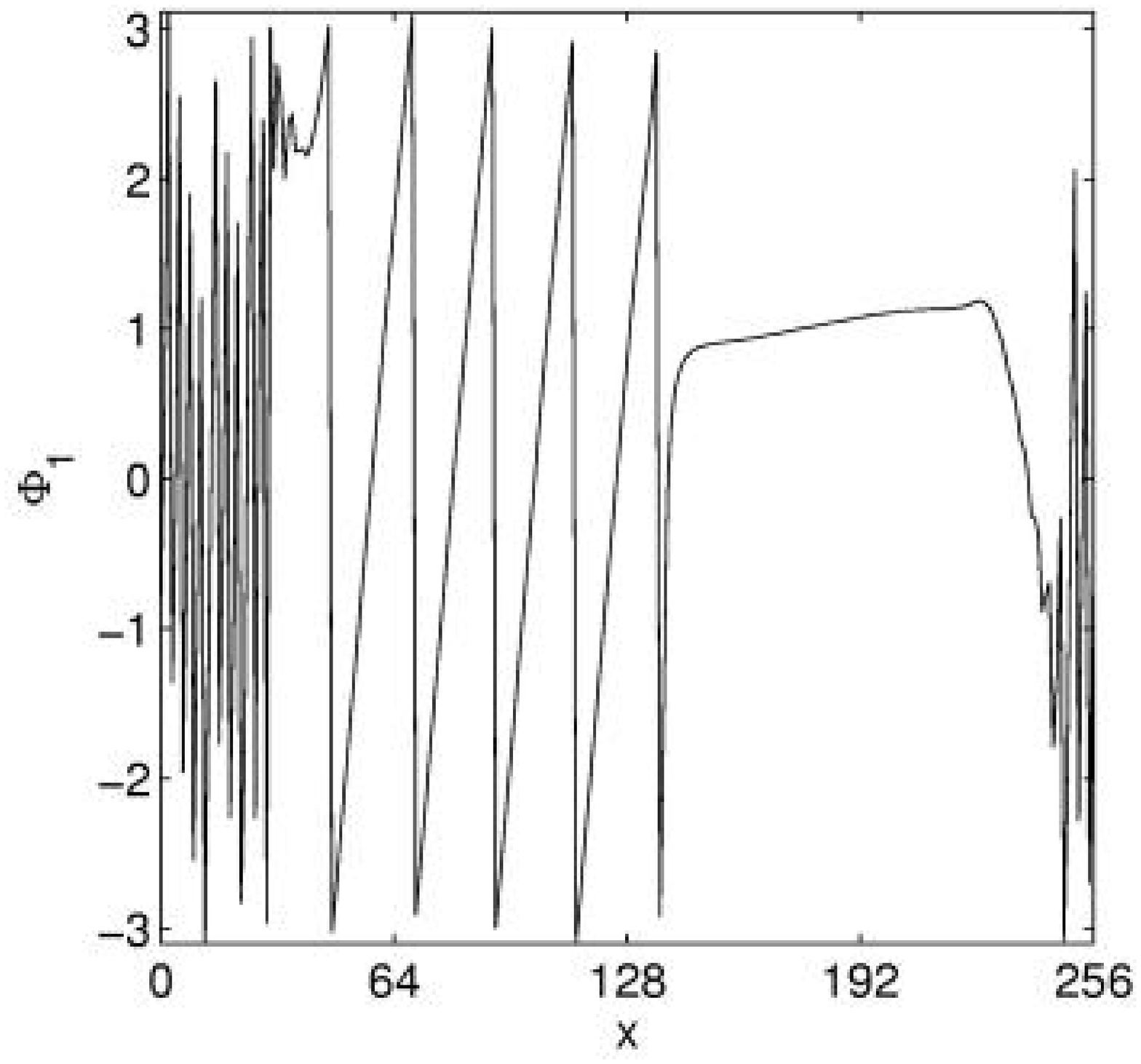}}
    \end{subfigure}
    \caption{\label{fig:phaseamp_issues}(Color online) (a) and (c): The
field $\Psi_1$, smooth everywhere except near interfaces and defects.
(b) and (d) The field $\Phi_1$, which is computed naively as
$\arctan(\Im(A_1)/\Re(A_1))$ is periodic and discontinuous. The chaotic
fluctuations in $\Phi_1$ in the regions outside the crystals correspond
to the liquid phase where $\Phi_1$ has no physical meaning. The rapid,
but periodic, variations of $\Phi_1$ in the left grain is due to its
large misorientation angle of $\pi/6$. In contrast, the grain on the
right is oriented along $\mathbf{k}_j$ causing $\Phi_1$ to vary much
more smoothly.} \end{figure}

The field $\Psi_j$ is nearly constant within the individual
grains and varies sharply only near grain boundaries, rendering its
equation ideally suited for solution on adaptive meshes. The field
$\Phi_j$ on the other hand, if computed naively as
$\arctan(\Im(A_j)/\Re(A_j))$, is a periodic and discontinuous
function \footnote{$\Phi_j\approx\mathbf{q}(\theta)\cdot\mathbf{x}$,
$\Phi_j\in[-\pi,\pi]$, where $\mathbf{q}(\theta)$ is the phase
vector, constant for a particular orientation of the grain, and
$\theta$ is the misorientation angle of the grain. Thus $\Phi_j$,
roughly speaking, has the structure of a sawtooth waveform.} bounded
between the values $-\pi$ and $\pi$, with a frequency that increases
with increasing grain misorientation. This poses a problem similar
to that previously posed by the beats, with the grid this time
having to resolve the fine scale structure of $\Phi_j$. Further, one
may need to resort to shock-capturing methods in order to correctly
evaluate higher order derivatives, and resolve jumps where $\Phi_j$
changes value from $\pi$ to $-\pi$ and vice-versa. Complications are
also caused by $\Phi_j$ being undefined in the liquid phase, and the
tendency for $\Psi_j$, which appears in the denominator on the right
hand side of Eq.~(\ref{eq:phase}), to approach zero at those
locations. This calls for some type of robust regularization scheme
\footnote{We have determined that simple tricks such as setting
$\Psi_j$ to some small non-zero value, or setting a heuristic upper
bound on higher-order derivatives, have the effect of destroying
defects and other topological features in the pattern.} for the
phase equations. These problems are clearly highlighted in
Fig.~\ref{fig:phaseamp_issues}, which shows the impingement of two
misaligned crystals and the corresponding values of $\Psi_1$ and $\Phi_1$.

Ideally, one would like to reconstruct from the periodic $\Phi_j$, a
continuous surface $\Phi_j+2n\pi$ (where $n$ is an integer) which
would be devoid of jumps, and therefore amenable to straightforward
resolution on  adaptive meshes. The implementation of such a
reconstruction algorithm however, even if possible, requires
information about individual crystal orientations, and the precise
location of solid/liquid interfaces, defects, and grain boundaries
at every time step, making it very computationally intensive.
Further, such an algorithm would be more appropriate in the
framework of an interface-tracking approach such as the level set
method \cite{sethian96}, rather than our phase-field modeling
approach.

Despite these issues with the polar (phase/amplitude) equations
progress can be made, under certain non-critical approximations, by
solving the phase/amplitude equations in the interior of crystalline
regions, in conjunction with the Cartesian complex amplitude equations
in regions closer to domain boundaries and topological defects.

\subsection{Reduced equations and the frozen phase gradient approximation}

The main idea that will be developed in this and subsequent sections is
that of evolving the phase/amplitude and complex amplitude equations
simultaneously in different parts of the domain, depending on where they
can most appropriately be applied.  The phase/amplitude formulation is
solved in the crystal interior, away from defects, interfacial regions, and
the liquid phase.  The complex amplitude equations are solved everywhere
else in the computational domain.  This does away with the need for
regularizing the phase equations where $\Psi_j\to 0$ (since $\Psi_j\gg0$
in the crystal interior) as well as the issue of the phase being undefined
in certain regions.  We overcome the remaining issues with the phase
equation, i.e. the difficulty of evaluating derivatives of the phase and
the need to resolve its periodic variations via certain controlled
approximations described next.

\begin{figure}
    \centering
    \begin{tabular}{ccc}
        \includegraphics[ height=2.0in]{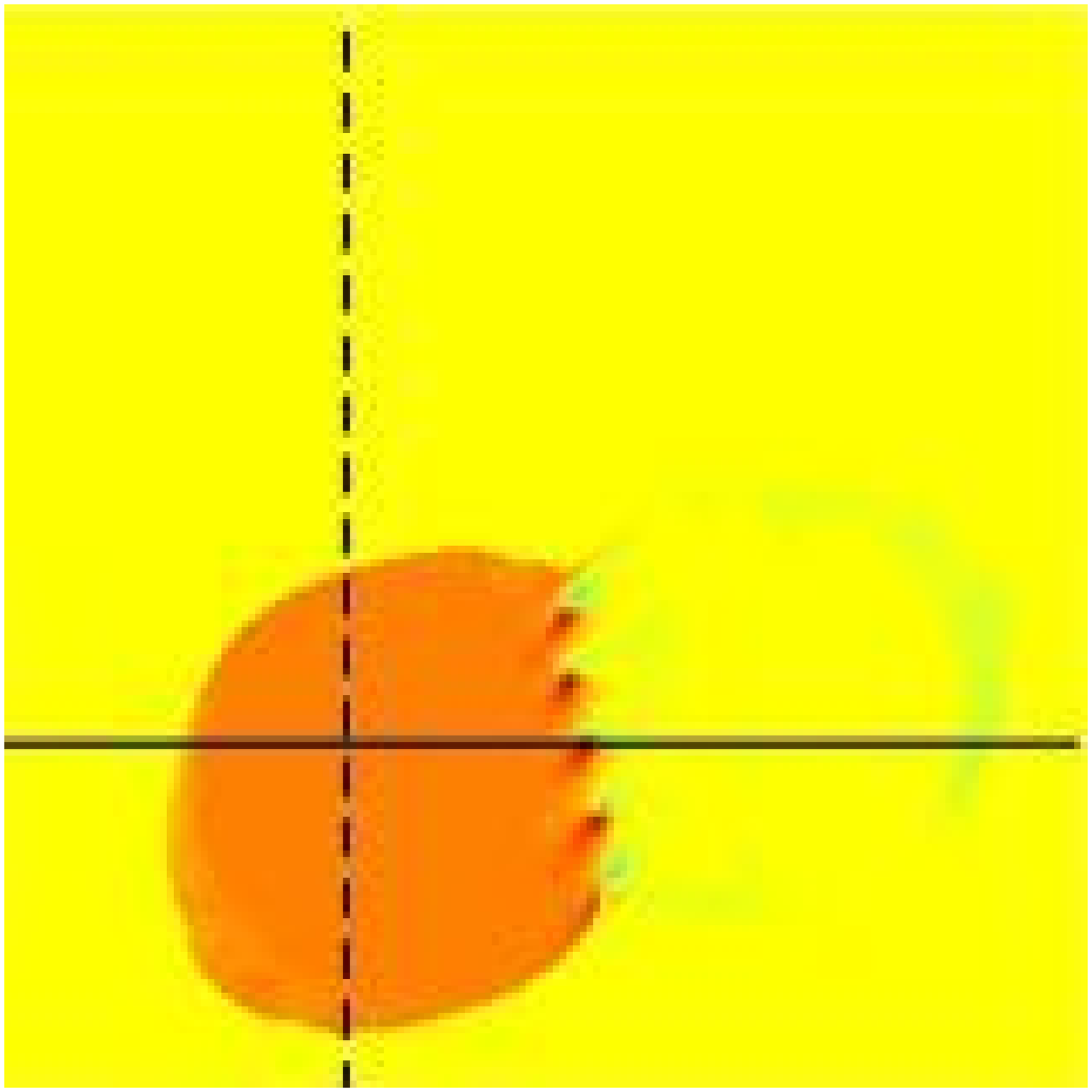} &
        \includegraphics[ height=2.0in]{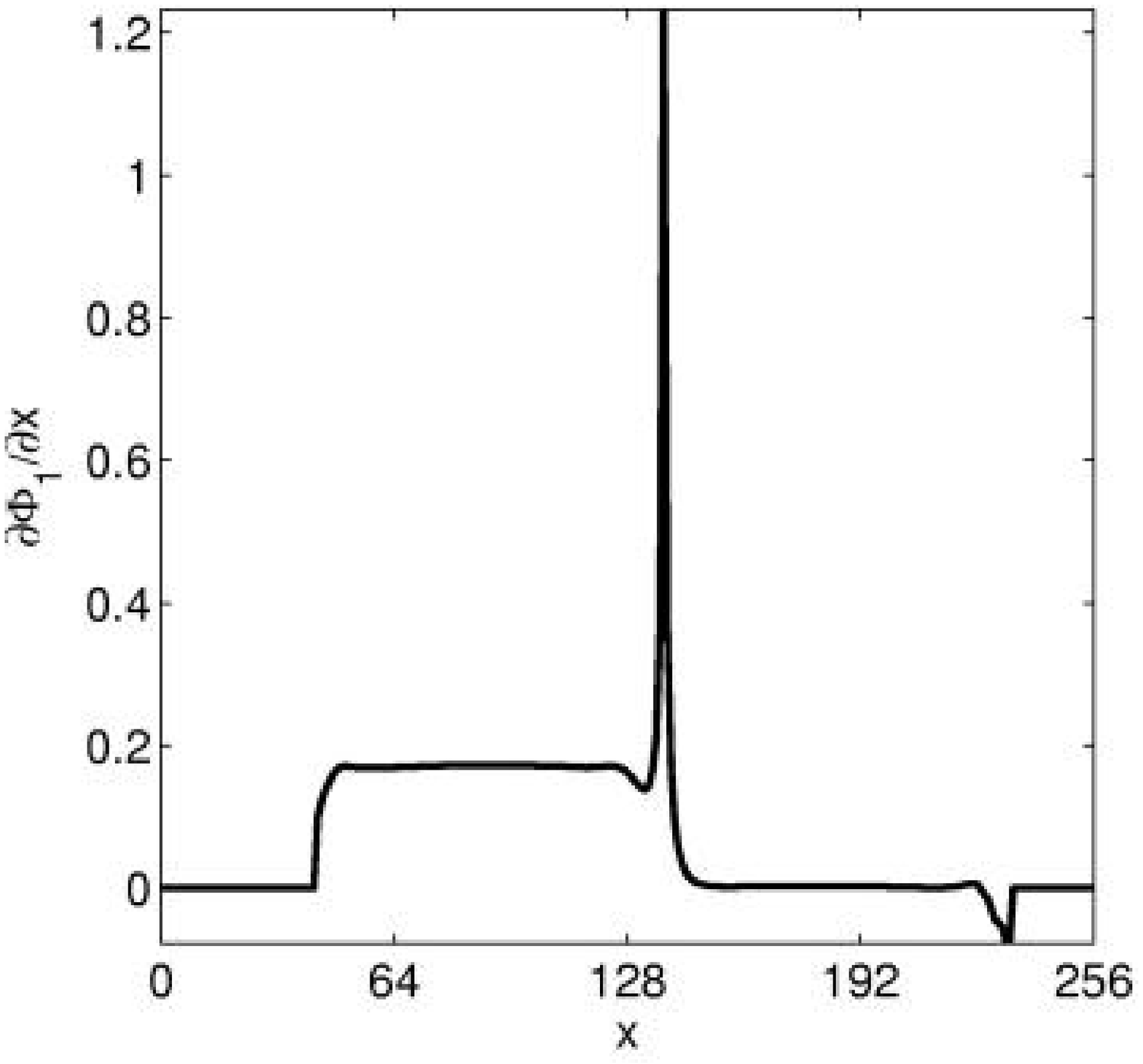} &
        \includegraphics[ height=2.0in]{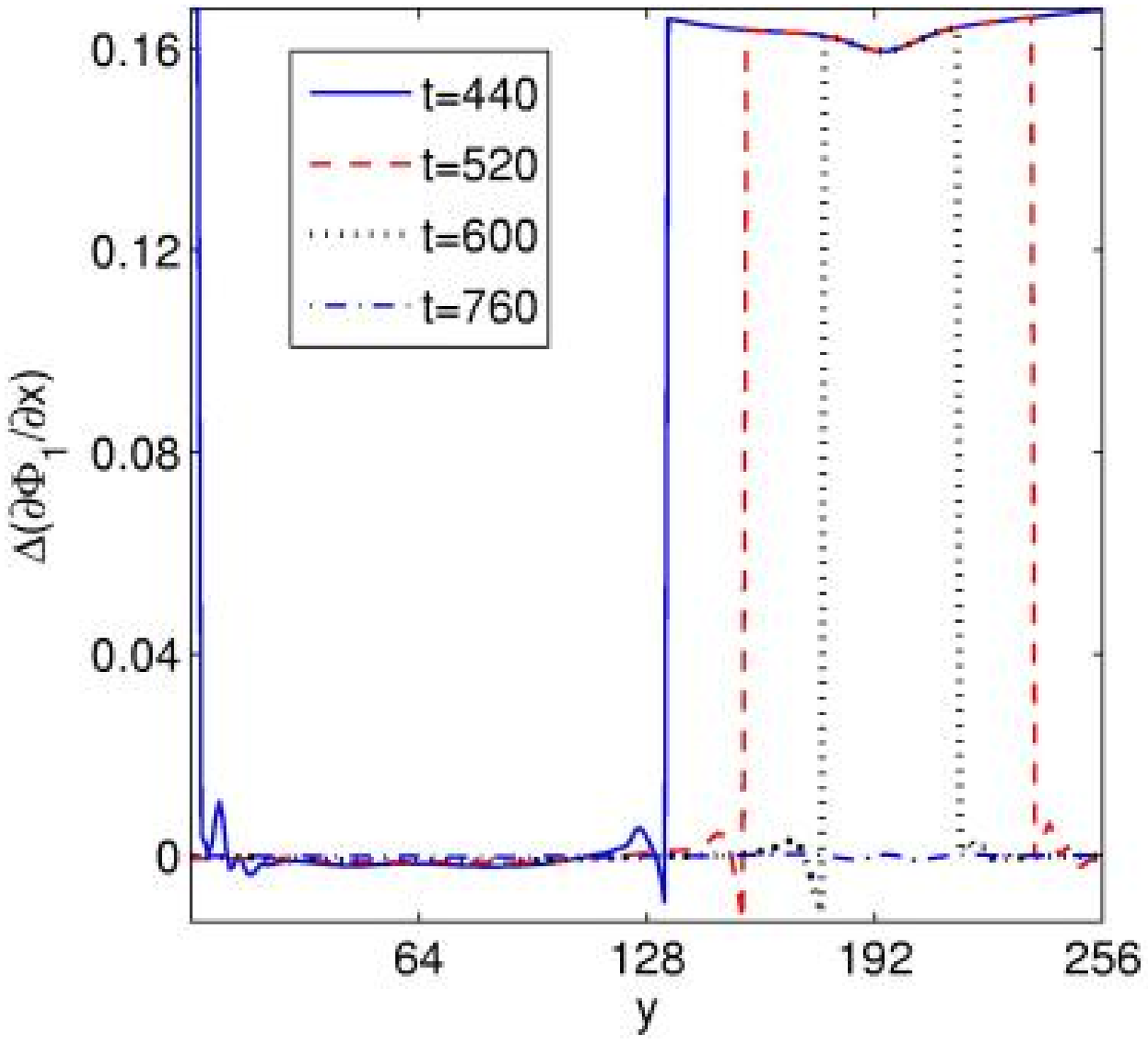}
    \\
    (a) & (b) & (c)\\
            \end{tabular}
    \caption{\label{fig:frozen_gradient} (Color online) $\Phi_1$ and its time evolution for the pair
    of crystals shown in Fig.~\ref{fig:phaseamp_issues}. (a) Contours of $\partial\Phi_1/\partial x$ at $t=520$.
    (b) Lineplot of $\partial\Phi_1/\partial x$ along solid line in
    (a).  (c) $\Delta(\partial\Phi_1/\partial x)$ along dashed line
    in (a).  Just as with $\Psi_j$,
    the components of $\nabla\Phi_j$ are also practically constant inside
    the individual crystals. The spike in (b) corresponds to a defect on
    the grain boundary. As seen from the time series in (c) for
    $\partial\Phi_1/\partial x$, $\nabla\Phi_j$ hardly changes in the
    crystal bulk during its evolution.}
\end{figure}

Let us examine the results of a fixed grid calculation performed using the
complex amplitude equations, illustrated in
Fig.~\ref{fig:frozen_gradient}, showing a sequence of line plots of the
quantity $\Delta(\partial \Phi_1/\partial x)=\left.\partial
\Phi_1/\partial x\right|_{t=840}-\left.\partial \Phi_1/\partial
x\right|_t$. $\nabla\Phi_j$ inside the growing crystal is seen to be
essentially time invariant. As the crystal on the left grows, it can be
seen that $\Delta(\partial \Phi_1/\partial x)$ stays close to zero inside.
We have verified that this is also true for the $y$ component of
$\nabla\Phi_1$, and both components of $\nabla\Phi_2$ and $\nabla\Phi_3$.

%\begin{figure}
%    \centering
%    \begin{tabular}{ccc}
%        \includegraphics[ height=2.6in]{phgradx2d_v2}
%    &
%        \includegraphics[ height=2.6in]{phgradx1dseries}
%    \\
%    (a) Contours of $\partial\Phi_1/\partial x$ at $t=520$. &
%    (c) $\Delta(\partial\Phi_1/\partial x)$ along dashed line
%    in (a). \vspace{0.5em} \\
%        \includegraphics[ height=2.6in]{phgradx1d}
%    &
%
%    \\
%    (b) Lineplot of $\partial\Phi_1/\partial x$ along solid line in (a).
%    &  \vspace{0.5em} \\
%    \end{tabular}
%    \caption{\label{fig:frozen_gradient} (Color online) $\Phi_1$ and its time evolution for the pair
%    of crystals shown in Fig.~\ref{fig:phaseamp_issues}. Like $\Psi_j$,
%    the components of $\nabla\Phi_j$ are also practically constant inside
%    the individual crystals. The spike in (b) corresponds to a defect on
%    the grain boundary. As seen from the time series in (c) for
%    $\partial\Phi_1/\partial x$, $\nabla\Phi_j$ hardly changes in the
%    crystal bulk during its evolution.}
%\end{figure}

These results suggest that we may employ a \emph{locally frozen phase
gradient}. {Note that the assumption of a frozen phase gradient does not
mean that $\Phi_j$ itself cannot change. $\Phi_j$ can continue to evolve
as per Eq.~(\ref{eq:reduced_phase}) under the constraint of a fixed
$\nabla\Phi_j$, although the changes may actually be quite small.}  On the
other hand, when similarly oriented crystals collide to form a small angle
grain boundary, it is energetically more favorable for grains to locally
realign (i.e. for $\nabla\Phi_j$ to change close to grain boundaries) in
order to reduce orientational mismatch
\cite{harris98,moldovan02,moldovan02_2,moldovan02_3}, rather than to
nucleate dislocations. Since such interaction effects originate at the
grain boundary, where the full complex RG equations will be solved, we
anticipate that our assumption will not lead to artificially ``stiff''
grains.

This approximation allows us to neglect third and higher order derivatives
of $\Psi_j$ and $\Phi_j$ \footnote{To consistent order, we can also
neglect second order derivatives of $\Psi_j$.}, which allows us to
reduce Eqs.~(\ref{eq:modamp}) and (\ref{eq:phase}) to the following
second order PDEs:
\begin{eqnarray}\label{eq:reduced_amp}
\frac{\partial\Psi_j}{\partial t} &=& (r+3\bar{\psi}^2)\left[-\Psi_j+
\mathcal{C}^{\Re}(\Psi_j,\Phi_j)\right]\nonumber\\
& &-3\Psi_j\left(\Psi_j^2+2\sum_{k\ne j}\Psi_k^2\right)-
6\frac{\bar{\psi}}{\Psi_j}\left(\prod_{k}\Psi_k\right)
\cos\left(\sum_k\Phi_k\right)\\
\label{eq:reduced_phase} \frac{\partial\Phi_j}{\partial t} &=&
\frac{(r+3\bar{\psi}^2)\mathcal{C}^{\Im}(\Psi_j,\Phi_j)}{\Psi_j}+
6\frac{\bar{\psi}}{\Psi_j^2}\left(\prod_{k}\Psi_k\right)\sin
\left(\sum_k\Phi_k\right),
\end{eqnarray}
where $\mathcal{C}^{\Re}$ and $\mathcal{C}^{\Im}$ contain only first and
second order derivatives of $\Psi_j$ and $\Phi_j$.
Eqs.~(\ref{eq:reduced_amp}) and (\ref{eq:reduced_phase}) are referred to as
the reduced phase/amplitude equations.

The task of evolving the phase/amplitude equations is now considerably
simplified, as only derivatives up to second order in $\Phi_j$ need to be
computed. While the Laplacian and gradient of $\Psi_j$ can be computed in
a straightforward manner using Eqs.~(\ref{eqn:lapdisc}) and
(\ref{eqn:graddisc}) respectively, the gradient of $\Phi_j$ needs to be
computed with a little more care (in order to avoid performing derivative
operations on a discontinuous function).  The result is that
\begin{equation}\label{eq:gradphase}
\nabla\Phi_j = \frac{\Re(A_j)\nabla\Im(A_j)-\Im(A_j)\nabla\Re(A_j)}{\Psi_j^2}.
\end{equation}
Thus, the gradient operation on a discontinuous function $\Phi_j$ is
now transformed into gradient operations on the smooth components of
the complex amplitude $A_j$. Further, $\nabla^2\Phi_j$ is computed
as $\nabla\cdot\nabla\Phi_j$, where the divergence operator is
discretized using a simple second order central difference scheme.

However, as can be seen from Eq.~(\ref{eq:gradphase}), $\nabla\Phi_j$ now
depends on gradients of the real and imaginary components of $A_j$, which
may not be properly resolved in the crystal bulk as we intend to coarsen
the mesh there. To address this point, we assume that $\nabla\Phi_j$ is
frozen temporally  in the crystal bulk. This assumption implies that once
$\nabla\Phi_j$ is accurately initialized in the crystal interior via
Eq.~(\ref{eq:gradphase}), after ensuring adequate resolution of the
components of $A_j$, it need not be computed again. For example, in
simulations of crystal growth from seeds, we can start with a mesh that is
initially completely refined inside the seeds, so that $\nabla\Phi_j$ is
correctly computed. Once initial transients disappear and the crystals
reach steady state evolution, the growth is monotonic in the outward
direction. From this point on, $\nabla\Phi_j$ hardly changes inside the
crystal bulk and the grid can unrefine inside the grains while correctly
preserving gradients in $\nabla\Phi_j$. Note that the apparent
discontinuities in $\Phi_j$ no longer need be resolved by the grid.

\section{\label{sec:hybrid_rg}A hybrid formulation}

\begin{figure}[htb]
\begin{center}
{\includegraphics[height=2.5in,angle=0]{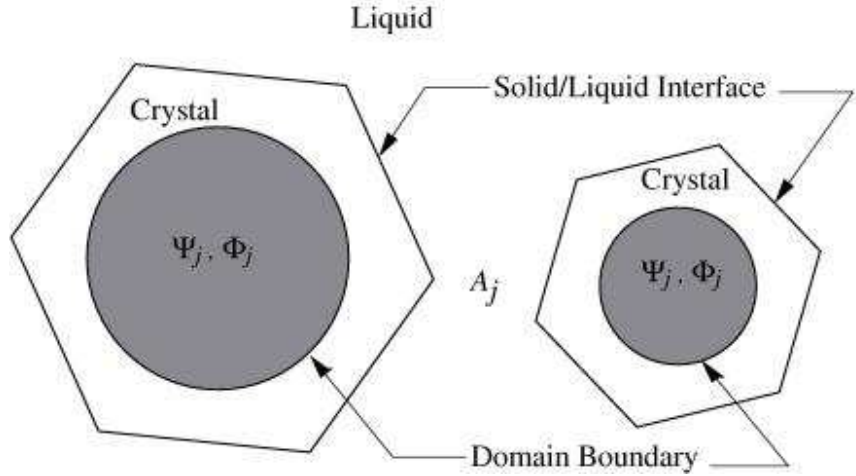}}
\caption{\label{fig:domain_div} Sketch illustrating the idea of
selectively evolving the complex amplitude and phase/amplitude
equations in different regions of the computational domain. $\Psi_j$
and $\Phi_j$ are evolved inside the shaded circles, which fall well
inside the crystalline phase, while the real and imaginary
components of $A_j$ are evolved everywhere else.}
\end{center}
\end{figure}

In order to implement our idea of evolving Eq.~(\ref{eq:cmplx_rg}), and
Eqs.~(\ref{eq:reduced_amp}) and (\ref{eq:reduced_phase}) selectively
within different regions, we begin by dividing the computational domain
into two regions where each set of equations may be evolved
simultaneously in a stable fashion. The region where $A_j$ is computed
in terms of its real and imaginary parts is called X, and the region
where $\Psi_j$ and $\Phi_j$ are computed is called P. We
ensure that subdomain P is well separated from locations with sharp
gradients, such as interfaces and defects. Otherwise, errors resulting
from our approximations may grow rapidly, causing X to invade P, which
will in turn require us to solve the complex equations everywhere. We
will further assume that the decomposition algorithm is implemented
after a sufficient time, when initial transients have passed, and that
the crystals are evolving steadily, which implies that $\Psi_j$ inside
the crystals has reached some maximum saturation value $\Psi_j^{max}$.
The scenario we have in mind is sketched in Fig.~\ref{fig:domain_div},
with P constituting the shaded regions and all other regions correspond
to X.

The pseudo-code shown in Algorithm \ref{alg:divide_domain} presents a
simple algorithm to achieve this decomposition. The algorithm first
determines nodes with $\Psi_j$ exceeding some minimum value
$\gamma\Psi_j^{max}$, and $|\nabla\Psi_j|$ beneath some limit
$\epsilon_1$. The nodes satisfying these conditions constitute domain P,
while those failing to, constitute X. The P nodes are then checked again
to see if the quantity $|\nabla(|\nabla\Phi_j|)|$ is under some limit
$\epsilon_2$. Nodes in set P that fail to satisfy this condition are
placed in set X. The parameters $\gamma$, $\epsilon_1$, and $\epsilon_2$
are chosen to ensure the largest possible size of set P. A
small problem is caused by the fields $\Psi_j$ and $|\nabla\Phi_j|$ not
being perfectly monotonic. As the limits $\epsilon_1$ and $\epsilon_2$ are
sharp, several small islands (clusters of grid points)  of X or P can be
produced, which are detrimental to numerical stability. We have resolved
this issue via a coarsening algorithm that eliminates very small clusters
of X and P.

\begin{algorithm}[htbp]
\caption{\label{alg:divide_domain} (Color online) Domain decomposition. The parameters
$\gamma$, $\epsilon_1$, and $\epsilon_2$ are
heuristic.}
\begin{algorithmic}
\STATE Compute $\Psi_j^{max}$
\STATE $\Psi_j^{max} = \gamma\times\Psi_j^{max}$
\STATE \COMMENT{Split domain based on the magnitude of $\Psi_j$ and $|\nabla\Psi_j|$}
\FOR[loop over all nodes]{i = 1 to maxnode}
\STATE count = 0
\FOR[loop over amplitude components]{j = 1 to 3}
\IF{$\Psi_j \ge \Psi_j^{max}$ and $|\nabla\Psi_j| \le \epsilon_1$}
\STATE count++
\ENDIF
\ENDFOR
\IF{count = 3}
\STATE domain = P   \COMMENT{passed test, solve phase/amplitude equations}
\ELSE
\STATE domain = X   \COMMENT{failed test, solve complex equations}
\ENDIF
\ENDFOR
\STATE \COMMENT{Split domain based on $\left|\nabla\left(|\nabla\Phi_j|\right)\right|$}
\FOR[loop over all nodes]{i = 1 to maxnode}
\STATE count = 0
\IF[check only nodes that passed previous test]{domain = P}
\FOR[loop over amplitude components]{j = 1 to 3}
\IF{$|\nabla(|\nabla\Phi_j|)| \le \epsilon_2$}
\STATE count++
\ENDIF
\ENDFOR
\IF{count $\ne$ 3}
\STATE domain = X   \COMMENT{failed test, solve complex equations}
\ENDIF
\ENDIF
\ENDFOR
\end{algorithmic}
\end{algorithm}

%\begin{figure}[h]
%    \centerline{\includegraphics*[height=6in,width=5in]{algorithms/algo1}}
%    \caption{ Domain decomposition Algorithm. The parameters $\gamma$,
%$\epsilon_1$, and $\epsilon_2$ are heuristic.}
%\label{alg:divide_domain}
%\end{figure}

\begin{figure}[htb]
\begin{center}
\subfigure[\label{dom1}$t=120$]
{\includegraphics[height=2.5in,angle=0]{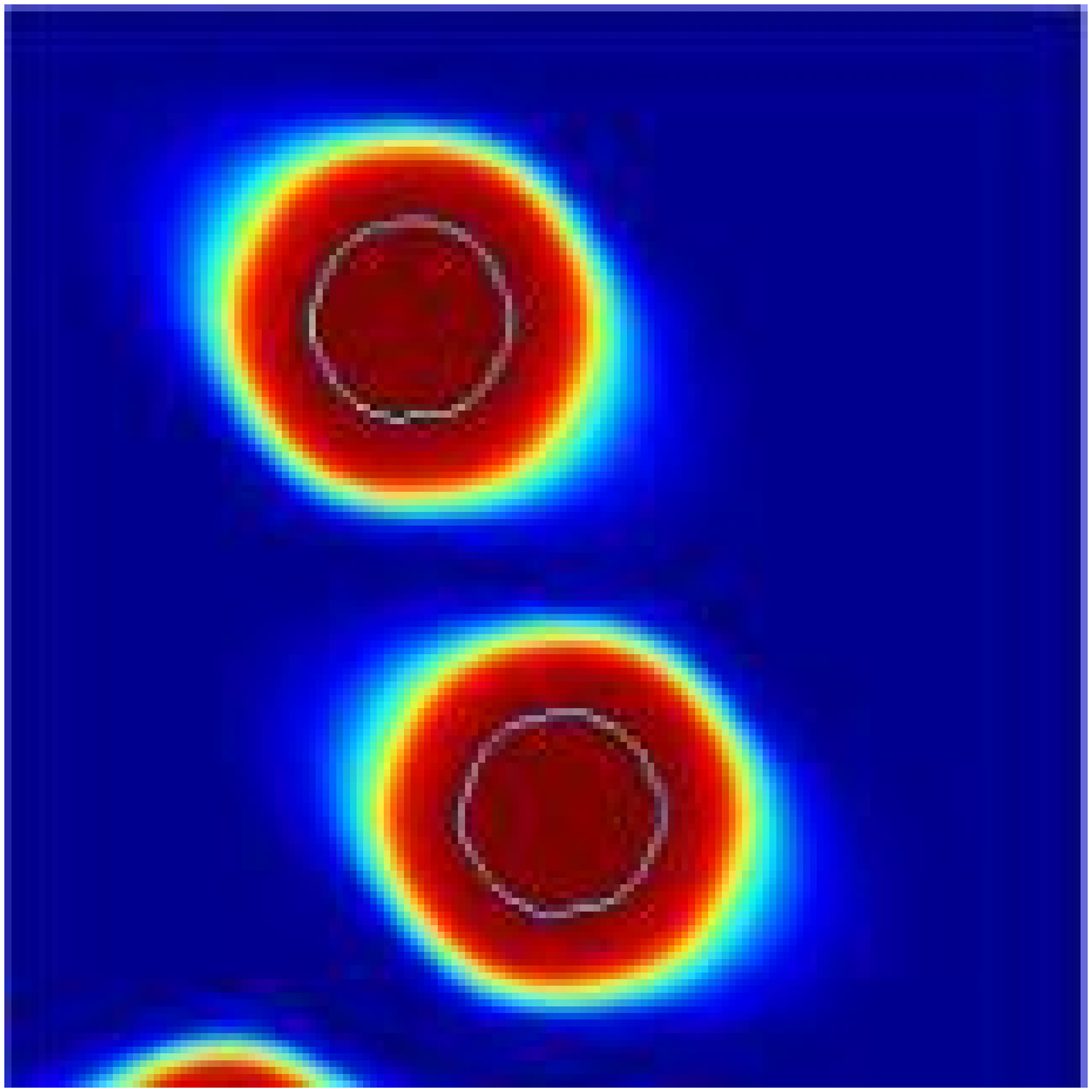}}
\subfigure[\label{dom2}$t=200$]
{\includegraphics[height=2.5in,angle=0]{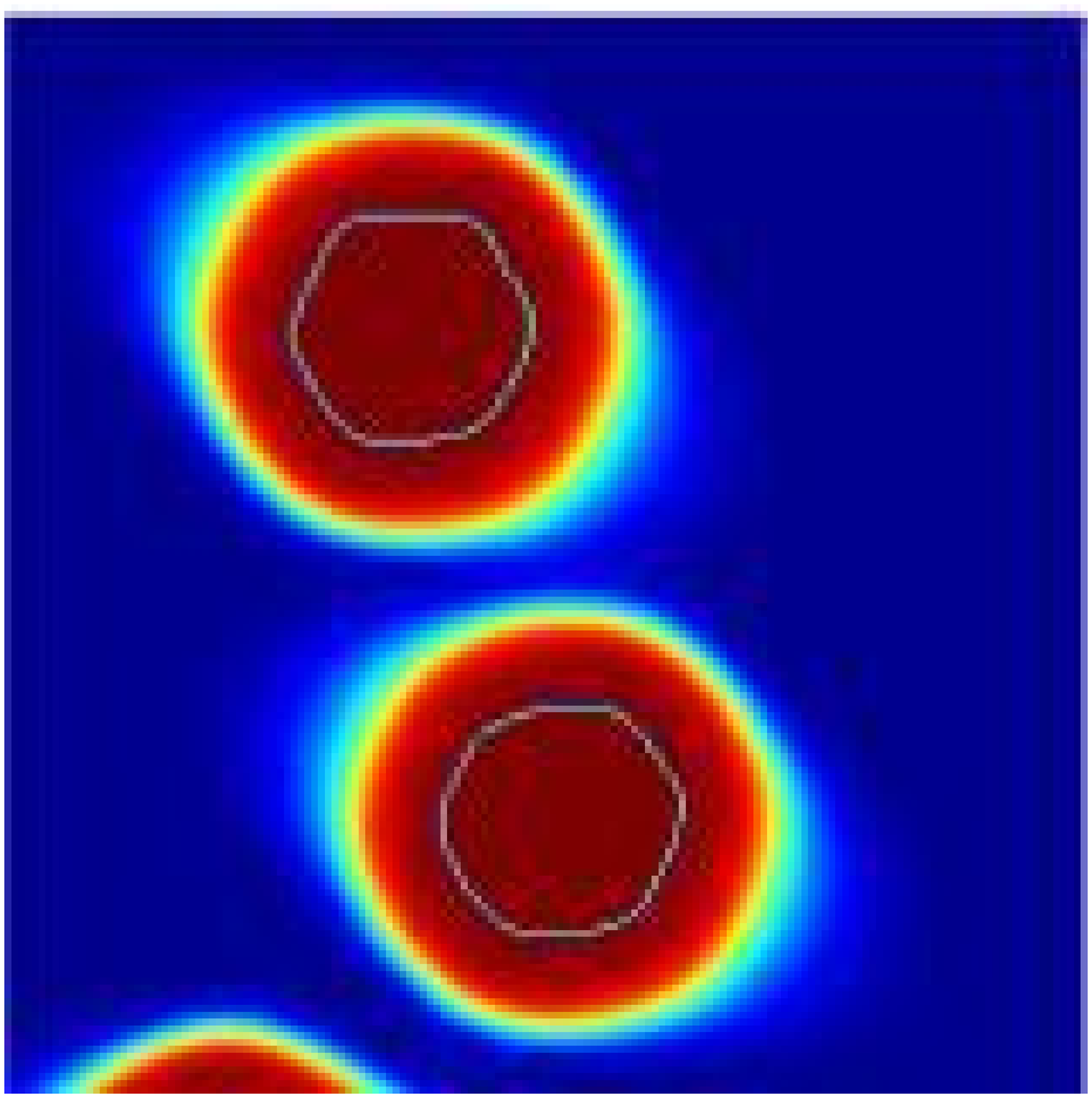}} \\
\subfigure[\label{dom3}$t=280$]
{\includegraphics[height=2.5in,angle=0]{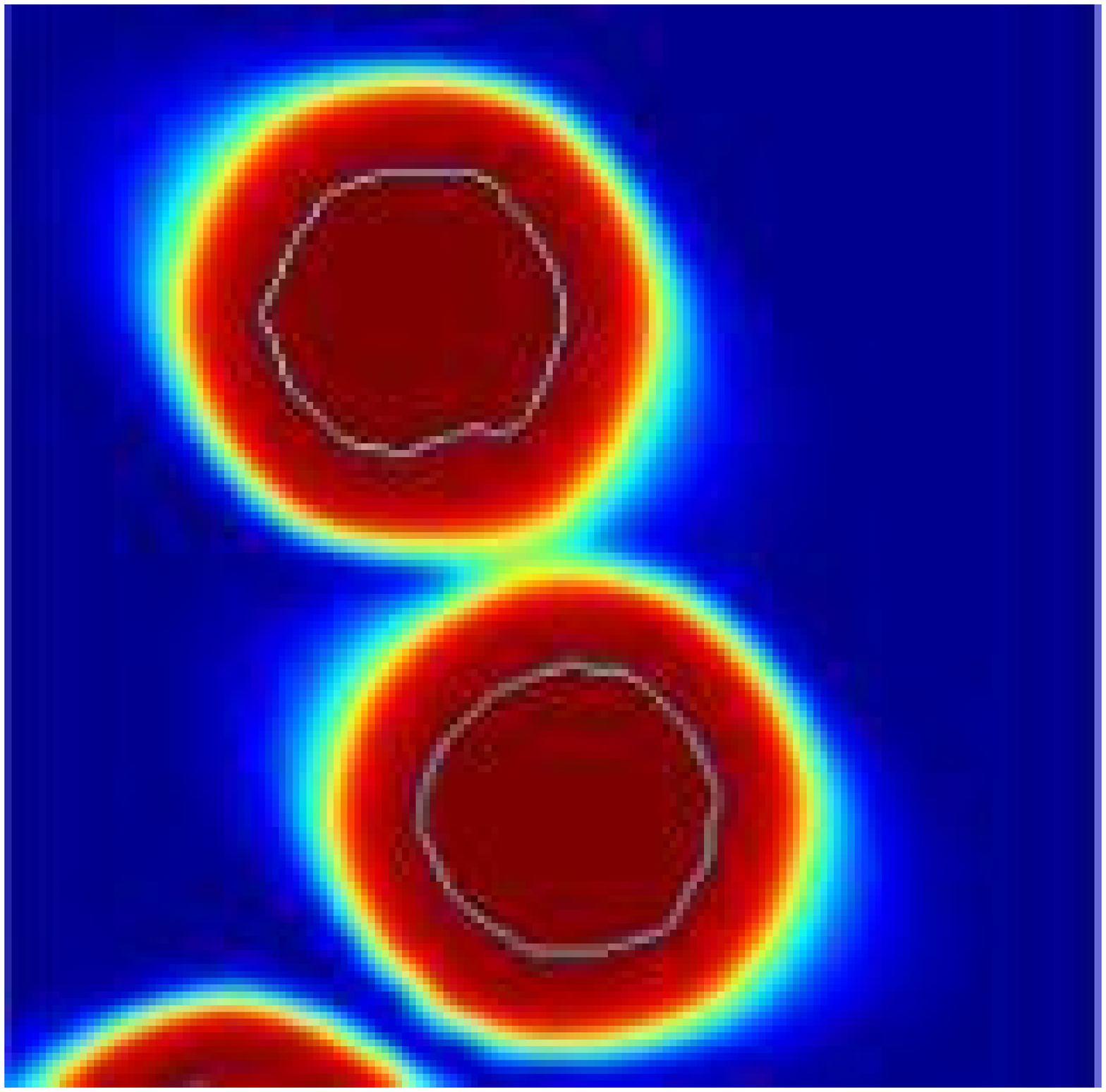}}
\subfigure[\label{dom4}$t=360$]
{\includegraphics[height=2.5in,angle=0]{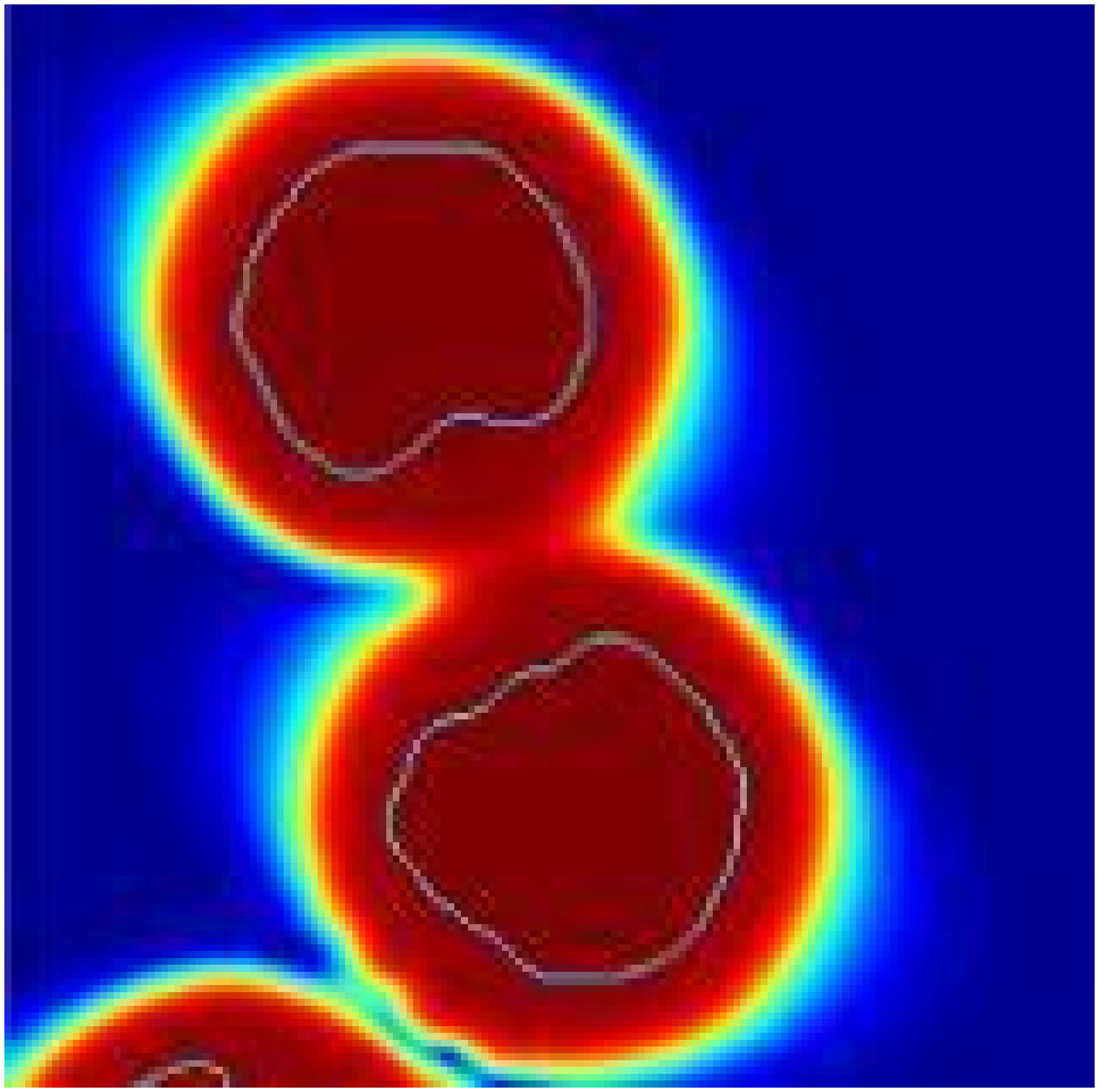}}
\caption{\label{fig:dom_div_uni} (Color online) Filled contour plot showing the
time evolution of three misoriented crystals. The field plotted is
$\Psi_3$. Superimposed on the plot as solid curves are the
boundaries that separate domains X and P, with P being enclosed by
the curves.}
\end{center}
\end{figure}

Fig.~\ref{fig:dom_div_uni} shows results from a uniform grid
implementation of Algorithm \ref{alg:divide_domain}. No islands
are present, as the algorithm decomposes the domain in an
unsupervised manner. It is noteworthy that domain boundaries are
distorted in Figs.~\ref{dom3} and \ref{dom4} in response to the
formation of a grain boundary between the two crystals, after being
roughly hexagonal at earlier times. The fact that the domain
separatrices maintain a safe distance from the grain boundary
ensures that the phase/amplitude equations are not evolved in
regions containing sharp gradients in $\nabla\Phi_j$. Parameter
values used were $\gamma=0.85$, $\epsilon_1=0.0005$, and
$\epsilon_2= 0.003$.

The remarkable feature of our numerical scheme is that solving
different sets of equations in X and P does not require doing
anything special near the domain boundaries, such as creating
``ghost'' nodes outside each domain, or constraining solutions to
match at the boundaries. Both sets of variables, \{$\Psi_j,\Phi_j$\}
and \{$A_j$\}, are maintained at all grid points irrespective of the
domain they belong to, with one set allowing easy computation of the
other \footnote{For example in domain X where \{$A_j$\} is the field
variable, $\Psi_j=|A_j|$ and $\Phi_j=\Im(A_j)/\Re(A_j)$, whereas in
domain P where \{$\Psi_j,\Phi_j$\} are the field variables,
$\Re(A_j)=\Psi_j\cos(\Phi_j)$ and $\Im(A_j)=\Psi_j\sin(\Phi_j)$.}.
Therefore the transition between the two domains is a continuous one
in terms of field variables, which allows the finite difference
stencils in Eqs.~(\ref{eqn:lapdisc}) and (\ref{eqn:graddisc}) to be
applied to the respective fields without any modification near
domain boundaries.

\section{\label{sec:AMR}Solving the RG Equations with Adaptive Mesh Refinement}

As highlighted in earlier work \cite{Pro98a,Pro99c,Jeong1,Jeong2}
the use of dynamic adaptive mesh refinement alters the numerical
mesh resolution dynamically such as to place high resolution near
phase boundaries and a very low resolution in bulk regions where
there is little activity. This dramatically reduces computer memory
requirements, allowing larger systems to be simulated. It also
significantly reduces overall simulation times. The determining
factor guiding the use or otherwise of an AMR technique to solve a
particular problem is the simple criterion
\begin{equation}
\frac{{\rm Interface\,length\,(or Area)}}{{\rm Domain\, area\,(or Volume)}} \ll 1
\end{equation}

The phase/amplitude RG equations discussed above are precisely in
the class of problems that can benefit from and is amenable to
adaptive mesh refinement. Indeed, as will be shown below, the
speedup in time contained intrinsically by the physics of the PFC
equation is complemented by the concomitant bridging of length
scales afforded by the RG equations solved adaptively.

We solved the RG equations discussed above using a new C++ adaptive
mesh refinement (AMR) algorithm that uses finite differences (FD) to
resolve spatial gradients \cite{Jun2006}. While it is typical to use
the finite element method (FEM) in situations involving non-structured
meshes, adaption using finite differences schemes allows approximately
a 5-10 fold increase in simulation speed (measured as CPU time per
node) over previous (FEM) formulations involving traditional phase
field models \cite{Pro98a,Pro99c}. This improvement increases further
still in cases where model equations contain spatial gradients or order
higher than two, such as in the case of
Eqs.~(\ref{eq:cmplx_rg}),~(\ref{eq:reduced_amp}) and
(\ref{eq:reduced_phase}). The basic reason for the difference in speeds
is that FEM formulations generally have more overhead due to their
reliance on local matrix multiplication at multiple Gauss quadrature
points. This overhead time becomes even more pronounced when using
elements of order higher than two, as is required if an FEM formulation
is to be used to resolve the spatial derivatives involved with the RG
equations in this work.

\subsection{AMR Algorithm}

\begin{figure}[h]
    \centerline{\includegraphics*[width=4in]{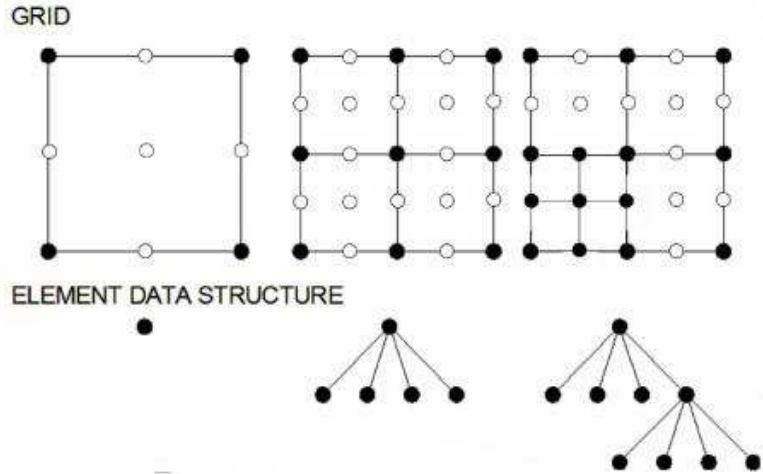}}
    \caption[]{\linespread{0.75} Schematic of a quad-tree data structure
     used in adaptive meshing.  An element splits and creates 'children'
     beneath it in the tree structure. Nodes are created at the corner of
     each element and 'ghost' nodes are placed in the center of elements
     and along edges that have no real nodes, to accommodate resolution
     mismatches. Ghost nodes are approximated by interpolation from the
     element to which they belong. }
\label{fig23}
\end{figure}

At the heart of our algorithm is a routine that creates a non-uniform
mesh that increases nodal density in specific regions according to a
local error estimator. Nodes are grouped into pseudo-elements, managed
by dynamic tree data structures, as used in a finite element
formulation by Provatas et al \cite{Pro99c}. The quad-tree structure
illustrated in Figure \ref{fig23} is a hierarchy of elements where
every level deeper in the tree results in elements of higher
refinement. Every element has associated with it 4 corner `nodes' and 5
`ghost' nodes: 4 in the center of each edge and 1 in the center of the
element. The ghost nodes facilitate interpolation when neighboring
elements are at different levels of refinement. Each node is a
structure containing field values, such as phase, amplitude and the
real and complex components of $A_j$. A node also contains information
about its nearest neighbors. Edge ghost nodes can serve as field nodes
if a resolution mismatch occurs across neighbor elements. This can be
seen in the schematic in figure \ref{fig23}. The field equations are
not solved at ghost nodes, but are instead interpolated  linearly from
the nodal values of the element (or edge) to which they belong.  The
inclusion of ghosts nodes simplifies the calculation of local
derivatives.

An adapter algorithm refines/unrefines the tree structure by using a
user-defined error estimator computed for each element.  Refinement is
done by bisection as shown in Figure \ref{fig23}, and unrefinement is
done by fusing four child elements into their parent element.
Once refinement/unrefinement of all elements is complete, the adapter
produces an array of nodes, each of which is the center of a local grid.
The equation solver accepts this array as input and then solves the
equations at these nodal points. This method of node organization
modularizes our algorithm and allows the solver and adapter to be
separately parallelized.

The mesh data structure contains nodes with detailed knowledge of their
local neighbors, each of which exist at the center of a $5\times 5$
uniform mesh (See Fig.~\ref{fig:unstructured}). During adaptation, the
data structure applies rules that either increase (if higher accuracy
is needed) or decrease (to decrease memory requirements) the size of
this local nodal mesh. Also during adaptation, the data structure and
its associated elements and nodes respect the following six rules (3
applied to each element and 3 to each node).  Rule 1 ensures mesh
cohesion and maintains accuracy in the solution of the PDEs.

\begin{description}
\item[RULE 1:] Neighboring elements can vary by at most one level

\item[RULE 2:] All elements contain 9 nodes, real or ghost

\item[RULE 3:] Element neighbors are all at the same level
(Therefore elements may have $NULL$ as a neighbor)
% JON From Jon: I don't understand what this means.

\item[RULE 4:] Node neighbors are all at the same level (Nodes
will never have $NULL$ as a neighbor, but may instead have a ghost
as a neighbor)

\item[RULE 5:] Each node is at the center of what is defined as a
uniform mini-mesh

\item[RULE 6:] Each node is assigned the resolution ($\Delta x$) of
the most refined element attached to it.

\end{description}

The adaptive process is controlled primarily at the tree level, but invokes
function calls inside of the element and node structures. Its basic flow
is illustrated in Algorithm \ref{alg:Regrid}. This process allows an
element-by-element examination using recursion to maintain the rules
above.  Once the process of adaptation is complete the adapter creates an
array of nodes each with the index of its neighbors in the array, which is
then used to solve the equations before adapting again. We also note that
element `leaves' are stored by their resolution level in an array of
element lists. Element resolution is restricted to vary by at most one
level compared to the element neighbors (Rule 1).

\begin{algorithm}[htb]
\caption{\label{alg:Regrid} Regridding algorithm}
\begin{algorithmic}
\STATE Delete the Ghost Pointer List
  \WHILE{Some elements have split or unsplit}
    \STATE Check all elements for unsplitting criteria
    \STATE Check all elements for splitting criteria
  \ENDWHILE
\STATE Update node neighbors
\STATE Build Ghost List
\STATE Set Ghost Averaging Information
\end{algorithmic}
\end{algorithm}

%\begin{figure}[h]
%    \centerline{\includegraphics*[height=2in,width=5in]{algorithms/algo3}}
%    \caption[]{\linespread{0.75}\em{Regridding Algorithm }}
%\label{alg:Regrid}
%\end{figure}

Element splitting (refinement) is the dominant process
in the algorithm, taking precedence over element coarsening (i.e. fusing
four children elements into their parent). Elements are searched one
refinement level at a time, starting from the second highest level of
resolution. Each element is considered for splitting using an error
criterion computed for that element. If splitting is required, an element
data structure it is pushed onto a stack, where its neighbors are
subsequently checked against Rule 1. If splitting will violate Rule 1, the
neighbors are recursively split until all refined elements satisfy Rule
1. When an element is split, it and its updated neighbor elements generate
new real and ghost nodes, as well as information about their neighbors.  The
splitting algorithm is illustrated in Algorithm \ref{alg:Split}.

\begin{algorithm}
\caption{\label{alg:Split} Splitting Algorithm}
    \begin{algorithmic}
        \FOR{ElementLevel=maxLevel-1 to 0}
            \WHILE{Level is not empty}
                \IF{Element Splitting criteria is met}
                    \STATE PUSH Element onto stack
                    \WHILE{Stack is not empty}
                        \IF{Top of Stack's neighbors need splitting}
                            \STATE Push Needed neighbors onto stack
                        \ELSE
                            \STATE Pop the Element from the Stack
                            \STATE Create 4 new children elements
                            \STATE Push children onto ElementLevel+1
                            \STATE Add New Nodes to the Node List
                            \STATE Remove former Ghosts from Ghost List
                        \ENDIF
                    \ENDWHILE
                \ENDIF
            \ENDWHILE
        \ENDFOR
    \end{algorithmic}
\end{algorithm}

The unrefinement algorithm  starts at the lowest level of refinement.
Again, Rule 1 above must be imposed. The elements are examined to see if a
parent requires splitting. If it does not, the parent has its four
child elements eliminated, assuming their relationship with the
neighbors allows it (i.e. Rule 1). Recursive unsplitting of elements is
not allowed. As in the case of refinement, during unrefinement, element
neighbors and node neighbors are updated. The flow of unsplitting is
shown in Algorithm \ref{alg:UnSplit}.

\begin{algorithm}
\caption{\label{alg:UnSplit} UnSplitting Algorithm}
    \begin{algorithmic}
        \FOR{ElementLevel=maxLevel-1 to 0}
            \WHILE{Level is not empty}
                \IF{Element Splitting criteria is NOT met}
                    \IF{Element's Parent's Splitting criteria is NOT met}
                        \IF{Parent's Children Have NO Children}
                            \IF{Parent's neighbor's are Same level or One level Lower}
                                \STATE ADD Parent to ElementLevel+1
                                \STATE REMOVE Parent's Children from ElementLevel
                                \STATE REMOVE Unneeded nodes from node list
                                \STATE REMOVE Ghosts from Ghost list
                                \STATE Center Node Becomes a Ghost
                                \STATE Edge Nodes Become Ghosts depending on neighbor levels
                                \STATE DELETE Children
                            \ENDIF
                        \ENDIF
                    \ENDIF
                \ENDIF
            \ENDWHILE
        \ENDFOR
    \end{algorithmic}
\end{algorithm}

As discussed above, nodes follow rules 4, 5 and 6.  This is
enforced by the introduction of ghost nodes in elements and by
maintaining rule 1. Rule 5 is maintained by creating a list of local
nodes (ghost or real) and by maintaining rule 1 during splitting. Rule
6 determines which node neighbors are chosen and the local node spacing
($\Delta x$). A real node will contain the minimesh on which is applied
the governing equations; a ghost node provides instructions on how to
interpolate the values needed for real node calculations.

%Elements follow rules 1, 2 and 3.  Rule 1 is the core to the adaptive
%process and would be very difficult to maintain nodal cohesion if it were
%not followed.  Rule 1 also provides stability to the solutions of PDEs on
%the non-uniform mesh.  This rule states that no element can be more than
%one level of refinement different from any of its immediate neighbors.
%
%The structure of the element contains pointers to its 4 children, its
%parent, to 9 nodes and to its 9 immediate element neighbors. Rule 2 states
%that all of these neighbors must be of the same level. Each element also
%determines whether it should be split, the user can provide a variety of
%methods based on the information provided within a certain element.

\subsection{Handling of Ghost Nodes in The Hybrid Formulation}

\begin{figure}[htb]
\begin{center}
{\includegraphics[height=2.5in,angle=0]{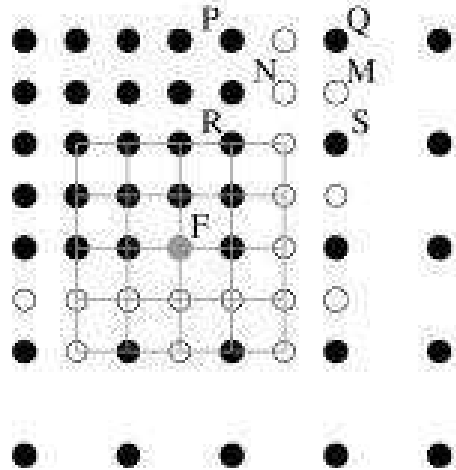}}
\caption{\label{fig:unstructured} Schematic showing a portion of the
adaptive grid where the refinement level changes. Filled circles (and
node F) are real nodes where the fields are computed, whereas the open
circles are non-computational ghost nodes where the fields are
interpolated.}
\end{center}
\end{figure}

A very useful feature of the AMR implementation outlined in the
previous section is that each node sits at the center of a uniform
$5\times 5$ mini-grid, illustrated in Fig.~\ref{fig:unstructured}.  Let
us focus on the node represented by the lightly shaded circle (labeled
F).  The data structure ensures that node F has access to all other
nodes on the wireframe. The advantage of this construction is that it
allows us to use the uniform grid finite difference stencils for the
Laplacian and gradient operators in Eqs.~(\ref{eqn:lapdisc}) and
(\ref{eqn:graddisc}) respectively, instead of modifying them node-wise
to accommodate variations in grid spacing.  This requires the
introduction of ghost nodes, shown as open circles in
Figure~\ref{fig:unstructured}.

The scheme used to interpolate values at the ghost nodes is a potential
source of error in the numerical solution, and must be chosen
carefully. $\Psi_j$, $\partial\Phi_j/\partial x$,
$\partial\Phi_j/\partial y$, $\Re(A_j)$ and $\Im(A_j)$ are very
smoothly varying functions, and therefore we \emph{linearly}
interpolate their values to the ghost nodes. Values at ghosts residing
on element \footnote{Here, we define an element as a square with real
corner nodes.} edges, for example node M in
Fig.~\ref{fig:unstructured}, are obtained by averaging values of the
two end nodes Q and S, whereas values at  ghosts residing at the center
of an element, node N for example, are obtained by averaging the values
at the four corner nodes, P, Q, R, and S. We have found this
interpolation scheme to be quite stable. We note however that, given
the near-periodic variations in $\Re(A_j)$ and $\Im(A_j)$, especially
in misoriented grains, higher order interpolation functions (such as
cubic splines) could improve solution accuracy, while strongly
enforcing continuity of fields across elements. This issue will be
examined in future work.

The interpolation of $\Phi_j$ at the ghost nodes is a little more delicate.
Since $\Phi_j$ is a discontinuous function, a simple average of the values
at the neighboring real nodes may not always give the correct answer,
especially because the grid does not resolve the discontinuities. Even if
it did, a simple average could lead to the wrong result. As an example,
consider the two real nodes Q and S in Fig.~\ref{fig:unstructured}, with
values $\Phi_1^Q=\pi-\delta_1$ and $\Phi_1^S=-\pi+\delta_2$ where
$\delta_1$ and $\delta_2$ are very small but positive real numbers, on
either side of a discontinuity in $\Phi_1$. We wish to determine the value
at the ghost node M that lies between Q and S. Although the values of
$\Phi_1$ at Q and S are essentially equivalent in phase space, differing
in magnitude by approximately $ 2\pi$, a simple average gives
$\Phi_1^M=(\delta_2-\delta_1)/2\approx0$, which is quite wrong.

In order to interpolate correctly we need to make use of
$\nabla\Phi_j$. For example in the above case, the total change in
the phase from Q to S is obtained by integrating the directional
derivative of $\Phi_1$ along the edge QS, i.e.
\begin{equation}\label{eq:phase_inter}
\Delta\Phi_1^{QS} = \int_Q^S\nabla\Phi_1\cdot d\mathbf{r} =
\int_{y=y^Q}^{y^S}\frac{\partial\Phi_1}{\partial y}dy.
\end{equation}
Eq.~(\ref{eq:phase_inter}) can be evaluated numerically, and the
accuracy of the result depends on how well $\partial\Phi_1/\partial
y$ is approximated. Consistent with our earlier assumptions, we
approximate $\partial\Phi_1/\partial y$ as piecewise constant where
\begin{equation}
\frac{\partial\Phi_1}{\partial y} =
\frac{1}{2}\left(\left.\frac{\partial\Phi_1}{\partial
y}\right|_S+\left.\frac{\partial\Phi_1}{\partial y}\right|_Q\right)
\end{equation}
which leads to
\begin{equation}
\Delta\Phi_1^{QS} =
\frac{1}{2}\left(\left.\frac{\partial\Phi_1}{\partial
y}\right|_S+\left.\frac{\partial\Phi_1}{\partial
y}\right|_Q\right)\left(y^S-y^Q\right).
\end{equation}
Since $\partial\Phi_1/\partial y$ is constant along the edge QS,
$\Phi_1$ must vary linearly along QS. Hence at node M,
\begin{equation}
\Phi_1^{M}=\Phi_1^{S}+\frac{1}{2}\Delta\Phi_1^{QS}\;:\;\Phi_1^{M}\in[-\pi,\pi].
\end{equation}
Interpolation of $\Phi_j$ at element center ghost nodes, such as N,
is done in a similar manner by interpolating linearly from ghost
nodes at the centers of opposite element edges. Once again, this
scheme might be improved by choosing higher order polynomials to
approximate  $\nabla\Phi_j$ inside elements.

\subsection{Refinement Criteria in the Hybrid Formulation}

Traditionally, AMR algorithms rely on some kind of local error
estimation procedure to provide a criterion for grid refinement.
Zienkiewicz and Zhu \cite{Zei87} developed a simple scheme for finite
element discretization of elliptic and parabolic PDEs by computing the
error in the gradients of the fields using higher order interpolation
functions. Berger and Oliger \cite{berger84} on the other hand
estimated the local truncation error of their finite difference
discretization of hyperbolic PDEs via Richardson extrapolation.
Depending on the equations being solved and the numerical methods being
used, one scheme may be more effective than another. We use a very
simple and computationally inexpensive refinement criterion that works
nicely for our equations, based purely on gradients in the various
fields.

The outline of the algorithm used to decide whether or not to split an
element is given in Algorithm \ref{alg:refine}. The algorithm initially
computes absolute changes in the real and imaginary parts of $A_j$, and
the $x$ and $y$ components of $\nabla\Phi_j$ in each element. We use
absolute differences in place of derivatives in order for the refinement
criterion to be independent of element size. We note that prior to
implementing this algorithm, the domain decomposition algorithm described
above (see Figure \ref{alg:divide_domain}) needs to be called first in
order to split the computational domain into subdomains X and P.

The process begins by examining an element flag to see if the element
lies on the separatrix between X and P, or in $k$ layers from the
boundary, within the P subdomain. If so this element is split.
This ensures that the fields are always resolved on the interface
between X and P, and just within the boundary on the P side. The latter
is required because of the higher order derivative operations that need
to be performed while evolving the complex amplitude equations in X.

\begin{algorithm}[htb]
\caption{\label{alg:refine} Criteria for element splitting}
\begin{algorithmic}
\STATE \COMMENT{N1, N2, N3 and N4 are the element nodes in clockwise manner}
\FOR[loop over amplitude components]{i = 1 to 3}
\STATE \COMMENT{Change in real part of $A_i$ over element}
\STATE $DR_i = |\mbox{N1}\rightarrow\Re(A_i)-\mbox{N2}\rightarrow\Re(A_i)| +
|\mbox{N2}\rightarrow\Re(A_i)-\mbox{N3}\rightarrow\Re(A_i)|$
\STATE $+ |\mbox{N3}\rightarrow\Re(A_i)-\mbox{N4}\rightarrow\Re(A_i)| +
|\mbox{N4}\rightarrow\Re(A_i)-\mbox{N1}\rightarrow\Re(A_i)|$
\STATE \COMMENT{Change in imaginary part of $A_i$ over element}
\STATE $DI_i = |\mbox{N1}\rightarrow\Im(A_i)-\mbox{N2}\rightarrow\Im(A_i)| +
|\mbox{N2}\rightarrow\Im(A_i)-\mbox{N3}\rightarrow\Im(A_i)|$
\STATE $+ |\mbox{N3}\rightarrow\Im(A_i)-\mbox{N4}\rightarrow\Im(A_i)| +
|\mbox{N4}\rightarrow\Im(A_i)-\mbox{N1}\rightarrow\Im(A_i)|$
\STATE \COMMENT{Change in $x$ component of $\nabla\Phi_i$ over element}
\STATE $DGPX_i = |\mbox{N1}\rightarrow\partial\Phi_1/\partial x-\mbox{N2}\rightarrow\partial\Phi_1/\partial x| +
|\mbox{N2}\rightarrow\partial\Phi_1/\partial x-\mbox{N3}\rightarrow\partial\Phi_1/\partial x|$
\STATE $+ |\mbox{N3}\rightarrow\partial\Phi_1/\partial x-\mbox{N4}\rightarrow\partial\Phi_1/\partial x| +
|\mbox{N4}\rightarrow\partial\Phi_1/\partial x-\mbox{N1}\rightarrow\partial\Phi_1/\partial x|$
\STATE \COMMENT{Change in $y$ component of $\nabla\Phi_i$ over element}
\STATE $DGPY_i = |\mbox{N1}\rightarrow\partial\Phi_1/\partial y-\mbox{N2}\rightarrow\partial\Phi_1/\partial y| +
|\mbox{N2}\rightarrow\partial\Phi_1/\partial y-\mbox{N3}\rightarrow\partial\Phi_1/\partial y|$
\STATE $+ |\mbox{N3}\rightarrow\partial\Phi_1/\partial y-\mbox{N4}\rightarrow\partial\Phi_1/\partial y| +
|\mbox{N4}\rightarrow\partial\Phi_1/\partial y-\mbox{N1}\rightarrow\partial\Phi_1/\partial y|$
\ENDFOR
\IF{element on X/P boundary OR k layers inside P}
\STATE Split element and exit
\ELSIF{element inside X}
\STATE count=0
\FOR[loop over amplitude components]{i = 1 to 3}
\IF{$DR_i\ge\epsilon_1$ OR $DI_i\ge\epsilon_1$}
\STATE count++
\ENDIF
\ENDFOR
\IF{count $\ne$ 0}
\STATE Split element and exit
\ENDIF
\ELSE[element is inside P]
\STATE count = 0
\FOR[loop over amplitude components]{i = 1 to 3}
\IF{$DGPX_i\ge\epsilon_2$ OR $DGPY_i\ge\epsilon_2$}
\STATE count++
\ENDIF
\ENDFOR
\IF{count $\ne$ 0}
\STATE Split element and exit
\ENDIF
\ENDIF
\end{algorithmic}
\end{algorithm}

%\begin{figure}[h]
%    \centerline{\includegraphics*[height=6.1in,width=5in]{algorithms/algo5}}
%    \caption[]{\linespread{0.75}\em{Algorithm: Criteria for element splitting }}
%\label{alg:refine}
%\end{figure}

If the element does not split and belongs to X (where $A_j$ are the field
variables), the variations in the real and imaginary parts of $A_j$ are
checked to see if they exceed a certain bound $\epsilon_1$. If any one of
them does, this element is split. If, on the other hand, the element
belongs to P where the phase/amplitude equations are solved, variations in
the $x$ and $y$ components of $\nabla\Phi_j$ are checked to see if they
exceed another limit $\epsilon_2$. If they do, this element is split. If
none of the above criteria are satisfied, the element is not split and is
placed in the list of elements to be checked for coarsening.
Since refinement criteria are recursively applied to the quadtree, the
finest elements are automatically placed around domain separatrices,
solid/liquid interfaces, and defects.

\section{\label{sec:results}Results and computational efficiency}

Using the various approximations and algorithms described in the
previous sections we solved the phase/amplitude and complex equations
simultaneously in different parts of our computational domain using
adaptive mesh refinement. Algorithm \ref{alg:main} shows the flow of
control in the main routine. The complex amplitude equations,
Eq.~(\ref{eq:cmplx_rg}), are initially evolved everywhere until time
$N_{tr}$, when initial transients have dissipated, and the crystals
evolve steadily outward. The domain is then split into subdomains X and
P, following which the reduced phase/amplitude equations,
Eqs.~(\ref{eq:reduced_amp}) and (\ref{eq:reduced_phase}), are evolved
using a forward Euler time stepping scheme in subdomain P. The grid is
refined after a predetermined number of time steps $N_{adapt}$, which
is chosen heuristically. We note that the current implementation can
handle only periodic boundary conditions. Work is currently underway to
enable handling of more general boundary conditions.

\begin{algorithm}[htb]
\caption{\label{alg:main} Flow of control}
\begin{algorithmic}
\STATE InitVar() \COMMENT{Initialize program variables and parameters}
\STATE InitGrid() \COMMENT{decide where to initially refine/coarsen based on initial condition}
\STATE UpdateGhostsFV() \COMMENT{interpolate $A_j$, $\Psi_j$, $\Phi_j$ at ghost nodes}
\STATE ComputePhaseGradients(1) \COMMENT{compute $\nabla\Phi_j$ everywhere}
\STATE UpdateGhostsPG() \COMMENT{interpolate $\nabla\Phi_j$ at ghost nodes}
\FOR[evolve until initial transients subside]{i = 1 to $N_{tr}-1$}
\IF{i $\bmod$ $N_{adapt}$ = 0 OR i = 1}
\STATE AdaptGrid() \COMMENT{uses {\bf Algorithm \ref{alg:refine}}}
\ENDIF
\STATE EvolveComplexAmp() \COMMENT{evolve Eq.~(\ref{eq:cmplx_rg}) everywhere}
\STATE UpdateAllFields() \COMMENT{compute $\Psi_j$ and $\Phi_j$ from $A_j$}
\STATE UpdateGhostsFV()
\STATE ComputePhaseGradients(1)
\STATE UpdateGhostsPG()
\ENDFOR
\STATE DivideDomain() \COMMENT{call {\bf Algorithm \ref{alg:divide_domain}} to split domain into X and P}
\FOR[evolve after transients subside]{i = $N_{tr}$ to $N_{end}$}
\IF{i $\bmod$ $N_{adapt}$ = 0}
\STATE DivideDomain()
\STATE AdaptGrid()
\ENDIF
\STATE EvolveComplexAmp() \COMMENT{evolve Eq.~(\ref{eq:cmplx_rg}) in X}
\STATE EvolvePhaseAmp() \COMMENT{evolve Eqs.~(\ref{eq:reduced_amp}) and (\ref{eq:reduced_phase}) in P}
\STATE UpdateAllFields() \COMMENT{evaluate $\Psi_j$ and $\Phi_j$ in X, evaluate $A_j$ in P}
\STATE UpdateGhostsFV()
\STATE ComputePhaseGradients(i) \COMMENT{compute $\nabla\Phi_j$ in X only, frozen gradient approx.}
\STATE UpdateGhostsPG()
\ENDFOR
\end{algorithmic}
\end{algorithm}

%\begin{figure}[h]
%    \centerline{\includegraphics*[height=5in,width=5in]{algorithms/algo6}}
%    \caption[]{\linespread{0.75}\em{Algorithm: Flow of control }}
%\label{alg:main}
%\end{figure}

\begin{figure}[h]
    \centering
    \begin{subfigure}[t=0]
    {\includegraphics[width=0.3\textwidth]{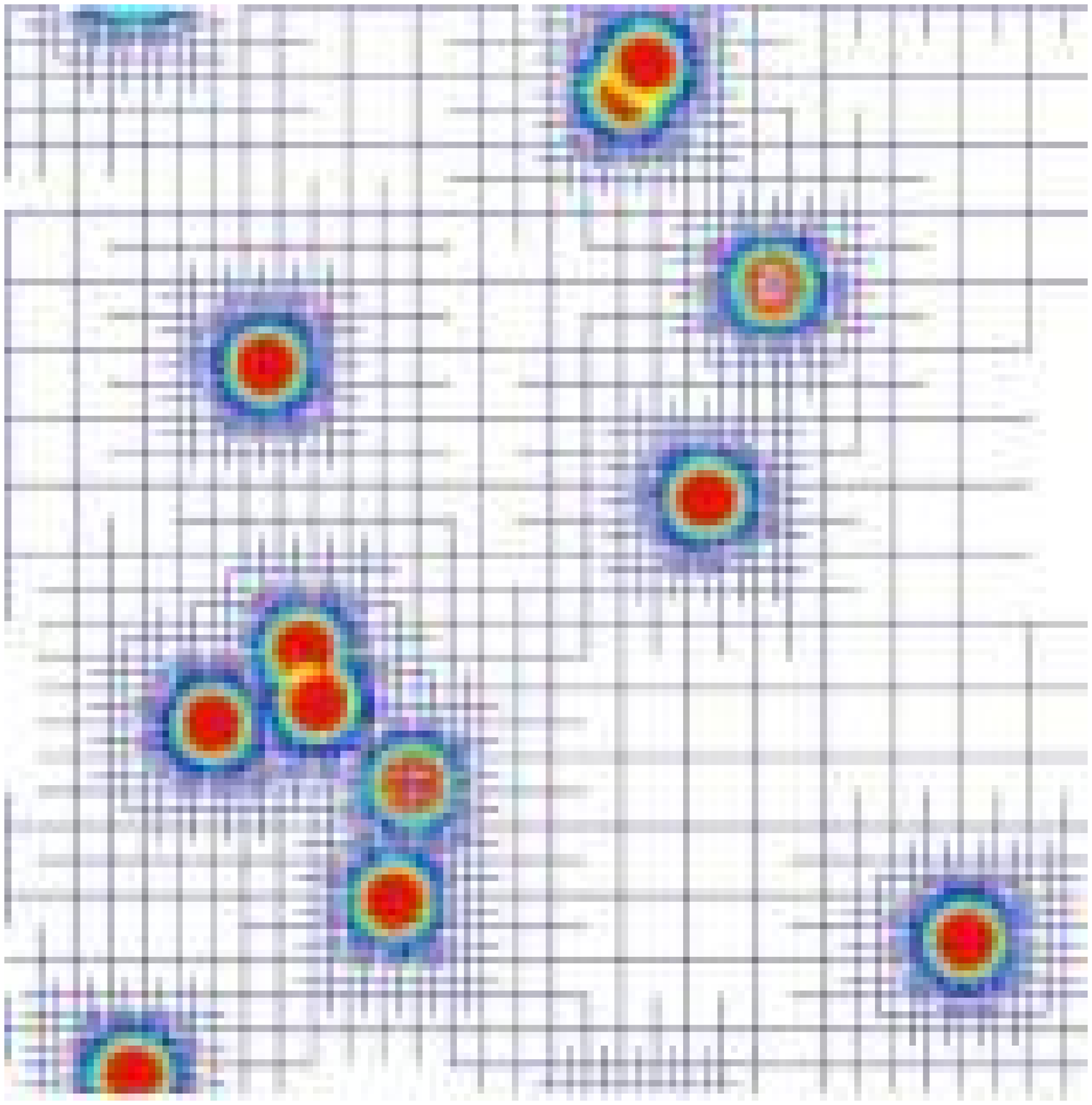}}
    \end{subfigure}
    \begin{subfigure}[t=88]
    {\includegraphics[width=0.3\textwidth]{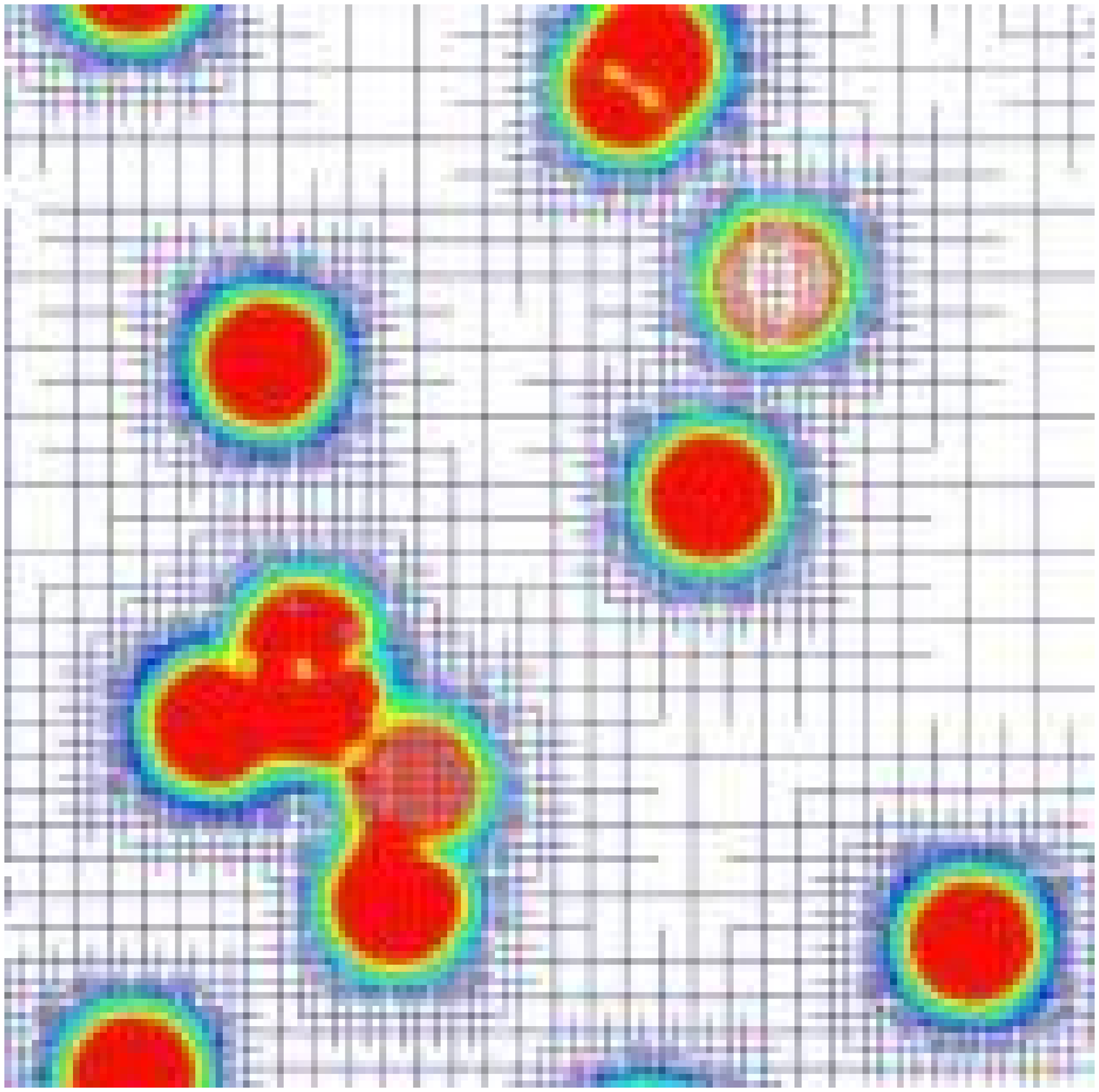}}
    \end{subfigure}
    \begin{subfigure}[t=168]
    {\includegraphics[width=0.3\textwidth]{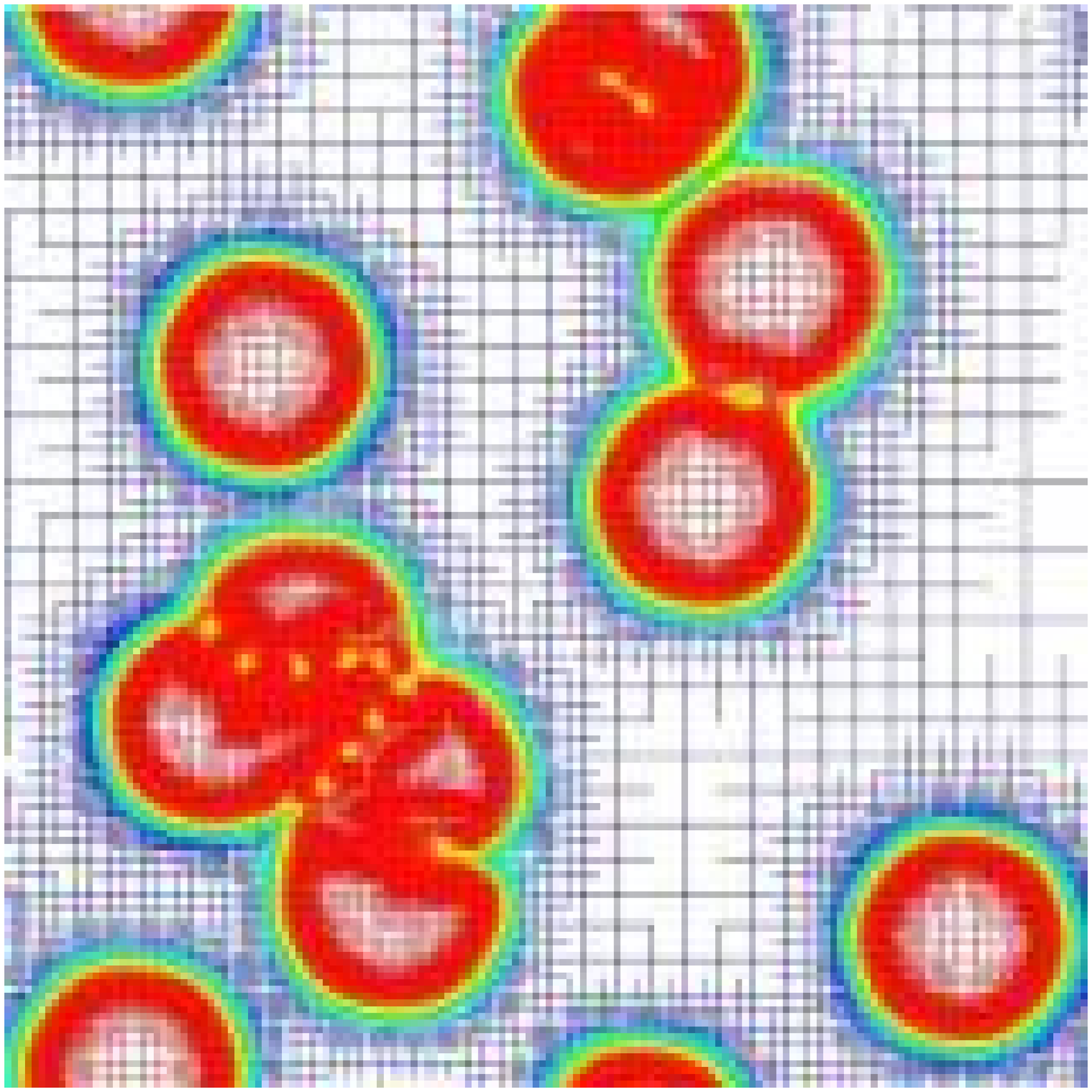}}
    \end{subfigure}
    \begin{subfigure}[t=248]
    {\includegraphics[width=0.3\textwidth]{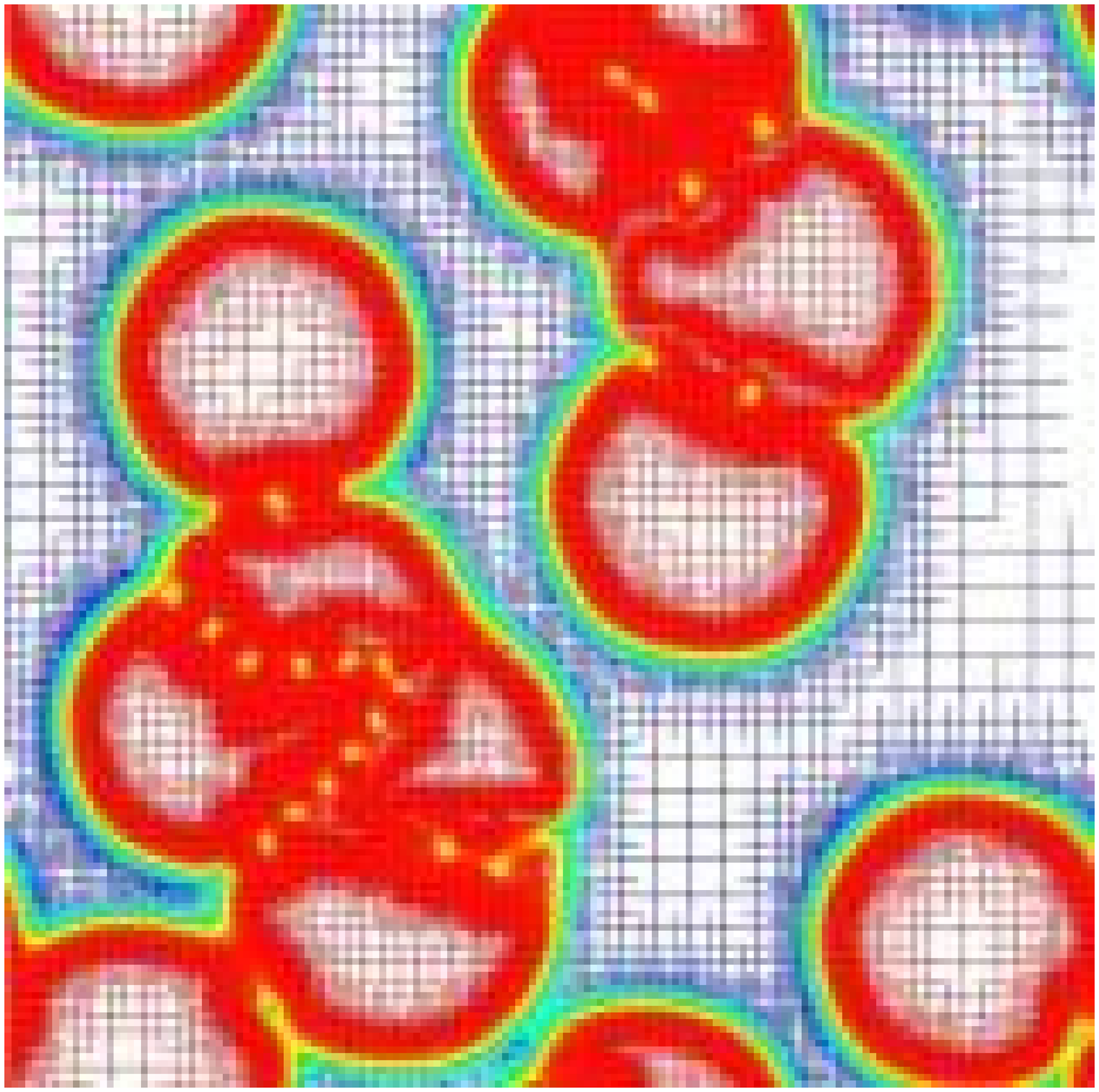}}
    \end{subfigure}
    \begin{subfigure}[t=320]
    {\includegraphics[width=0.3\textwidth]{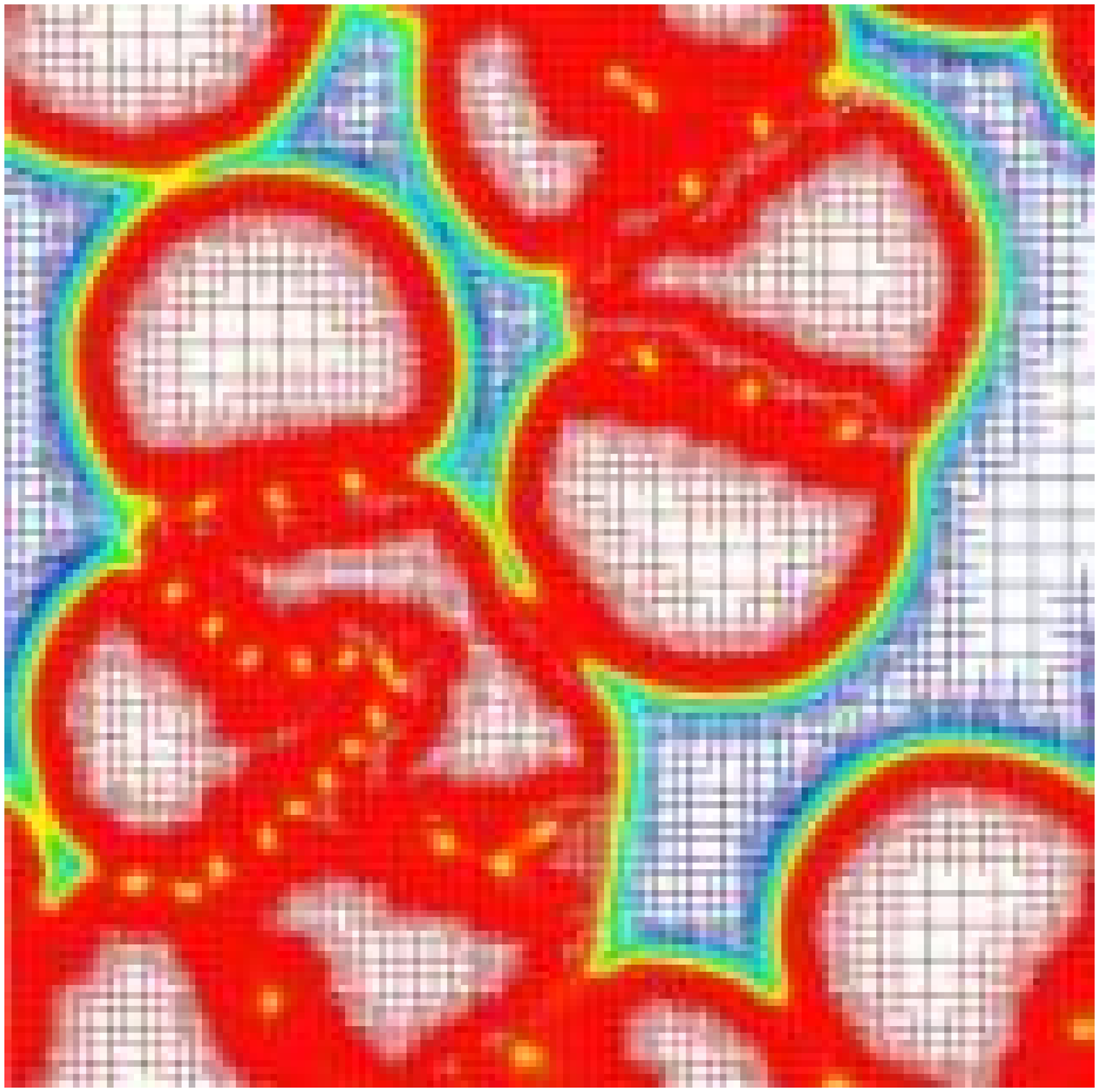}}
    \end{subfigure}
    \begin{subfigure}[t=552]
    {\includegraphics[width=0.3\textwidth]{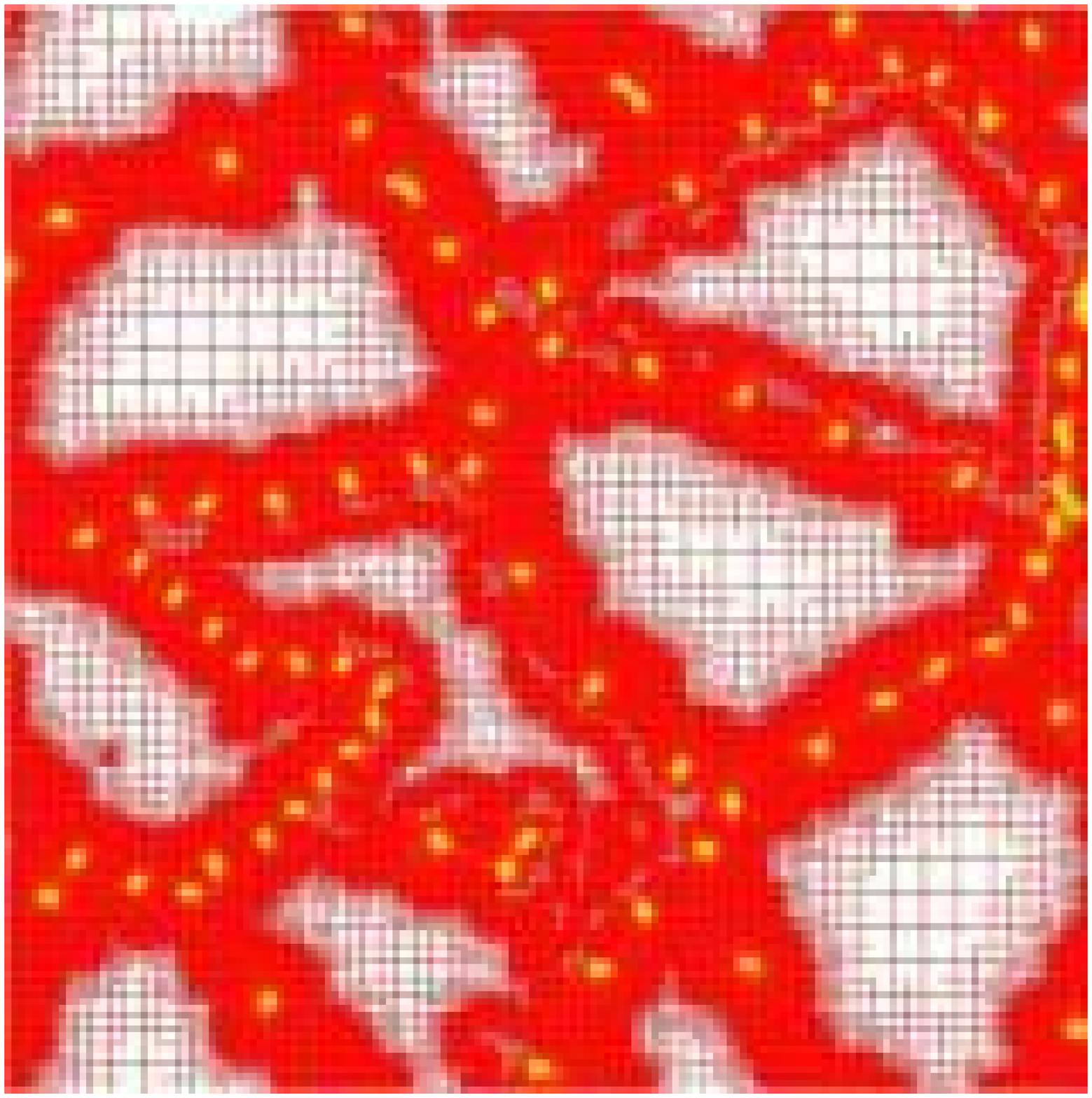}}
    \end{subfigure}
%    \begin{tabular}{ccc}
%        \includegraphics[ height=0.25\textheight]{maeq1}
%    &
%        \includegraphics[ height=0.25\textheight]{maeq4}
%    \\
%    (a) $t=0$ & (d) $t=248$ \vspace{0.5em} \\
%        \includegraphics[ height=0.25\textheight]{maeq2}
%    &
%        \includegraphics[ height=0.25\textheight]{maeq5}
%    \\
%    (b) $t=88$   & (e) $t=320$ \vspace{0.5em} \\
%        \includegraphics[ height=0.25\textheight]{maeq3}
%    &
%        \includegraphics[ height=0.25\textheight]{maeq6}
%    \\
%    (c) $t=168$  & (f) $t=552$ \vspace{0.5em} \\
%    \end{tabular}
%
    \caption{\label{fig:mixed_adaptive} (Color online) Evolution of a polycrystalline
    film simulated with Eq.~(\ref{eq:cmplx_rg}), and
    Eqs.~(\ref{eq:reduced_amp}) and (\ref{eq:reduced_phase}), using our
    adaptive mesh refinement algorithm. The conditions in this simulation
    are identical to those in section \ref{sec:polycryst_cart} and
    Fig.~\ref{fig:cmplx_adaptive}. Note that the grid now coarsens inside
    grains that are misoriented with respect to $\mathbf{k}_j$, and
    ``beats'' are no longer a limitation. The colored field plotted is the
    average amplitude modulus, which is ``red'' inside the  crystal phase,
    ``blue'' in the liquid phase, ``green'' at the crystal/liquid
    interface, and ``yellow'' near defects.   }
\end{figure}

Using this implementation, we simulated the same problem (same initial and
boundary conditions and problem parameters) that was solved adaptively in
section \ref{sec:polycryst_cart} using only the complex amplitude
equations. Figure~\ref{fig:mixed_adaptive} shows the crystal boundaries
and grid structure at various times during the simulation. $N_{tr}$ was
chosen to be 3000 for this simulation. With $\Delta t = 0.04$, this
implies that this simulation is identical to the previous one until
$t=N_{tr}\times\Delta t = 120$. Thus, Figs.~\ref{fig:mixed_adaptive}(a)
and \ref{fig:mixed_adaptive}(b) are identical to
Figs.~\ref{fig:cmplx_adaptive}(a) and \ref{fig:cmplx_adaptive}(b). The
advantage of the hybrid implementation starts to appear from
Fig.~\ref{fig:mixed_adaptive}(c), whenceforth, unlike in
Fig.~\ref{fig:cmplx_adaptive}, even grains that are misoriented with
respect to the basis $\mathbf{k}_j$ show grid unrefinement within. It is
also noteworthy that the grid remains refined near solid/liquid
interfaces, grain boundaries and defects, ensuring that key topological
features are correctly resolved.

\begin{figure}[htb]
\begin{center}
\subfigure[\label{xcut168} $t=168$]
{\includegraphics[width=0.45\textwidth]{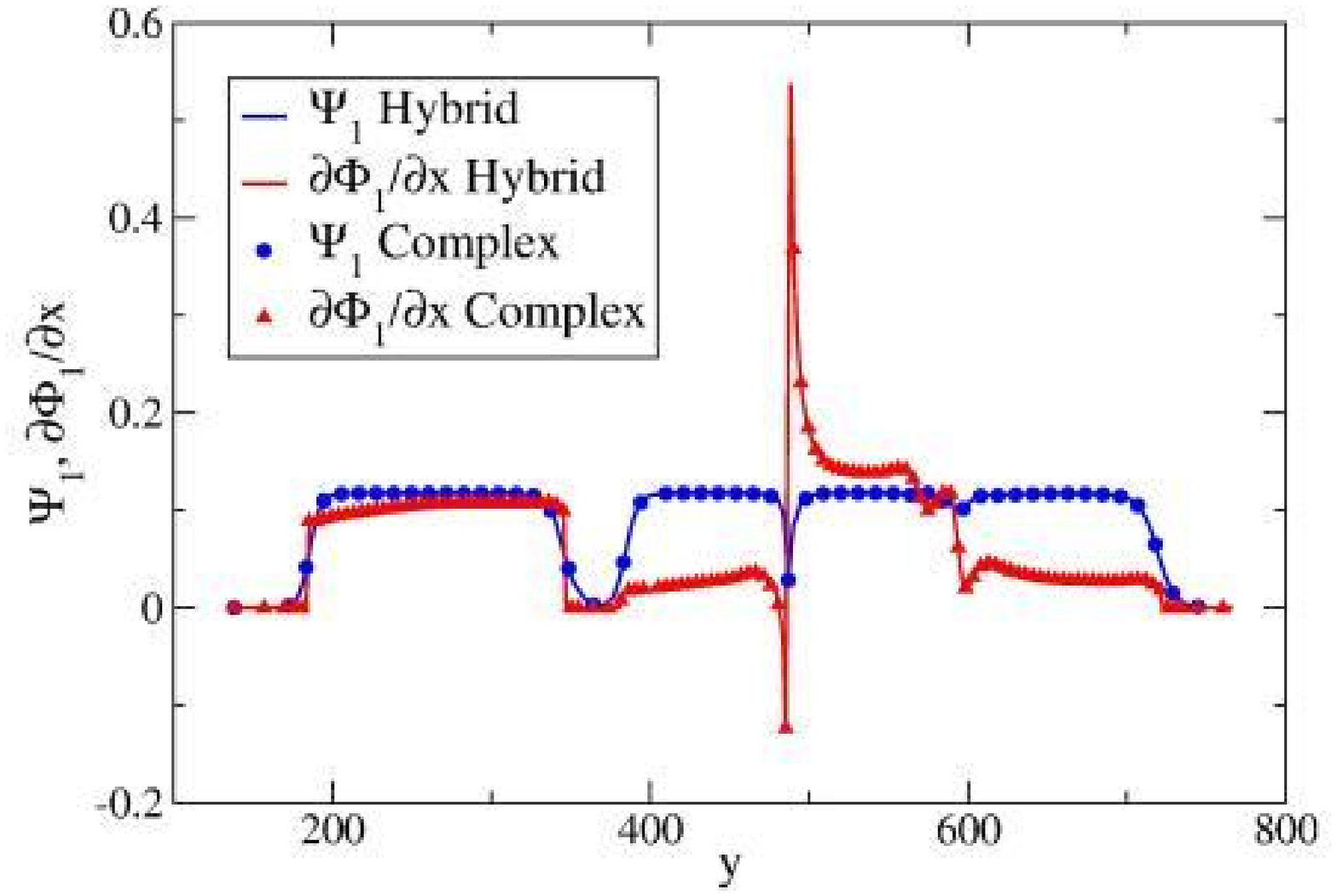}}
\subfigure[\label{xcut552} $t=552$]
{\includegraphics[width=0.45\textwidth]{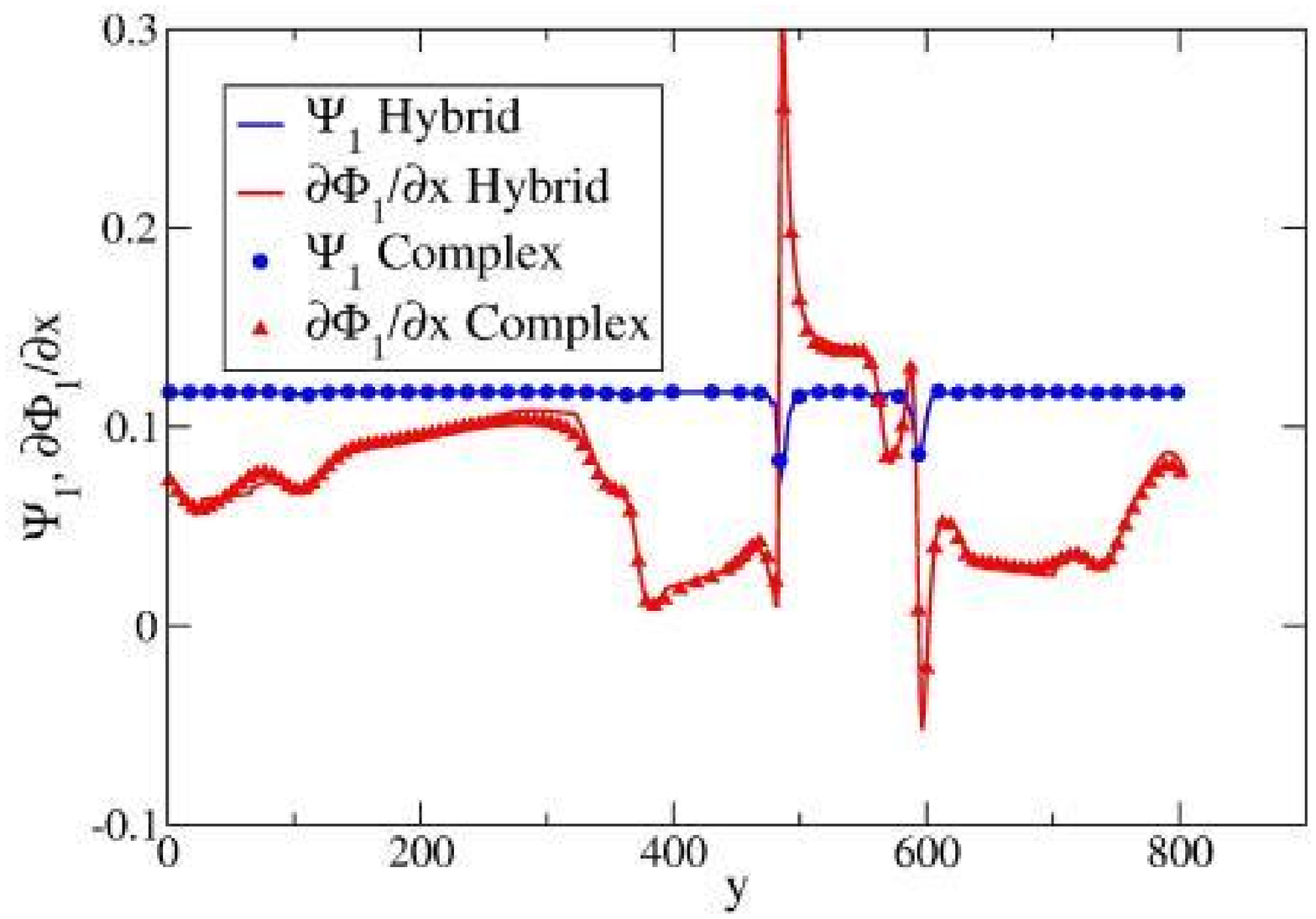}}
\caption{\label{fig:xcut} (Color online) Numerical
solution along the line $x=70\pi$ in Fig.~\ref{fig:mixed_adaptive}
compared to the results using the hybrid scheme.  Some of the data points in the
complex solution were omitted for clarity of presentation.}
\end{center}
\end{figure}

\begin{figure}[htb]
\begin{center}
\subfigure[\label{ycut168} $t=168$]
{\includegraphics[width=0.45\textwidth]{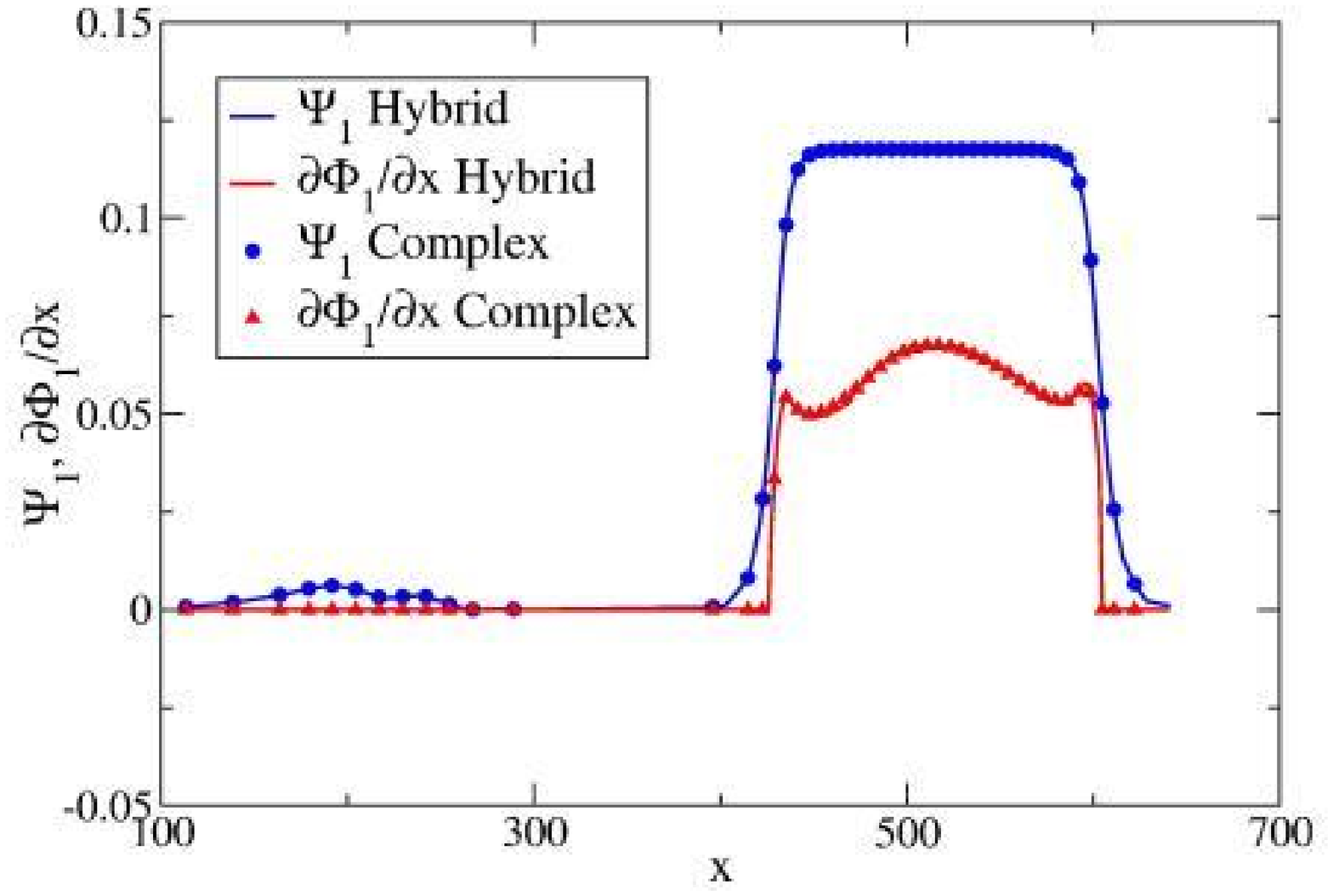}}
\subfigure[\label{ycut552} $t=552$]
{\includegraphics[width=0.45\textwidth]{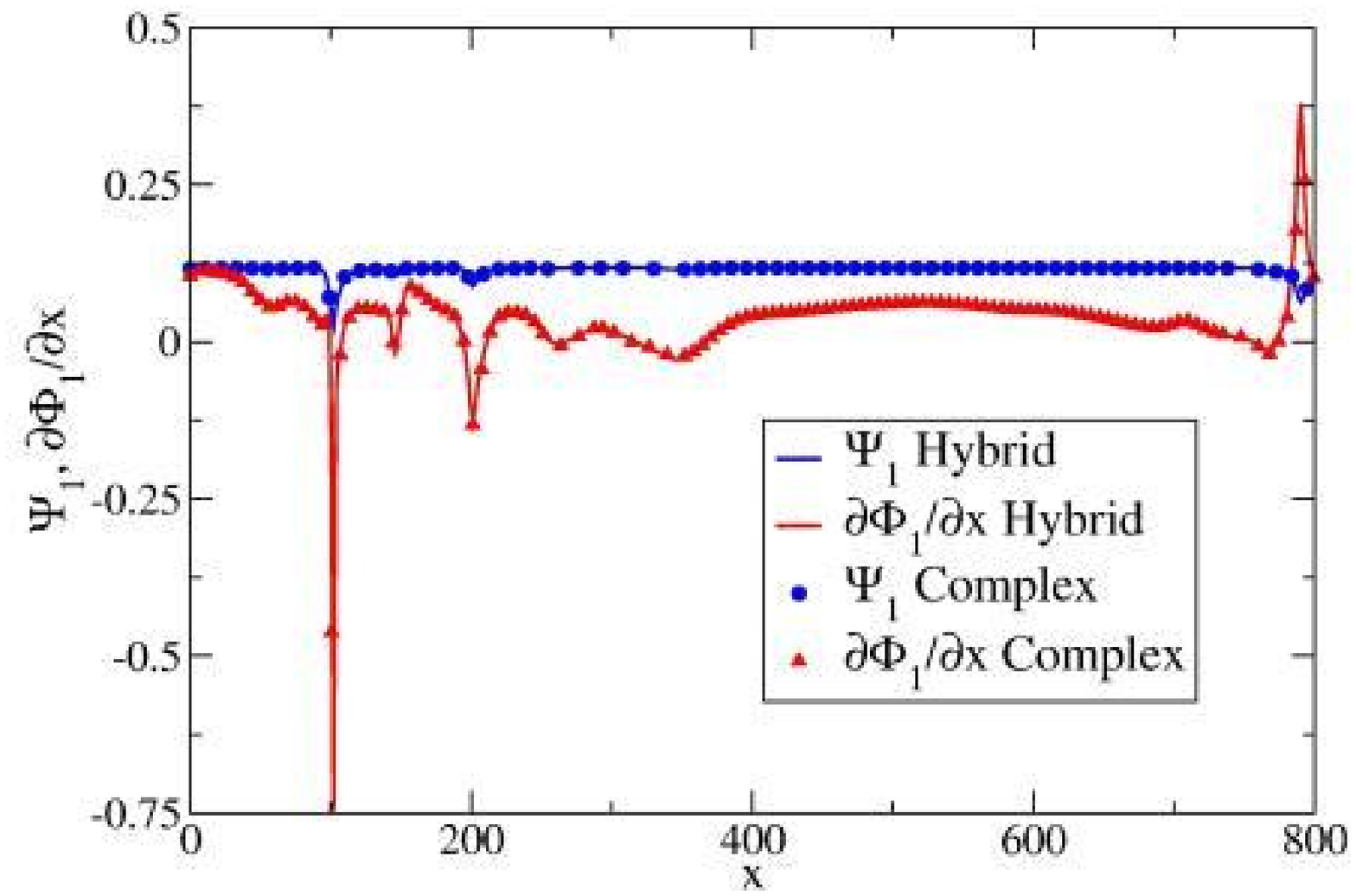}}
\caption{\label{fig:ycut} (Color online) Numerical
solution along the line $y=118\pi$ in Fig.~\ref{fig:mixed_adaptive}
compared to the results using the hybrid scheme.  Some of the data points in the
complex solution were omitted for clarity of presentation.
}
\end{center}
\end{figure}

We now compare solutions from the two simulations quantitatively. We find
it more informative to make a pointwise comparison of the two solutions
along cross sections of the domain, rather than comparing solution norms,
as we believe that this is a more stringent test of our implementation. We
choose two random cuts, one running parallel to the  $y$ axis at
$x_{cut}=70\pi$, and the other parallel to the $x$ axis at
$y_{cut}=118\pi$. The solutions are compared along these cuts at two
different times, $t=168$ and $t=552$ in Figs.~\ref{fig:xcut} and
\ref{fig:ycut} respectively. The solid curves in the figures (labeled
``hybrid'') are variations in $\Psi_1$ and $\partial\Phi_1/\partial x$
along the entire length of the domain as computed with the current
(``hybrid") implementation, whereas the symbols (labeled ``complex'') are
variations in the same variables as computed using fully complex equations
(section \ref{sec:polycryst_cart}). The agreement is essentially perfect,
indicating that our simplifications based on approximations in the
preceding sections work reasonably well.

\begin{figure}[H]
\begin{center}
{\includegraphics[width=0.4\textwidth]{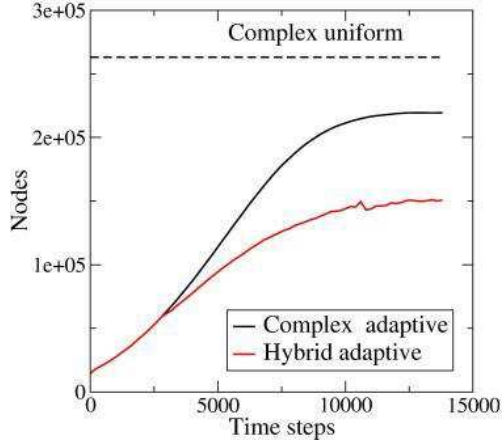}}
\caption{\label{fig:12seedsnodes} (Color online) Number of computational nodes in the
grid as a function of time, for simulations in
Fig.~\ref{fig:cmplx_adaptive} (black curve) and
Fig.~\ref{fig:mixed_adaptive} (red curve). The number of nodes reaches a
constant value after all the liquid freezes. The dashed line shows the
number of nodes required by a uniform grid implementation of the complex
amplitude equations for the same problem.}
\end{center}
\end{figure}

Because the performance of our algorithm is sensitively tied to the type
of problem that is being solved, it is difficult to come up with a simple
metric that quantifies its computational efficiency. The difficulty lies
in accounting for the change in CPU time per time step, which increases
with the number of mesh points. For example, Fig.~\ref{fig:12seedsnodes}
shows the number of nodes in this simulation over time. Clearly, an
adaptive grid implementation has a significant computational
advantage over an equivalent fixed grid implementation at the early
stages of the simulation.

One performance measure is the projected speed of our implementation
compared to a uniform grid implementation of the PFC equation. This
speedup is estimated by the simple formula,
\begin{equation}\label{eq:speedup_form}
S = \frac{N_{PFC}}{N_{RG-AG}}\times\frac{\Delta t_{RG-AG}}{\Delta
t_{PFC}}\times\frac{1}{6}\times\beta
\end{equation}
where $N_{PFC}$ is the number of grid points required to solve the PFC
equation, $N_{RG-AG}$ is the number of grid points required in a hybrid
implementation of the amplitude/RG equations, $\Delta t_{PFC}$ and
$\Delta t_{RG-AG}$ are the time steps used in the respective
implementations, the factor 1/6 comes from solving six RG equations in
place of the [one] PFC equation directly, and $\beta\in[0,1]$ is the
overhead of the AMR algorithm. The difficulty lies in  fixing
$N_{RG-AG}$ which is constantly changing with time. One estimate for
$N_{RG-AG}$ is the number of nodes averaged over the entire simulation.
This can be computed easily by dividing the area under the hybrid curve
in Fig.~\ref{fig:12seedsnodes} by the total number of time steps taken,
which gives $N_{RG-AG}=104,747$. Further, based on heuristics collected
while running our code, we conservatively estimate mesh
refinement/coarsening to constitute about 3\% of the CPU time, which
gives $\beta = 0.97$. Therefore, from Eq.~(\ref{eq:speedup_form}) we
have
\begin{equation}
S =
\frac{1,050,625}{104,747}\times\frac{0.04}{0.008}\times\frac{1}{6}\times0.97
= 8.1.
\end{equation}
We do recognize that for a more accurate estimate of $S$ we would
also need to consider overhead costs that may come from sub-optimal
cache and memory usage owing to the data structures used. Hence
these numbers should only be considered as rough estimates of true
speedup.

\begin{figure}[htb]
    \centering
    \begin{subfigure}[t=0]
    {\includegraphics[width=0.3\textwidth,angle=270]{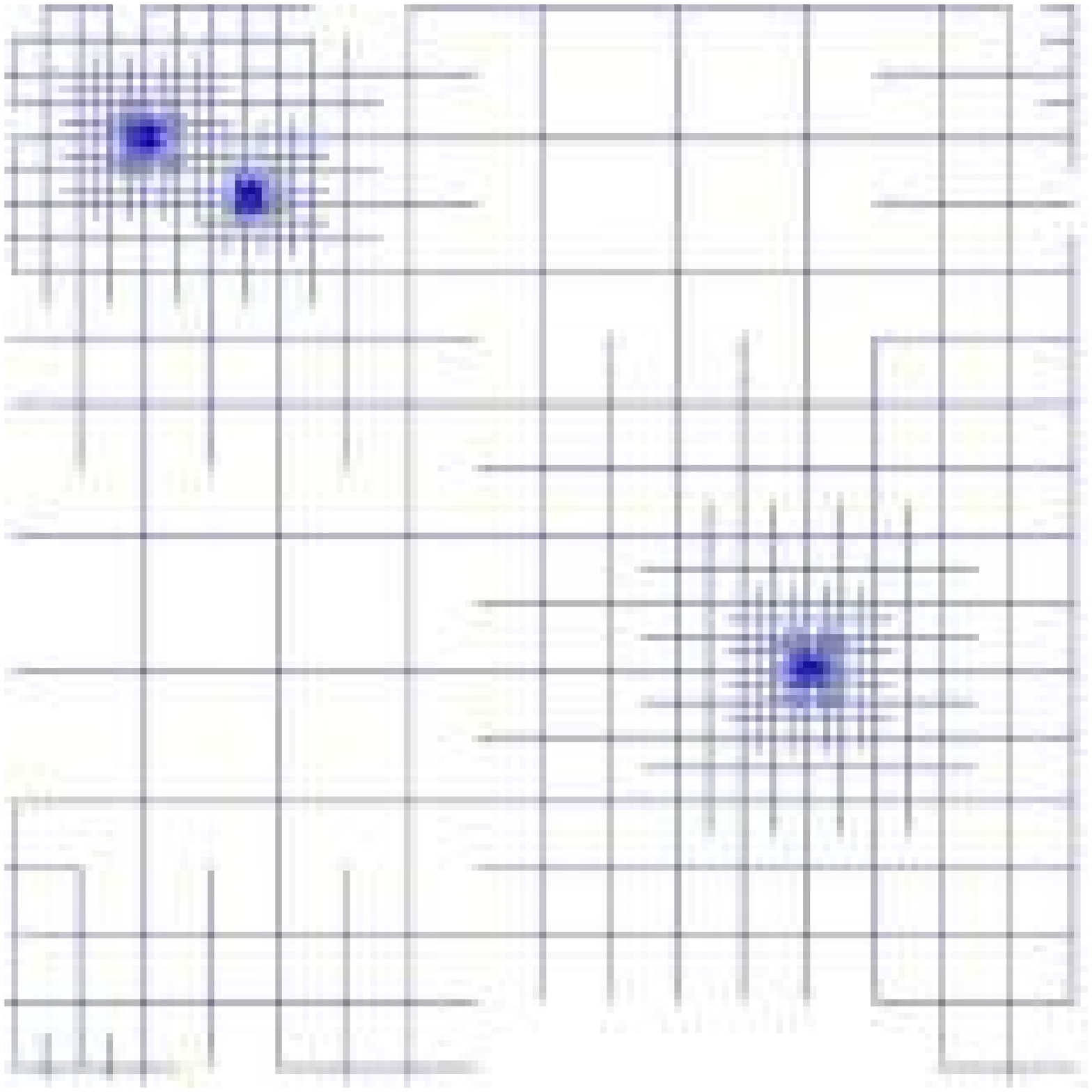}}
    \end{subfigure}
    \begin{subfigure}[t=840]
    {\includegraphics[width=0.3\textwidth,angle=270]{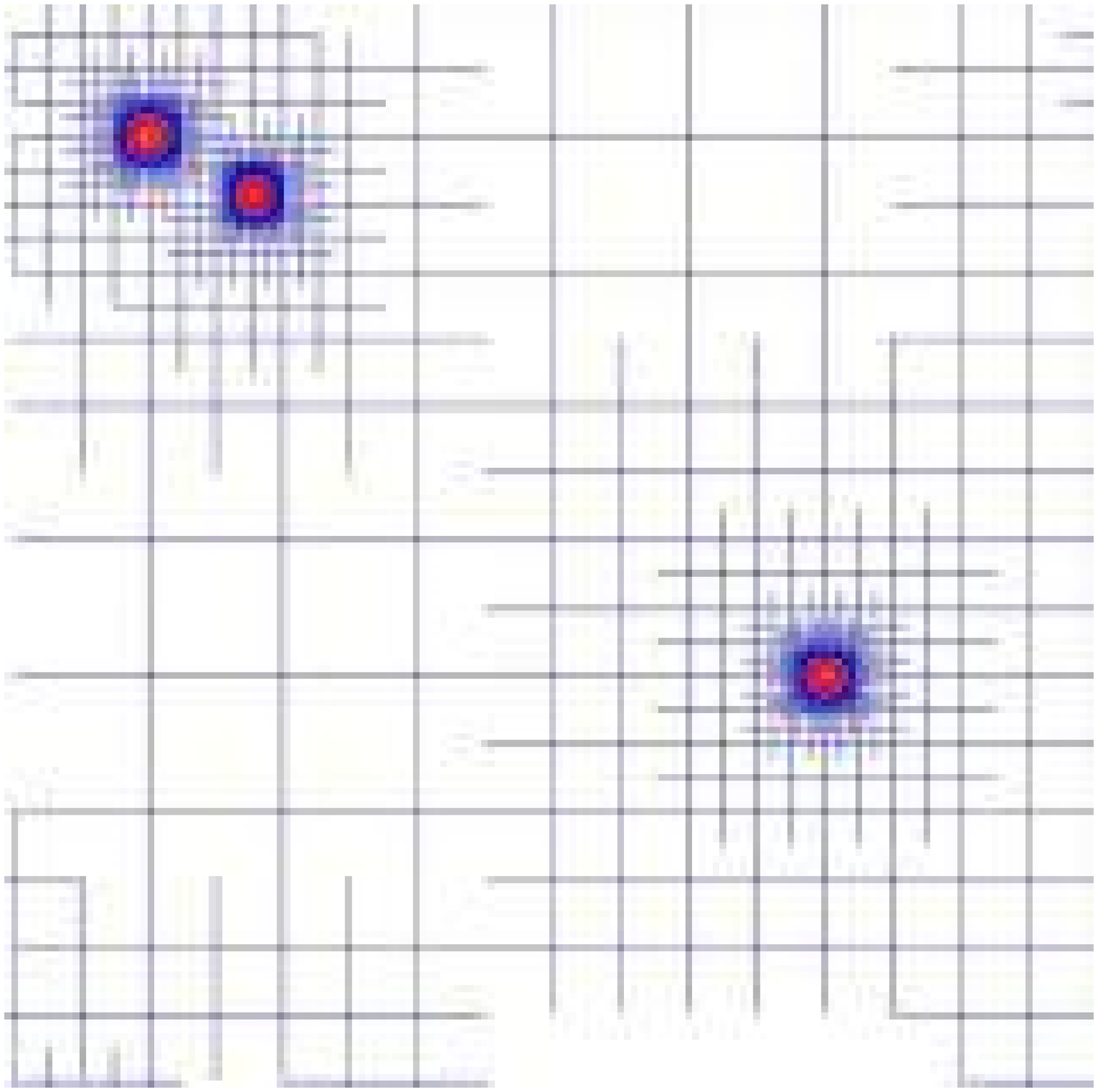}}
    \end{subfigure}
    \begin{subfigure}[t=1120]
    {\includegraphics[width=0.3\textwidth,angle=270]{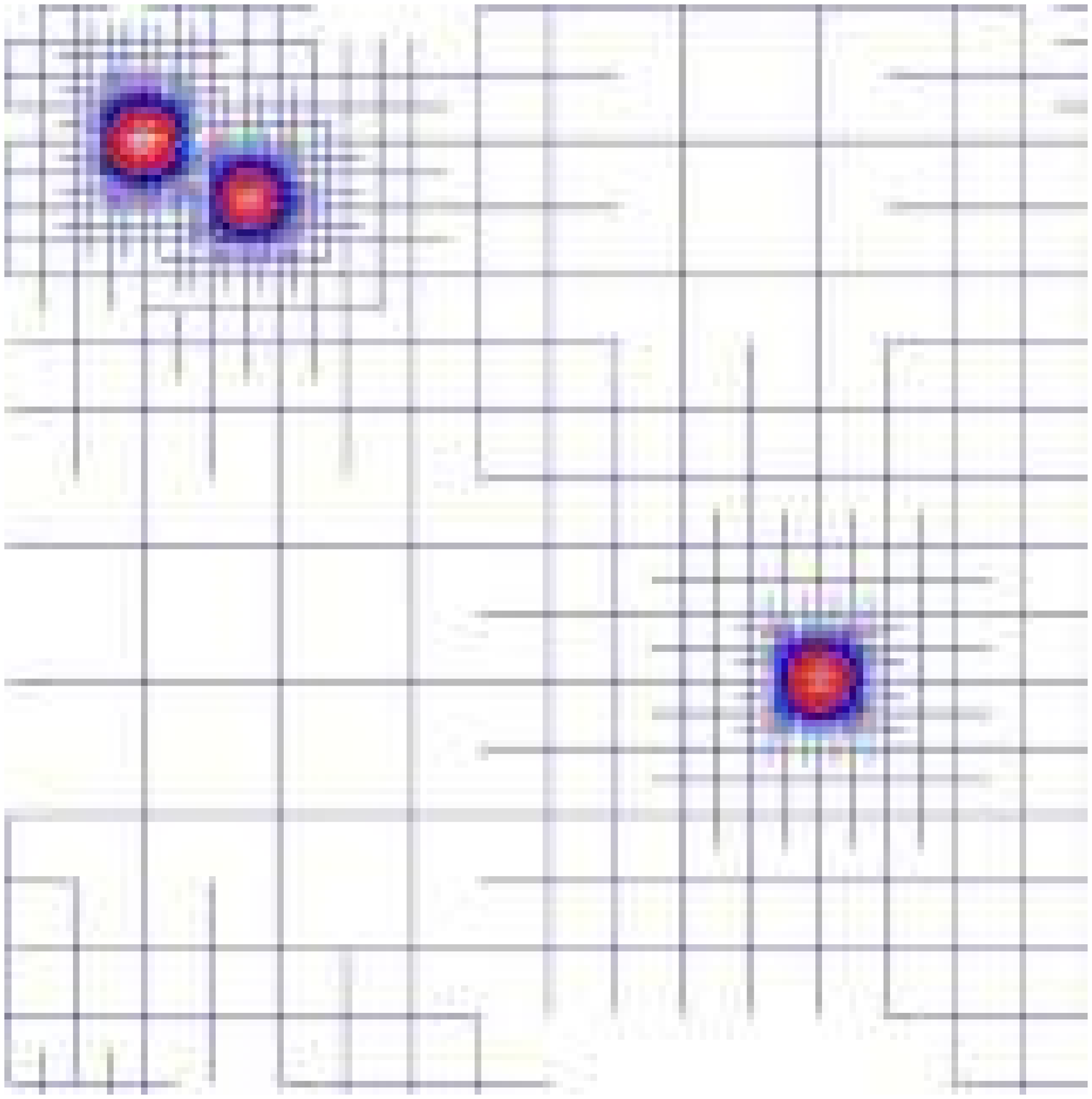}}
    \end{subfigure}
    \begin{subfigure}[t=1620]
    {\includegraphics[width=0.3\textwidth,angle=270]{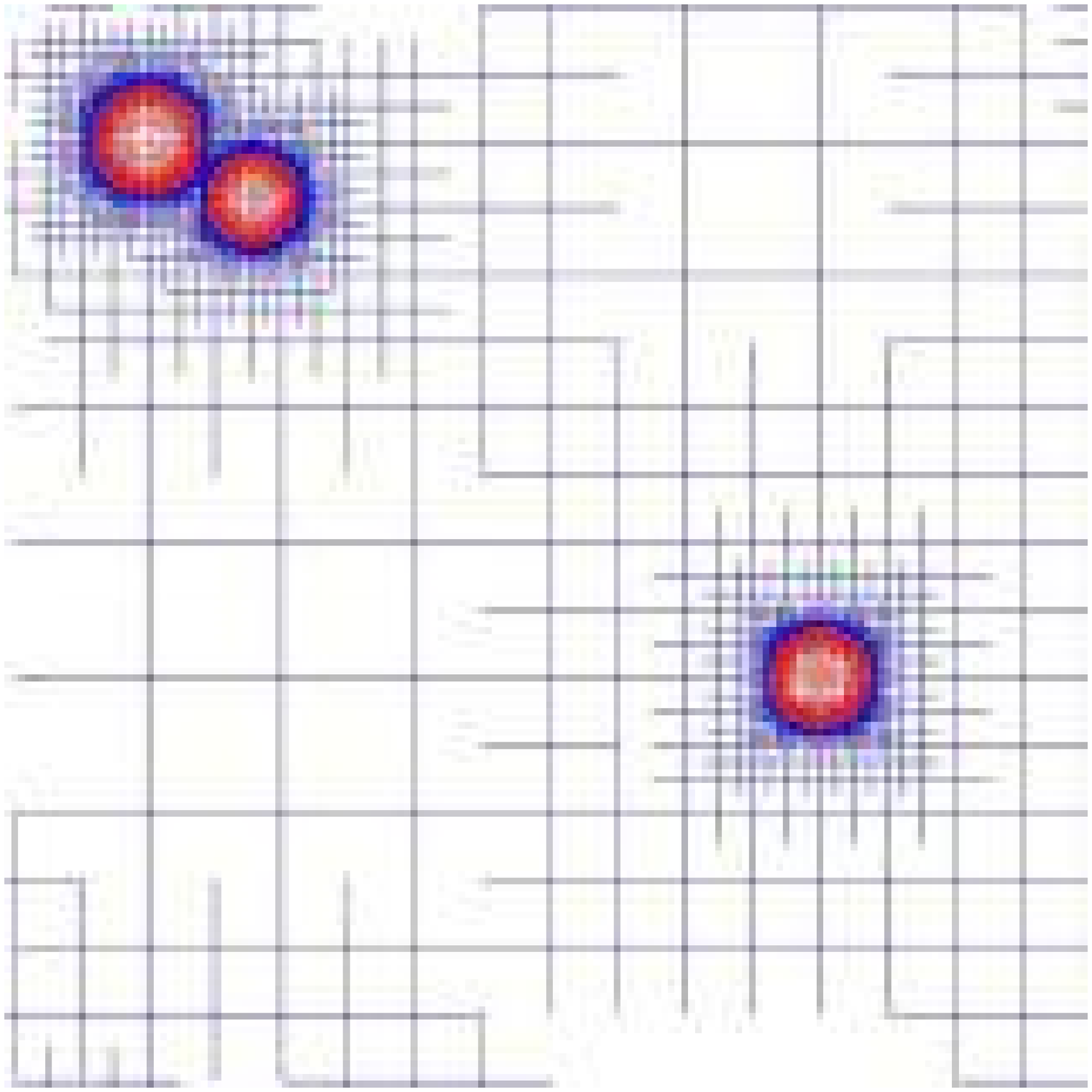}}
    \end{subfigure}
    \begin{subfigure}[t=2080]
    {\includegraphics[width=0.3\textwidth,angle=270]{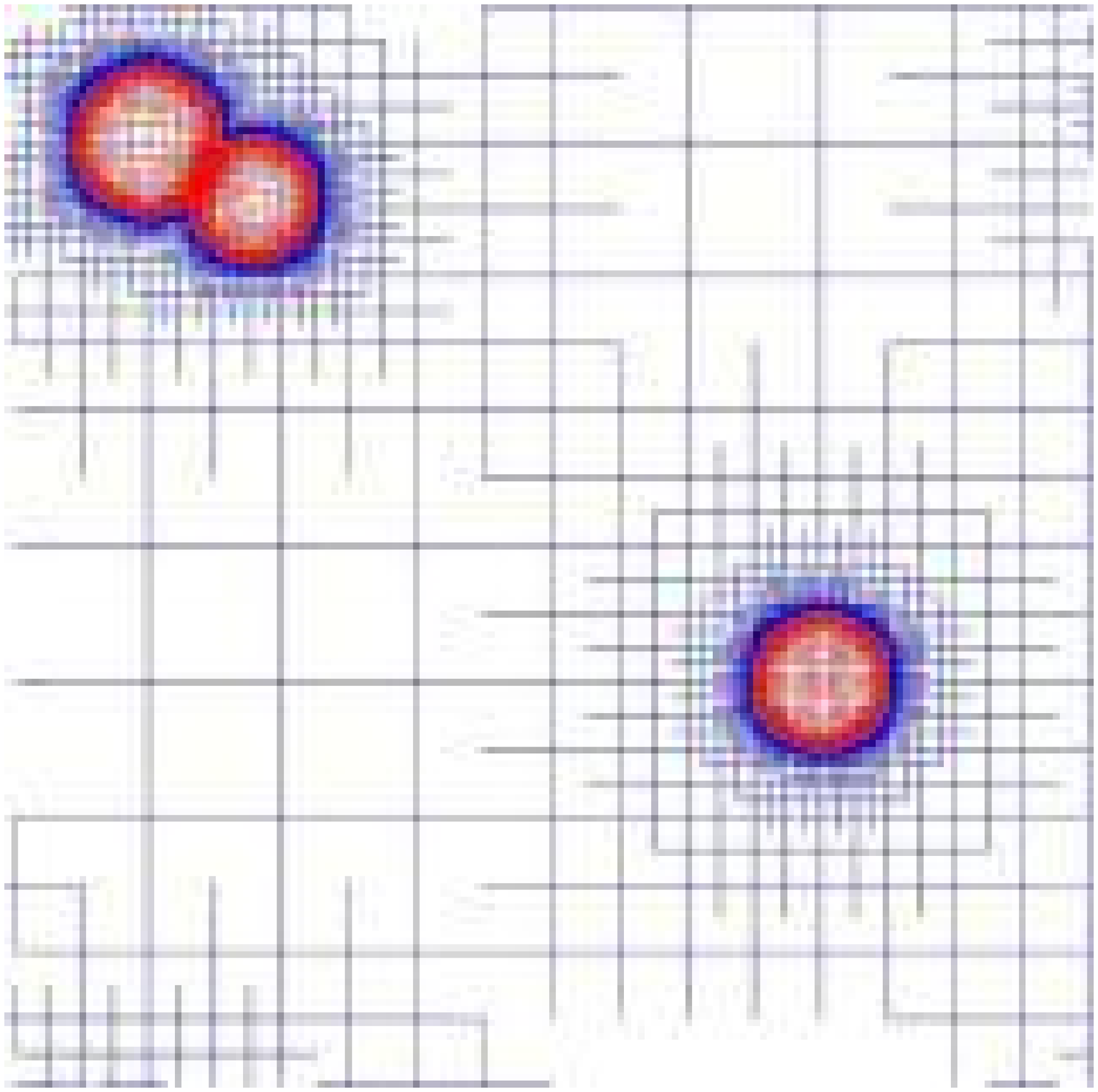}}
    \end{subfigure}
    \begin{subfigure}[t=2800]
    {\includegraphics[width=0.3\textwidth,angle=270]{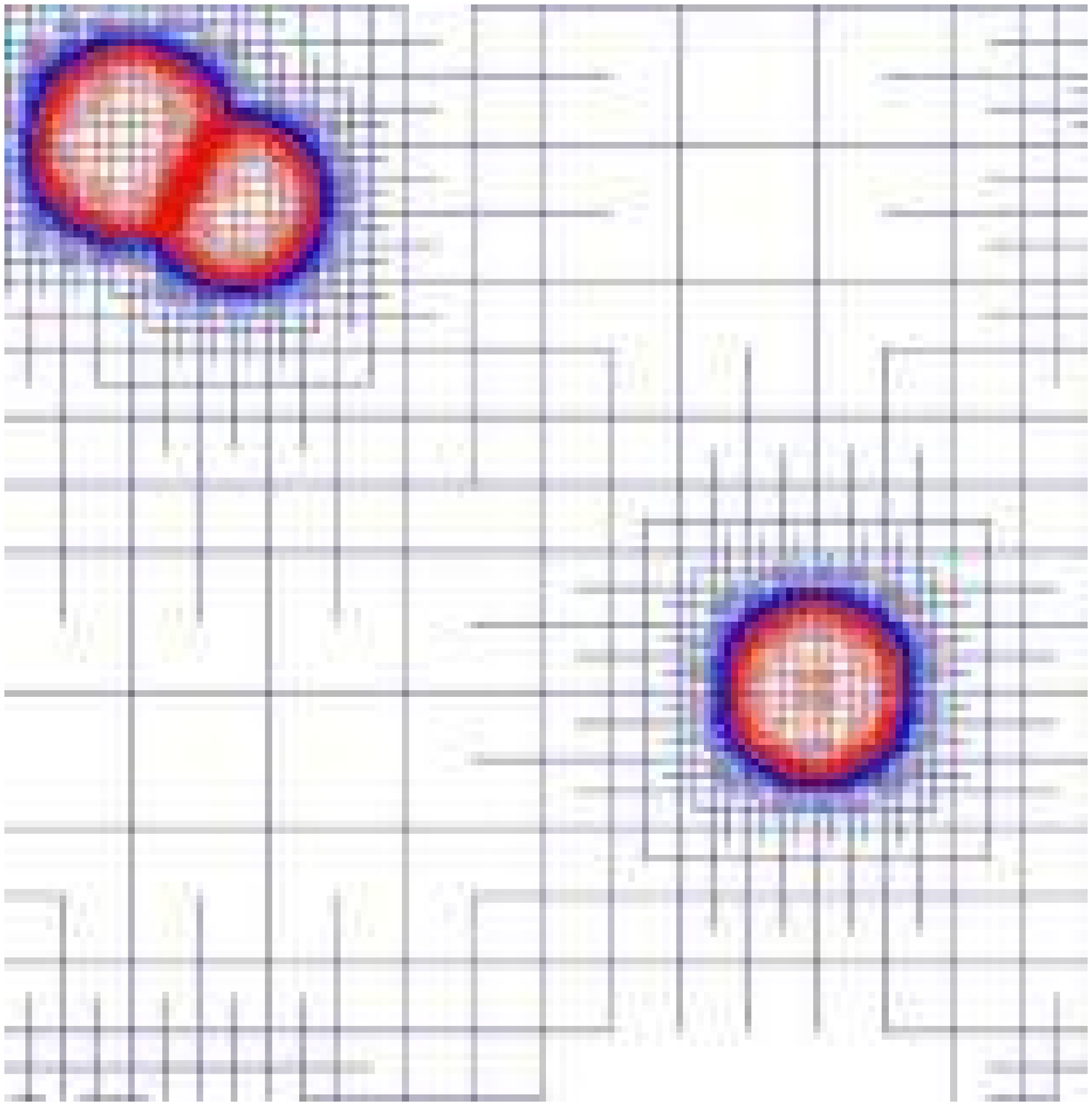}}
    \end{subfigure}
%    \begin{tabular}{ccc}
%        \includegraphics[ height=0.25\textheight]{mabig1}
%    &
%        \includegraphics[ height=0.25\textheight]{mabig5}
%    \\
%    (a) $t=0$ & (d) $t=1620$ \vspace{0.5em} \\
%       \includegraphics[ height=0.25\textheight]{mabig3}
%    &
%        \includegraphics[ height=0.25\textheight]{mabig6}
%    \\
%    (b) $t=840$   & (e) $t=2080$ \vspace{0.5em} \\
%        \includegraphics[ height=0.25\textheight]{mabig4}
%    &
%        \includegraphics[ height=0.25\textheight]{mabig7}
%    \\
%    (c) $t=1120$  & (f) $t=2800$ \vspace{0.5em} \\
%    \end{tabular}
%
    \caption{\label{fig:micro} (Color online) Micro-scale simulation of two dimensional
    crystal growth with amplitude equations using AMR.}
\end{figure}

While a speedup factor of 8 may not seem to be a great improvement in
computational efficiency, one should bear in mind that the number of nodes
in the AMR algorithm scales (roughly) linearly with interface/grain
boundary length, which is quite substantial in the system we just
simulated. Thus, one should not expect to derive the maximum computational
benefit when simulating small systems with large numbers of grains. On the
other hand, with this new method, we can now simulate the growth of a few
crystals in a much larger system. We choose a square domain of side
$4096\pi$, which in physical dimensions translates to 0.722 $\mu$m, if we
assume an interatomic spacing of 4 {\AA} \footnote{This is the
interatomic spacing in Aluminum \cite{callister}, which has a face
centered cubic lattice.}. We initiate three randomly oriented crystals,
two a little closer together than the third, so that a grain boundary
forms quickly. The crystals are shown at different times in
Fig.~\ref{fig:micro}. The simulation was terminated at $t=3960$ when
memory requirements exceeded 1 GB, after running on a dedicated 3.06 GHz
Intel Xeon processor for about one week.

\begin{figure}[htb]
\begin{center}
{\includegraphics[width=0.5\textwidth]{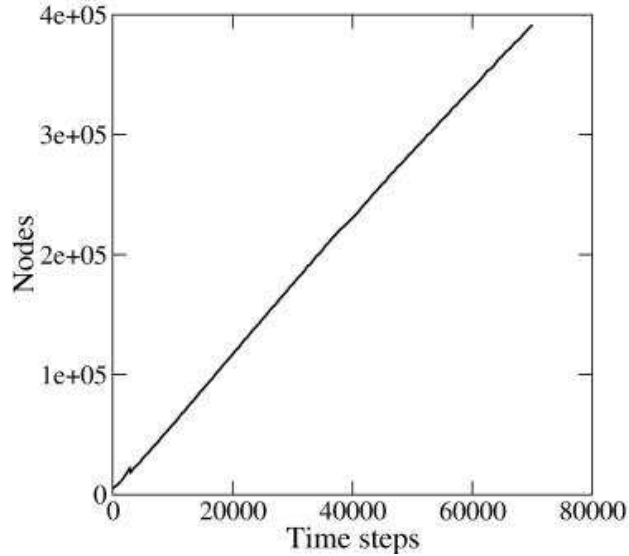}}
\caption{\label{fig:bigproblem_nodes} Number of computational nodes
in the grid as a function of time for the 1 $\mu$m $\times$ 1$\mu$m
domain. The growth is almost linear.}
\end{center}
\end{figure}

\begin{figure}[htb]
\begin{center}
{\includegraphics[width=0.8\textwidth]{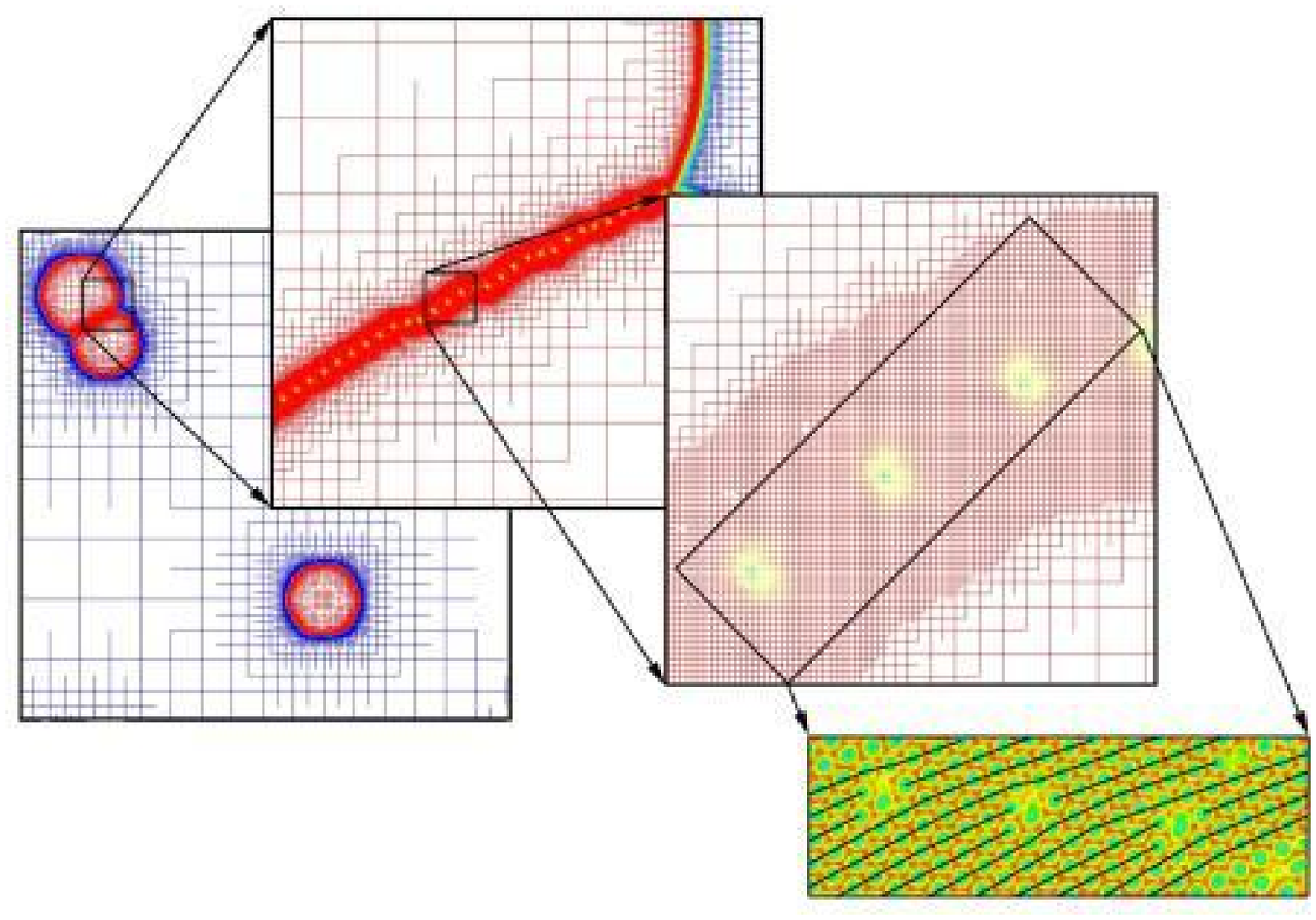}}
\caption{\label{fig:multiscale} (Color online) The above grid spans roughly three
orders of magnitude in length scales, from a nanometer up to a
micron. The leftmost box resolves the entire computational domain
whereas the rightmost resolves dislocations at the atomic scale.}
\end{center}
\end{figure}

Let us calculate the speedup factor for this simulation as we did
previously, after $70,000$ time steps ($t=2800$,
Fig.~\ref{fig:micro}(f)). Fig.~\ref{fig:bigproblem_nodes} shows that the
number of nodes in the adaptive grid varies nearly linearly
with the number of time steps, and we estimate the average number of
nodes $N_{AG-RG}$ to be $200,721$. The same simulation on a uniform
grid using the PFC equation would have required $268,435,456$ nodes
(not possible on our computers). We estimate $\beta = 0.98$. In this
case the speedup is about three orders of magnitude,
\begin{equation}
S =
\frac{268,435,456}{200,721}\times\frac{0.04}{0.008}\times\frac{1}{6}\times0.98
= 1091.
\end{equation}
Fig.~\ref{fig:multiscale} shows vividly  the range
of length scales from nanometers to microns spanned by our grid in
this simulation, highlighting its ``multiscale'' capability.

We would like to emphasize that as with any adaptive grid implementation,
refinement criteria can change $S$ by a  factor that is approximately
constant. In order to enable testing our implementation on a much larger
domain subject to the available memory  resources, the criteria were
relaxed. Note however, that even if we had roughly doubled the number of
finely spaced nodes near the interfaces and the grain boundary, which
would lead to a significantly more accurate calculation, $S$ would still
be about 500 times faster than an equivalent implementation of the PFC
equation on a uniform grid.

\section{\label{sec:conc}Concluding remarks}

In this article, we have presented an efficient hybrid numerical
implementation that combines Cartesian and polar representations of the
complex amplitude with adaptive mesh refinement, and allows the modeling
capabilities of  the PFC equation to be extended to microscopic length
scales. Depending on the choice of application, we have shown that our
scheme can be anywhere from $1-3$ orders of magnitude times faster than
an equivalent uniform grid implementation of the PFC equation, on a single
processor machine. We anticipate that this advantage will be preserved when
both implementations are migrated to a parallel computer, which is an
important next step required to give the RG extension of the PFC model
full access to micro- and meso- scale phenomena.

In conclusion, we have shown that multiscale modeling of complex
polycrystalline materials microstructure is possible using a
combination of continuum modeling at the nanoscale using the PFC
model, RG and related techniques from spatially-extended dynamical
systems theory and adaptive mesh refinement.

We regard this work as only a first step for our modeling approach with the RG extension of the PFC
to be successfully applied for studying important engineering and
materials science applications. We have identified a few issues that
require immediate attention. The first, although an implementation
issue, is critical, and has to do with using amplitude equations for
applications involving externally applied loads and displacements to
a polycrystal that has been evolved with our equations. Simple
applications could be, subjecting the polycrystal to shear,
uniaxial, or biaxial loading states \cite{eg04,BGE05}. Such boundary
conditions are difficult enough to apply to the scalar field $\psi$
in the PFC equation \cite{Stefanovic06}. Meaningful translation to equivalent
boundary conditions on the amplitudes and phases of $\psi$ can be a
very difficult task, requiring the solution of systems of
nonlinearly coupled equations at the boundaries.We have
not yet investigated this issue in any detail. 

Our derivation of the amplitude equations \cite{AGD06_1} was based on a
one mode approximation to the triangular lattice, and as we always
chose parameters fairly close to the boundary between the triangular
phase and coexisting triangular and constant phases,
i.e.~$|r+3\bar{\psi}^2| \ll 1$, the amplitude equations we derived were
within their domain of validity and our results were quite accurate. It
is almost certain that a one-mode approximation will not give similarly
accurate results when
$|r+3\bar{\psi}^2|\sim\mathcal{O}(1)$ (although it would be interesting
to see how much the error actually is). It is not clear if this in any
way precludes certain phenomena from being studied with our equations,
as we can always choose parameters to stay in the regime where the
one-mode approximation is valid, but if it does, amplitude equations
for dominant higher modes need to be systematically developed.

An important assumption made in the derivation of our so called
``hybrid'' formulation of the complex amplitude equations is that of
locally freezing the phase gradient vector $\nabla\Phi_j$. In fact, it
is this assumption that allows us to effectively unrefine the interior
of grains and gain significant speedup over the PFC equation. If for
example, the problem we are studying involves the application of a
large external shear strain that could change $\nabla\Phi_j$ in the
grain interior via grain rotation, it is uncertain whether our
algorithm would continue to maintain its computational efficiency over
the PFC. This is again a matter worth investigating.

\begin{acknowledgments}

We thank Ken Elder and Nicholas Guttenberg for several useful discussions. This work
was partially supported by the National Science Foundation through grant number
NSF-DMR-01-21695. One of the authors (NP) wishes to acknowledge support from the
National Science and Engineering Research Council of Canada.   

% JON Nik - do you need to insert a grant acknowledgement here?

\end{acknowledgments}

\appendix
\section{Discretization of Operators}\label{app:discoper}
\subsection{Laplacian}
The Laplacian of a function $f(x,y)$ is discretized at point
$(x_i,y_j)=(i\Delta x,j\Delta x)$ using a nine point finite
difference stencil as shown below, where $\Delta x$ is the mesh
spacing.
\begin{eqnarray}\label{eqn:lapdisc}
\left.\nabla^2f\right|_{i,j} &=&
\frac{f_{i+1,j}+f_{i-1,j}+f_{i,j+1}+f_{i,j-1}}{2\Delta x^2} +
\frac{f_{i+1,j+1}+f_{i-1,j-1}+f_{i-1,j+1}+f_{i+1,j-1}}{4\Delta x^2}\nonumber\\
& &- \frac{3f_{i,j}}{\Delta x^2} + \mathcal{O}(\Delta x^2).
\end{eqnarray}
A Fourier transform of this \emph{isotropic} discretization,
described by Tomita in \cite{tomita91}, is shown to very nearly
follow the $-k^2$ isocontours.

\subsection{Gradient}
The gradient of a function $f(x,y)$ is discretized at point
$(x_i,y_j)=(i\Delta x,j\Delta x)$ using a nine point second order
finite difference stencil as shown below, where $\Delta x$ is the
mesh spacing. The stencil is designed to minimize effects of grid
anisotropy which can introduce artifacts in the solution, especially
on adaptive grids. We have
\begin{eqnarray}\label{eqn:graddisc1}
\left.\nabla f\right|_{i,j} &=& \left.\widetilde{\nabla}_{\oplus}f\right|_{i,j} + \mathcal{O}(\Delta x^2)\nonumber\\
 &=& \left(\frac{f_{i+1,j}-f_{i-1,j}}{2\Delta
x}\right)\vec{i} + \left(\frac{f_{i,j+1}-f_{i,j-1}}{2\Delta
x}\right)\vec{j} + \mathcal{O}(\Delta x^2).
\end{eqnarray}
But
\begin{equation}
\nabla f =
\left(\frac{f_x+f_y}{\sqrt{2}}\right)\left(\frac{\vec{i}+\vec{j}}
{\sqrt{2}}\right)
+
\left(\frac{-f_x+f_y}{\sqrt{2}}\right)\left(\frac{-\vec{i}+\vec{j}}
{\sqrt{2}}\right)
\end{equation}
and hence we also have
\begin{eqnarray}\label{eqn:graddisc2}
\left.\nabla f\right|_{i,j} &=& \left.\widetilde{\nabla}_{\otimes}
f\right|_{i,j} + \mathcal{O}(\Delta x^2)\nonumber\\
 &=& \left(\frac{f_{i+1,j+1}-f_{i-1,j-1}}{2\sqrt{2}\Delta x}\right)
 \left(\frac{\vec{i}+\vec{j}}{\sqrt{2}}\right) +
\left(\frac{f_{i-1,j+1}-f_{i+1,j-1}}{2\sqrt{2}\Delta x}\right)
\left(\frac{-\vec{i}+\vec{j}}{\sqrt{2}}\right) + \mathcal{O}(\Delta x^2)
\nonumber\\
&=&\left(\frac{f_{i+1,j+1}-f_{i-1,j-1}-f_{i-1,j+1}+f_{i+1,j-1}}
{4\Delta x}\right)\vec{i}\nonumber\\
&
&+\left(\frac{f_{i+1,j+1}-f_{i-1,j-1}+f_{i-1,j+1}-f_{i+1,j-1}}{4\Delta
x}\right)\vec{j} + \mathcal{O}(\Delta x^2).
\end{eqnarray}
Using the discrete forms for the gradient in
Eqs.~(\ref{eqn:graddisc1}) and (\ref{eqn:graddisc2}) we can write
the isotropic second order discretization as
\begin{equation}\label{eqn:graddisc}
\left.\nabla f\right|_{i,j} =
\frac{1}{2}\left(\left.\widetilde{\nabla}_{\oplus}f\right|_{i,j} +
\left.\widetilde{\nabla}_{\otimes}f\right|_{i,j}\right) +
\mathcal{O}(\Delta x^2).
\end{equation}
A discretization scheme similar to Eq.~(\ref{eqn:graddisc}) is given
by Sethian and Strain \cite{sethian92}.

%\bibliography{biblio,jeong,rg_pfc}

\begin{thebibliography}{54}
\expandafter\ifx\csname natexlab\endcsname\relax\def\natexlab#1{#1}\fi
\expandafter\ifx\csname bibnamefont\endcsname\relax
  \def\bibnamefont#1{#1}\fi
\expandafter\ifx\csname bibfnamefont\endcsname\relax
  \def\bibfnamefont#1{#1}\fi
\expandafter\ifx\csname citenamefont\endcsname\relax
  \def\citenamefont#1{#1}\fi
\expandafter\ifx\csname url\endcsname\relax
  \def\url#1{\texttt{#1}}\fi
\expandafter\ifx\csname urlprefix\endcsname\relax\def\urlprefix{URL }\fi
\providecommand{\bibinfo}[2]{#2}
\providecommand{\eprint}[2][]{\url{#2}}

\bibitem[{\citenamefont{Phillips}(2001)}]{Phillipsbook}
\bibinfo{author}{\bibfnamefont{R.}~\bibnamefont{Phillips}},
  \emph{\bibinfo{title}{Crystals, defects and microstructures: modeling across
  scales}} (\bibinfo{publisher}{Cambridge University Press},
  \bibinfo{year}{2001}).

\bibitem[{\citenamefont{Tadmor et~al.}(1996)\citenamefont{Tadmor, Ortiz, and
  Phillips}}]{Tadmor}
\bibinfo{author}{\bibfnamefont{E.~B.} \bibnamefont{Tadmor}},
  \bibinfo{author}{\bibfnamefont{M.}~\bibnamefont{Ortiz}}, \bibnamefont{and}
  \bibinfo{author}{\bibfnamefont{R.}~\bibnamefont{Phillips}},
  \bibinfo{journal}{Phil. Mag. A} \textbf{\bibinfo{volume}{73}},
  \bibinfo{pages}{1529} (\bibinfo{year}{1996}).

\bibitem[{\citenamefont{Shenoy et~al.}(1998)\citenamefont{Shenoy, Miller,
  Tadmor, Phillips, and Ortiz}}]{Shenoy}
\bibinfo{author}{\bibfnamefont{V.~B.} \bibnamefont{Shenoy}},
  \bibinfo{author}{\bibfnamefont{R.}~\bibnamefont{Miller}},
  \bibinfo{author}{\bibfnamefont{E.~B.} \bibnamefont{Tadmor}},
  \bibinfo{author}{\bibfnamefont{R.}~\bibnamefont{Phillips}}, \bibnamefont{and}
  \bibinfo{author}{\bibfnamefont{M.}~\bibnamefont{Ortiz}},
  \bibinfo{journal}{Phys. Rev. Lett.} \textbf{\bibinfo{volume}{80}},
  \bibinfo{pages}{742} (\bibinfo{year}{1998}).

\bibitem[{\citenamefont{Knap and Ortiz}(2001)}]{Ortiz}
\bibinfo{author}{\bibfnamefont{J.}~\bibnamefont{Knap}} \bibnamefont{and}
  \bibinfo{author}{\bibfnamefont{M.}~\bibnamefont{Ortiz}}, \bibinfo{journal}{J.
  Mech. Phys. Solids} \textbf{\bibinfo{volume}{49}}, \bibinfo{pages}{1899}
  (\bibinfo{year}{2001}).

\bibitem[{\citenamefont{Miller and Tadmor}(2002)}]{Miller}
\bibinfo{author}{\bibfnamefont{R.~E.} \bibnamefont{Miller}} \bibnamefont{and}
  \bibinfo{author}{\bibfnamefont{E.~B.} \bibnamefont{Tadmor}},
  \bibinfo{journal}{Journal of Computer-Aided Materials Design}
  \textbf{\bibinfo{volume}{9}}, \bibinfo{pages}{203} (\bibinfo{year}{2002}).

\bibitem[{\citenamefont{E et~al.}(2003)\citenamefont{E, Enquist, and
  Huang}}]{Weinan1}
\bibinfo{author}{\bibfnamefont{W.}~\bibnamefont{E}},
  \bibinfo{author}{\bibfnamefont{B.}~\bibnamefont{Enquist}}, \bibnamefont{and}
  \bibinfo{author}{\bibfnamefont{Z.}~\bibnamefont{Huang}},
  \bibinfo{journal}{Phys. Rev. B} \textbf{\bibinfo{volume}{67}},
  \bibinfo{pages}{092101:1} (\bibinfo{year}{2003}).

\bibitem[{\citenamefont{E and Huang}(2001)}]{Weinan2}
\bibinfo{author}{\bibfnamefont{W.}~\bibnamefont{E}} \bibnamefont{and}
  \bibinfo{author}{\bibfnamefont{Z.}~\bibnamefont{Huang}},
  \bibinfo{journal}{Phys. Rev. Lett.} \textbf{\bibinfo{volume}{87}},
  \bibinfo{pages}{135501:1} (\bibinfo{year}{2001}).

\bibitem[{\citenamefont{Rudd and Broughton}(1998)}]{Rudd}
\bibinfo{author}{\bibfnamefont{R.~E.} \bibnamefont{Rudd}} \bibnamefont{and}
  \bibinfo{author}{\bibfnamefont{J.}~\bibnamefont{Broughton}},
  \bibinfo{journal}{Phys. Rev. B} \textbf{\bibinfo{volume}{58}},
  \bibinfo{pages}{R5893} (\bibinfo{year}{1998}).

\bibitem[{\citenamefont{Broughton et~al.}(1998)\citenamefont{Broughton,
  Abraham, Bernstein, and Kaxiras}}]{Kaxiras}
\bibinfo{author}{\bibfnamefont{J.~Q.} \bibnamefont{Broughton}},
  \bibinfo{author}{\bibfnamefont{F.~F.} \bibnamefont{Abraham}},
  \bibinfo{author}{\bibfnamefont{N.}~\bibnamefont{Bernstein}},
  \bibnamefont{and} \bibinfo{author}{\bibfnamefont{E.}~\bibnamefont{Kaxiras}},
  \bibinfo{journal}{Phys. Rev. B} \textbf{\bibinfo{volume}{60}},
  \bibinfo{pages}{2391} (\bibinfo{year}{1998}).

\bibitem[{\citenamefont{Denniston and Robbins}(2004)}]{Robbins}
\bibinfo{author}{\bibfnamefont{C.}~\bibnamefont{Denniston}} \bibnamefont{and}
  \bibinfo{author}{\bibfnamefont{M.~O.} \bibnamefont{Robbins}},
  \bibinfo{journal}{Phys. Rev. E} \textbf{\bibinfo{volume}{69}},
  \bibinfo{pages}{021505:1} (\bibinfo{year}{2004}).

\bibitem[{\citenamefont{Curtarolo and Ceder}(2002)}]{CURT02}
\bibinfo{author}{\bibfnamefont{S.}~\bibnamefont{Curtarolo}} \bibnamefont{and}
  \bibinfo{author}{\bibfnamefont{G.}~\bibnamefont{Ceder}},
  \bibinfo{journal}{Phys. Rev. Lett.} \textbf{\bibinfo{volume}{88}},
  \bibinfo{pages}{255504:1} (\bibinfo{year}{2002}).

\bibitem[{\citenamefont{Fish and Chen}(2004)}]{Fish}
\bibinfo{author}{\bibfnamefont{J.}~\bibnamefont{Fish}} \bibnamefont{and}
  \bibinfo{author}{\bibfnamefont{W.}~\bibnamefont{Chen}},
  \bibinfo{journal}{Comp. Meth. Appl. Mech. Eng.}
  \textbf{\bibinfo{volume}{193}}, \bibinfo{pages}{1693} (\bibinfo{year}{2004}).

\bibitem[{\citenamefont{Langer}(1986)}]{Langer86}
\bibinfo{author}{\bibfnamefont{J.~S.} \bibnamefont{Langer}}, in
  \emph{\bibinfo{booktitle}{Directions in Condensed Matter Physics}}, edited by
  \bibinfo{editor}{\bibfnamefont{G.}~\bibnamefont{Grinstein}} \bibnamefont{and}
  \bibinfo{editor}{\bibfnamefont{G.}~\bibnamefont{Mazenko}}
  (\bibinfo{publisher}{World Scientific}, \bibinfo{year}{1986}),
  vol.~\bibinfo{volume}{1}, p. \bibinfo{pages}{165}.

\bibitem[{\citenamefont{Karma and Rappel}(1998)}]{Karma}
\bibinfo{author}{\bibfnamefont{A.}~\bibnamefont{Karma}} \bibnamefont{and}
  \bibinfo{author}{\bibfnamefont{W.~J.} \bibnamefont{Rappel}},
  \bibinfo{journal}{Phys. Rev. E} \textbf{\bibinfo{volume}{57}},
  \bibinfo{pages}{4323} (\bibinfo{year}{1998}).

\bibitem[{\citenamefont{Beckermann et~al.}(1999)\citenamefont{Beckermann,
  H.-J.Diepers, Steinbach, Karma, and Tong}}]{Beckermann1}
\bibinfo{author}{\bibfnamefont{C.}~\bibnamefont{Beckermann}},
  \bibinfo{author}{\bibnamefont{H.-J.Diepers}},
  \bibinfo{author}{\bibfnamefont{I.}~\bibnamefont{Steinbach}},
  \bibinfo{author}{\bibfnamefont{A.}~\bibnamefont{Karma}}, \bibnamefont{and}
  \bibinfo{author}{\bibfnamefont{X.}~\bibnamefont{Tong}}, \bibinfo{journal}{J.
  Comp. Phys.} \textbf{\bibinfo{volume}{154}}, \bibinfo{pages}{468}
  (\bibinfo{year}{1999}).

\bibitem[{\citenamefont{Warren et~al.}(2003)\citenamefont{Warren, Kobayashi,
  Lobkovsky, and Carter}}]{warren03}
\bibinfo{author}{\bibfnamefont{J.~A.} \bibnamefont{Warren}},
  \bibinfo{author}{\bibfnamefont{R.}~\bibnamefont{Kobayashi}},
  \bibinfo{author}{\bibfnamefont{A.~E.} \bibnamefont{Lobkovsky}},
  \bibnamefont{and} \bibinfo{author}{\bibfnamefont{W.~C.}
  \bibnamefont{Carter}}, \bibinfo{journal}{Acta. Mater.}
  \textbf{\bibinfo{volume}{51}}, \bibinfo{pages}{6035} (\bibinfo{year}{2003}).

\bibitem[{\citenamefont{Vvedensky}(2004)}]{VVED04}
\bibinfo{author}{\bibfnamefont{D.~D.} \bibnamefont{Vvedensky}},
  \bibinfo{journal}{J. Phys.: Condens. Matter} \textbf{\bibinfo{volume}{16}},
  \bibinfo{pages}{R1537} (\bibinfo{year}{2004}).

\bibitem[{\citenamefont{Provatas et~al.}(2005)\citenamefont{Provatas,
  Greenwood, Athreya, Goldenfeld, and Dantzig}}]{provatasreview2005}
\bibinfo{author}{\bibfnamefont{N.}~\bibnamefont{Provatas}},
  \bibinfo{author}{\bibfnamefont{M.}~\bibnamefont{Greenwood}},
  \bibinfo{author}{\bibfnamefont{B.~P.} \bibnamefont{Athreya}},
  \bibinfo{author}{\bibfnamefont{N.}~\bibnamefont{Goldenfeld}},
  \bibnamefont{and} \bibinfo{author}{\bibfnamefont{J.~A.}
  \bibnamefont{Dantzig}}, \bibinfo{journal}{Int. J. Mod. Phys. B}
  \textbf{\bibinfo{volume}{19}}, \bibinfo{pages}{4525} (\bibinfo{year}{2005}).

\bibitem[{\citenamefont{Provatas
  et~al.}(1998{\natexlab{a}})\citenamefont{Provatas, Goldenfeld, and
  Dantzig}}]{Provatas1998}
\bibinfo{author}{\bibfnamefont{N.}~\bibnamefont{Provatas}},
  \bibinfo{author}{\bibfnamefont{N.}~\bibnamefont{Goldenfeld}},
  \bibnamefont{and} \bibinfo{author}{\bibfnamefont{J.}~\bibnamefont{Dantzig}},
  \bibinfo{journal}{Phys. Rev. Lett.} \textbf{\bibinfo{volume}{80}},
  \bibinfo{pages}{3308} (\bibinfo{year}{1998}{\natexlab{a}}).

\bibitem[{\citenamefont{Jeong et~al.}(2001)\citenamefont{Jeong, Goldenfeld, and
  Dantzig}}]{Jeong2}
\bibinfo{author}{\bibfnamefont{J.}~\bibnamefont{Jeong}},
  \bibinfo{author}{\bibfnamefont{N.}~\bibnamefont{Goldenfeld}},
  \bibnamefont{and} \bibinfo{author}{\bibfnamefont{J.}~\bibnamefont{Dantzig}},
  \bibinfo{journal}{Phys. Rev. E} \textbf{\bibinfo{volume}{64}},
  \bibinfo{pages}{041602:1} (\bibinfo{year}{2001}).

\bibitem[{\citenamefont{Kobayashi et~al.}(1998)\citenamefont{Kobayashi, Warren,
  and Carter}}]{Kobayashi98}
\bibinfo{author}{\bibfnamefont{R.}~\bibnamefont{Kobayashi}},
  \bibinfo{author}{\bibfnamefont{J.~A.} \bibnamefont{Warren}},
  \bibnamefont{and} \bibinfo{author}{\bibfnamefont{W.~C.}
  \bibnamefont{Carter}}, \bibinfo{journal}{Physica D}
  \textbf{\bibinfo{volume}{119}}, \bibinfo{pages}{415} (\bibinfo{year}{1998}).

\bibitem[{\citenamefont{Kobayashi et~al.}(2000)\citenamefont{Kobayashi, Warren,
  and Carter}}]{Kobayashi00}
\bibinfo{author}{\bibfnamefont{R.}~\bibnamefont{Kobayashi}},
  \bibinfo{author}{\bibfnamefont{J.~A.} \bibnamefont{Warren}},
  \bibnamefont{and} \bibinfo{author}{\bibfnamefont{W.~C.}
  \bibnamefont{Carter}}, \bibinfo{journal}{Physica D}
  \textbf{\bibinfo{volume}{140}}, \bibinfo{pages}{141} (\bibinfo{year}{2000}).

\bibitem[{\citenamefont{Onuki}(1989{\natexlab{a}})}]{onuki89_1}
\bibinfo{author}{\bibfnamefont{A.}~\bibnamefont{Onuki}}, \bibinfo{journal}{J.
  Phys. Soc. Jpn.} \textbf{\bibinfo{volume}{58}}, \bibinfo{pages}{3065}
  (\bibinfo{year}{1989}{\natexlab{a}}).

\bibitem[{\citenamefont{Onuki}(1989{\natexlab{b}})}]{onuki89_2}
\bibinfo{author}{\bibfnamefont{A.}~\bibnamefont{Onuki}}, \bibinfo{journal}{J.
  Phys. Soc. Jpn.} \textbf{\bibinfo{volume}{58}}, \bibinfo{pages}{3069}
  (\bibinfo{year}{1989}{\natexlab{b}}).

\bibitem[{\citenamefont{Muller and Grant}(1999)}]{mg99}
\bibinfo{author}{\bibfnamefont{J.}~\bibnamefont{Muller}} \bibnamefont{and}
  \bibinfo{author}{\bibfnamefont{M.}~\bibnamefont{Grant}},
  \bibinfo{journal}{Phys. Rev. Lett.} p. \bibinfo{pages}{1736}
  (\bibinfo{year}{1999}).

\bibitem[{\citenamefont{Kassner et~al.}(2001)\citenamefont{Kassner, Misbah,
  Muller, Kappey, and Kohlert}}]{kmmkk01}
\bibinfo{author}{\bibfnamefont{K.}~\bibnamefont{Kassner}},
  \bibinfo{author}{\bibfnamefont{C.}~\bibnamefont{Misbah}},
  \bibinfo{author}{\bibfnamefont{J.}~\bibnamefont{Muller}},
  \bibinfo{author}{\bibfnamefont{J.}~\bibnamefont{Kappey}}, \bibnamefont{and}
  \bibinfo{author}{\bibfnamefont{P.}~\bibnamefont{Kohlert}},
  \bibinfo{journal}{Phys. Rev. E} p. \bibinfo{pages}{036117}
  (\bibinfo{year}{2001}).

\bibitem[{\citenamefont{Karma et~al.}(2001)\citenamefont{Karma, Kessler, and
  Levine}}]{karmafracture2001}
\bibinfo{author}{\bibfnamefont{A.}~\bibnamefont{Karma}},
  \bibinfo{author}{\bibfnamefont{D.~A.} \bibnamefont{Kessler}},
  \bibnamefont{and} \bibinfo{author}{\bibfnamefont{H.}~\bibnamefont{Levine}},
  \bibinfo{journal}{Phys. Rev. Lett.} \textbf{\bibinfo{volume}{87}},
  \bibinfo{pages}{045501} (\bibinfo{year}{2001}).

\bibitem[{\citenamefont{Haataja et~al.}(2005)\citenamefont{Haataja, Mahon,
  Provatas, and L\'{e}onard}}]{Haa05}
\bibinfo{author}{\bibfnamefont{M.}~\bibnamefont{Haataja}},
  \bibinfo{author}{\bibfnamefont{J.}~\bibnamefont{Mahon}},
  \bibinfo{author}{\bibfnamefont{N.}~\bibnamefont{Provatas}}, \bibnamefont{and}
  \bibinfo{author}{\bibfnamefont{F.}~\bibnamefont{L\'{e}onard}},
  \bibinfo{journal}{App. Phys. Lett.} \textbf{\bibinfo{volume}{87}},
  \bibinfo{pages}{251901} (\bibinfo{year}{2005}).

\bibitem[{\citenamefont{Karma}(2001)}]{Karma2001}
\bibinfo{author}{\bibfnamefont{A.}~\bibnamefont{Karma}},
  \bibinfo{journal}{Phys. Rev. Lett.} \textbf{\bibinfo{volume}{87}},
  \bibinfo{pages}{115701:1} (\bibinfo{year}{2001}).

\bibitem[{\citenamefont{Echebarria et~al.}(2004)\citenamefont{Echebarria,
  Folch, Karma, and Plapp}}]{Echebarria04}
\bibinfo{author}{\bibfnamefont{B.}~\bibnamefont{Echebarria}},
  \bibinfo{author}{\bibfnamefont{R.}~\bibnamefont{Folch}},
  \bibinfo{author}{\bibfnamefont{A.}~\bibnamefont{Karma}}, \bibnamefont{and}
  \bibinfo{author}{\bibfnamefont{M.}~\bibnamefont{Plapp}},
  \bibinfo{journal}{Phys. Rev. E} \textbf{\bibinfo{volume}{70}},
  \bibinfo{pages}{061604} (\bibinfo{year}{2004}).

\bibitem[{\citenamefont{Elder et~al.}(2002)\citenamefont{Elder, Katakowski,
  Haataja, and Grant}}]{ekhg02}
\bibinfo{author}{\bibfnamefont{K.~R.} \bibnamefont{Elder}},
  \bibinfo{author}{\bibfnamefont{M.}~\bibnamefont{Katakowski}},
  \bibinfo{author}{\bibfnamefont{M.}~\bibnamefont{Haataja}}, \bibnamefont{and}
  \bibinfo{author}{\bibfnamefont{M.}~\bibnamefont{Grant}},
  \bibinfo{journal}{Phys. Rev. Lett.} \textbf{\bibinfo{volume}{88}},
  \bibinfo{pages}{245701:1} (\bibinfo{year}{2002}).

\bibitem[{\citenamefont{Elder and Grant}(2004)}]{eg04}
\bibinfo{author}{\bibfnamefont{K.~R.} \bibnamefont{Elder}} \bibnamefont{and}
  \bibinfo{author}{\bibfnamefont{M.}~\bibnamefont{Grant}},
  \bibinfo{journal}{Phys. Rev. E} \textbf{\bibinfo{volume}{70}},
  \bibinfo{pages}{051605:1} (\bibinfo{year}{2004}).

\bibitem[{\citenamefont{Berry et~al.}(2006)\citenamefont{Berry, Grant, and
  Elder}}]{BGE05}
\bibinfo{author}{\bibfnamefont{J.}~\bibnamefont{Berry}},
  \bibinfo{author}{\bibfnamefont{M.}~\bibnamefont{Grant}}, \bibnamefont{and}
  \bibinfo{author}{\bibfnamefont{K.~R.} \bibnamefont{Elder}},
  \bibinfo{journal}{Phys. Rev. E} \textbf{\bibinfo{volume}{73}},
  \bibinfo{pages}{031609} (\bibinfo{year}{2006}).

\bibitem[{\citenamefont{Stefanovic et~al.}(2006)\citenamefont{Stefanovic,
  Haataja, and Provatas}}]{Stefanovic06}
\bibinfo{author}{\bibfnamefont{P.}~\bibnamefont{Stefanovic}},
  \bibinfo{author}{\bibfnamefont{M.}~\bibnamefont{Haataja}}, \bibnamefont{and}
  \bibinfo{author}{\bibfnamefont{N.}~\bibnamefont{Provatas}},
  \bibinfo{journal}{Phys. Rev. Lett.} \textbf{\bibinfo{volume}{96}},
  \bibinfo{pages}{225504} (\bibinfo{year}{2006}).

\bibitem[{\citenamefont{Elder et~al.}(2006)\citenamefont{Elder, Provatas,
  Barry, Stefanovic, and Grant}}]{Eld06}
\bibinfo{author}{\bibfnamefont{K.}~\bibnamefont{Elder}},
  \bibinfo{author}{\bibfnamefont{N.}~\bibnamefont{Provatas}},
  \bibinfo{author}{\bibfnamefont{J.}~\bibnamefont{Barry}},
  \bibinfo{author}{\bibfnamefont{P.}~\bibnamefont{Stefanovic}},
  \bibnamefont{and} \bibinfo{author}{\bibfnamefont{M.}~\bibnamefont{Grant}},
  \bibinfo{journal}{Phys. Rev. E.}  (\bibinfo{year}{2006}), \bibinfo{note}{in
  press}.

\bibitem[{\citenamefont{Goldenfeld et~al.}(2005)\citenamefont{Goldenfeld,
  Athreya, and Dantzig}}]{GAD05_1}
\bibinfo{author}{\bibfnamefont{N.}~\bibnamefont{Goldenfeld}},
  \bibinfo{author}{\bibfnamefont{B.~P.} \bibnamefont{Athreya}},
  \bibnamefont{and} \bibinfo{author}{\bibfnamefont{J.~A.}
  \bibnamefont{Dantzig}}, \bibinfo{journal}{Phys. Rev. E}
  \textbf{\bibinfo{volume}{72}}, \bibinfo{pages}{020601(R)}
  (\bibinfo{year}{2005}).

\bibitem[{\citenamefont{Goldenfeld et~al.}(2006)\citenamefont{Goldenfeld,
  Athreya, and Dantzig}}]{GAD05_2}
\bibinfo{author}{\bibfnamefont{N.}~\bibnamefont{Goldenfeld}},
  \bibinfo{author}{\bibfnamefont{B.~P.} \bibnamefont{Athreya}},
  \bibnamefont{and} \bibinfo{author}{\bibfnamefont{J.~A.}
  \bibnamefont{Dantzig}}, \bibinfo{journal}{J. Stat. Phys.}
  \textbf{\bibinfo{volume}{\uppercase{O}nline \uppercase{F}irst}}
  (\bibinfo{year}{2006}).

\bibitem[{\citenamefont{Chen et~al.}(1996)\citenamefont{Chen, Goldenfeld, and
  Oono}}]{CGO2}
\bibinfo{author}{\bibfnamefont{L.}~\bibnamefont{Chen}},
  \bibinfo{author}{\bibfnamefont{N.}~\bibnamefont{Goldenfeld}},
  \bibnamefont{and} \bibinfo{author}{\bibfnamefont{Y.}~\bibnamefont{Oono}},
  \bibinfo{journal}{Phys. Rev. E} \textbf{\bibinfo{volume}{54}},
  \bibinfo{pages}{376} (\bibinfo{year}{1996}).

\bibitem[{\citenamefont{Nozaki and Oono}(2001)}]{Nozaki01}
\bibinfo{author}{\bibfnamefont{K.}~\bibnamefont{Nozaki}} \bibnamefont{and}
  \bibinfo{author}{\bibfnamefont{Y.}~\bibnamefont{Oono}},
  \bibinfo{journal}{Phys. Rev. E} \textbf{\bibinfo{volume}{63}},
  \bibinfo{pages}{046101} (\bibinfo{year}{2001}).

\bibitem[{\citenamefont{Athreya et~al.}(2006)\citenamefont{Athreya, Goldenfeld,
  and Dantzig}}]{AGD06_1}
\bibinfo{author}{\bibfnamefont{B.~P.} \bibnamefont{Athreya}},
  \bibinfo{author}{\bibfnamefont{N.}~\bibnamefont{Goldenfeld}},
  \bibnamefont{and} \bibinfo{author}{\bibfnamefont{J.~A.}
  \bibnamefont{Dantzig}}, \bibinfo{journal}{Phys. Rev. E}
  \textbf{\bibinfo{volume}{74}}, \bibinfo{pages}{011601}
  (\bibinfo{year}{2006}).

\bibitem[{\citenamefont{Sethian}(1996)}]{sethian96}
\bibinfo{author}{\bibfnamefont{J.~A.} \bibnamefont{Sethian}},
  \bibinfo{journal}{Proc. Nat. Acad. Sci.} \textbf{\bibinfo{volume}{93}},
  \bibinfo{pages}{1591} (\bibinfo{year}{1996}).

\bibitem[{\citenamefont{Harris et~al.}(1998)\citenamefont{Harris, Singh, and
  King}}]{harris98}
\bibinfo{author}{\bibfnamefont{K.~E.} \bibnamefont{Harris}},
  \bibinfo{author}{\bibfnamefont{V.~V.} \bibnamefont{Singh}}, \bibnamefont{and}
  \bibinfo{author}{\bibfnamefont{A.~H.} \bibnamefont{King}},
  \bibinfo{journal}{Acta. Mater.} \textbf{\bibinfo{volume}{46}},
  \bibinfo{pages}{2623} (\bibinfo{year}{1998}).

\bibitem[{\citenamefont{Moldovan
  et~al.}(2002{\natexlab{a}})\citenamefont{Moldovan, Yamakov, Wolf, and
  Phillpot}}]{moldovan02}
\bibinfo{author}{\bibfnamefont{D.}~\bibnamefont{Moldovan}},
  \bibinfo{author}{\bibfnamefont{V.}~\bibnamefont{Yamakov}},
  \bibinfo{author}{\bibfnamefont{D.}~\bibnamefont{Wolf}}, \bibnamefont{and}
  \bibinfo{author}{\bibfnamefont{S.~R.} \bibnamefont{Phillpot}},
  \bibinfo{journal}{Phys. Rev. Lett.} \textbf{\bibinfo{volume}{89}},
  \bibinfo{pages}{206101} (\bibinfo{year}{2002}{\natexlab{a}}).

\bibitem[{\citenamefont{Moldovan
  et~al.}(2002{\natexlab{b}})\citenamefont{Moldovan, Wolf, Phillpot, and
  Haslam}}]{moldovan02_2}
\bibinfo{author}{\bibfnamefont{D.}~\bibnamefont{Moldovan}},
  \bibinfo{author}{\bibfnamefont{D.}~\bibnamefont{Wolf}},
  \bibinfo{author}{\bibfnamefont{S.~R.} \bibnamefont{Phillpot}},
  \bibnamefont{and} \bibinfo{author}{\bibfnamefont{A.~J.}
  \bibnamefont{Haslam}}, \bibinfo{journal}{Acta. Mater.}
  \textbf{\bibinfo{volume}{50}}, \bibinfo{pages}{3397}
  (\bibinfo{year}{2002}{\natexlab{b}}).

\bibitem[{\citenamefont{Moldovan
  et~al.}(2002{\natexlab{c}})\citenamefont{Moldovan, Wolf, Phillpot, and
  Haslam}}]{moldovan02_3}
\bibinfo{author}{\bibfnamefont{D.}~\bibnamefont{Moldovan}},
  \bibinfo{author}{\bibfnamefont{D.}~\bibnamefont{Wolf}},
  \bibinfo{author}{\bibfnamefont{S.~R.} \bibnamefont{Phillpot}},
  \bibnamefont{and} \bibinfo{author}{\bibfnamefont{A.~J.}
  \bibnamefont{Haslam}}, \bibinfo{journal}{Philos. Mag. A}
  \textbf{\bibinfo{volume}{82}}, \bibinfo{pages}{1271}
  (\bibinfo{year}{2002}{\natexlab{c}}).

\bibitem[{\citenamefont{Provatas
  et~al.}(1998{\natexlab{b}})\citenamefont{Provatas, Dantzig, and
  Goldenfeld}}]{Pro98a}
\bibinfo{author}{\bibfnamefont{N.}~\bibnamefont{Provatas}},
  \bibinfo{author}{\bibfnamefont{J.}~\bibnamefont{Dantzig}}, \bibnamefont{and}
  \bibinfo{author}{\bibfnamefont{N.}~\bibnamefont{Goldenfeld}},
  \bibinfo{journal}{Phys. Rev. Lett.} \textbf{\bibinfo{volume}{80}},
  \bibinfo{pages}{3308} (\bibinfo{year}{1998}{\natexlab{b}}).

\bibitem[{\citenamefont{Provatas et~al.}(1999)\citenamefont{Provatas, Dantzig,
  and Goldenfeld}}]{Pro99c}
\bibinfo{author}{\bibfnamefont{N.}~\bibnamefont{Provatas}},
  \bibinfo{author}{\bibfnamefont{J.}~\bibnamefont{Dantzig}}, \bibnamefont{and}
  \bibinfo{author}{\bibfnamefont{N.}~\bibnamefont{Goldenfeld}},
  \bibinfo{journal}{J. Comp. Phys.} \textbf{\bibinfo{volume}{148}},
  \bibinfo{pages}{265} (\bibinfo{year}{1999}).

\bibitem[{\citenamefont{Jeong et~al.}(2003)\citenamefont{Jeong, Dantzig, and
  Goldenfeld}}]{Jeong1}
\bibinfo{author}{\bibfnamefont{J.}~\bibnamefont{Jeong}},
  \bibinfo{author}{\bibfnamefont{J.~A.} \bibnamefont{Dantzig}},
  \bibnamefont{and}
  \bibinfo{author}{\bibfnamefont{N.}~\bibnamefont{Goldenfeld}},
  \bibinfo{journal}{Met. Trans. A} \textbf{\bibinfo{volume}{34}},
  \bibinfo{pages}{459} (\bibinfo{year}{2003}).

\bibitem[{\citenamefont{J.~Fan and Provatas}(2006)}]{Jun2006}
\bibinfo{author}{\bibfnamefont{M.~H.} \bibnamefont{J.~Fan},
  \bibfnamefont{M.~Greenwood}} \bibnamefont{and}
  \bibinfo{author}{\bibfnamefont{N.}~\bibnamefont{Provatas}},
  \bibinfo{journal}{Phys. Rev. E} \textbf{\bibinfo{volume}{74}},
  \bibinfo{pages}{031602} (\bibinfo{year}{2006}).

\bibitem[{\citenamefont{Zienkiewicz and Zhu}(1987)}]{Zei87}
\bibinfo{author}{\bibfnamefont{O.~C.} \bibnamefont{Zienkiewicz}}
  \bibnamefont{and} \bibinfo{author}{\bibfnamefont{J.~Z.} \bibnamefont{Zhu}},
  \bibinfo{journal}{Int. J. Num. Meth. Eng.} \textbf{\bibinfo{volume}{24}},
  \bibinfo{pages}{337} (\bibinfo{year}{1987}).

\bibitem[{\citenamefont{Berger and Oliger}(1984)}]{berger84}
\bibinfo{author}{\bibfnamefont{M.~J.} \bibnamefont{Berger}} \bibnamefont{and}
  \bibinfo{author}{\bibfnamefont{J.~E.} \bibnamefont{Oliger}},
  \bibinfo{journal}{J. Comp. Phys.} \textbf{\bibinfo{volume}{53}},
  \bibinfo{pages}{484} (\bibinfo{year}{1984}).

\bibitem[{\citenamefont{Tomita}(1991)}]{tomita91}
\bibinfo{author}{\bibfnamefont{H.}~\bibnamefont{Tomita}},
  \bibinfo{journal}{Prog. Theor. Phys.} \textbf{\bibinfo{volume}{85}},
  \bibinfo{pages}{47} (\bibinfo{year}{1991}).

\bibitem[{\citenamefont{Sethian and Strain}(1992)}]{sethian92}
\bibinfo{author}{\bibfnamefont{J.~A.} \bibnamefont{Sethian}} \bibnamefont{and}
  \bibinfo{author}{\bibfnamefont{J.}~\bibnamefont{Strain}},
  \bibinfo{journal}{J. Comp. Phys.} \textbf{\bibinfo{volume}{98}},
  \bibinfo{pages}{231} (\bibinfo{year}{1992}).

\bibitem[{\citenamefont{Callister}(1997)}]{callister}
\bibinfo{author}{\bibfnamefont{W.~D.} \bibnamefont{Callister}},
  \emph{\bibinfo{title}{Materials science and engineering}}
  (\bibinfo{publisher}{Wiley}, \bibinfo{year}{1997}).

\end{thebibliography}

\end{document}